\documentclass[12pt]{article}
\usepackage[utf8]{inputenc}
\usepackage[T1]{fontenc}
\usepackage{amsmath,amssymb,amsthm,commath,enumitem,mathtools}
\usepackage{bm}
\usepackage{kbordermatrix}
\usepackage[singlelinecheck=false]{caption}
\usepackage[margin=2.54cm]{geometry}
\usepackage[normalem]{ulem}
\usepackage{natbib}
\usepackage{graphicx}
\usepackage{booktabs}
\usepackage{color} 
\usepackage[hyphens]{url} 
\usepackage[onehalfspacing]{setspace}
\usepackage[small]{titlesec}
\usepackage{enumitem}
\usepackage[ruled,vlined]{algorithm2e}
\usepackage[tableposition=top,labelfont=bf]{caption}
\usepackage{layouts}
\usepackage{xcolor}
\usepackage{makecell}
\usepackage{tabularray}
\usepackage{microtype}
\graphicspath{ {./figures/} }
\usepackage{fancyhdr}

\allowdisplaybreaks
\let\originalleft\left
\let\originalright\right
\renewcommand{\left}{\mathopen{}\mathclose\bgroup\originalleft}
\renewcommand{\right}{\aftergroup\egroup\originalright}
\makeatletter
\newcommand*\bigcdot{\mathpalette\bigcdot@{.5}}
\newcommand*\bigcdot@[2]{\mathbin{\vcenter{\hbox{\scalebox{#2}{$\m@th#1\bullet$}}}}}
\makeatother

\title{%
   A three-state coupled Markov switching model for COVID-19 outbreaks across Quebec based on hospital admissions \\[5pt]
  }
\author{Dirk Douwes-Schultz\footnote{{{\it Corresponding author}: Dirk Douwes-Schultz, Department of Epidemiology, Biostatistics and Occupational Health, McGill University, 2001 McGill College Avenue, Suite 1200, Montreal, QC, Canada, H3A 1G1. {\it E-mail}: {\tt
				dirk.douwes-schultz@mail.mcgill.ca}.}} \hspace{1mm} , Alexandra M. Schmidt, Yannan Shen and \\ David Buckeridge \\
				\textit{Department of Epidemiology, Biostatistics and Occupational Health} \\ \textit{McGill University, Canada }}
				
\date{\today}

\begin{document}

\maketitle

\begin{abstract}
Recurrent COVID-19 outbreaks have placed immense strain on the hospital system in Quebec. We develop a Bayesian three-state coupled Markov switching model to analyze COVID-19 outbreaks across Quebec based on admissions in the 30 largest hospitals. Within each catchment area, we assume the existence of three states for the disease: absence, a new state meant to account for many zeroes in some of the smaller areas, endemic and outbreak. Then we assume the disease switches between the three states in each area through a series of coupled nonhomogeneous hidden Markov chains. Unlike previous approaches, the transition probabilities may depend on covariates and the occurrence of outbreaks in neighboring areas, to account for geographical outbreak spread. Additionally, to prevent rapid switching between endemic and outbreak periods we introduce clone states into the model which enforce minimum endemic and outbreak durations. We make some interesting findings, such as that mobility in retail and recreation venues had a positive association with the development and persistence of new COVID-19 outbreaks in Quebec. Based on model comparison our contributions show promise in improving state estimation retrospectively and in real-time, especially when there are smaller areas and highly spatially synchronized outbreaks. Furthermore, our approach offers new and interesting epidemiological interpretations{\color{black}, such as being able to estimate the effect of covariates on disease extinction.}

 \end{abstract}

\section{Introduction} \label{section:intro}

Quebec has been the epicenter of the COVID-19 epidemic in Canada with 15,389 deaths and 52,788 hospitalizations through May 2022 \citep{inspqCOVID19DataQuebec2022}. As a result, immense strain has been placed on Quebec's hospital system. Due to the high demand for hospital beds, many elective surgeries have been delayed or canceled which can adversely affect outcomes in patients with conditions other than COVID-19 \citep{shinglerOverwhelmedCOVIDQuebec2022a}. Additionally, there have been concerning shortages of essential medical supplies, drugs and staff \citep{laframboiseQuebecCoronavirusCases2020,legaultConferencePresseFrancois2020}.

\begin{figure}[!t]
 	\centering
 	\includegraphics[width=\linewidth]{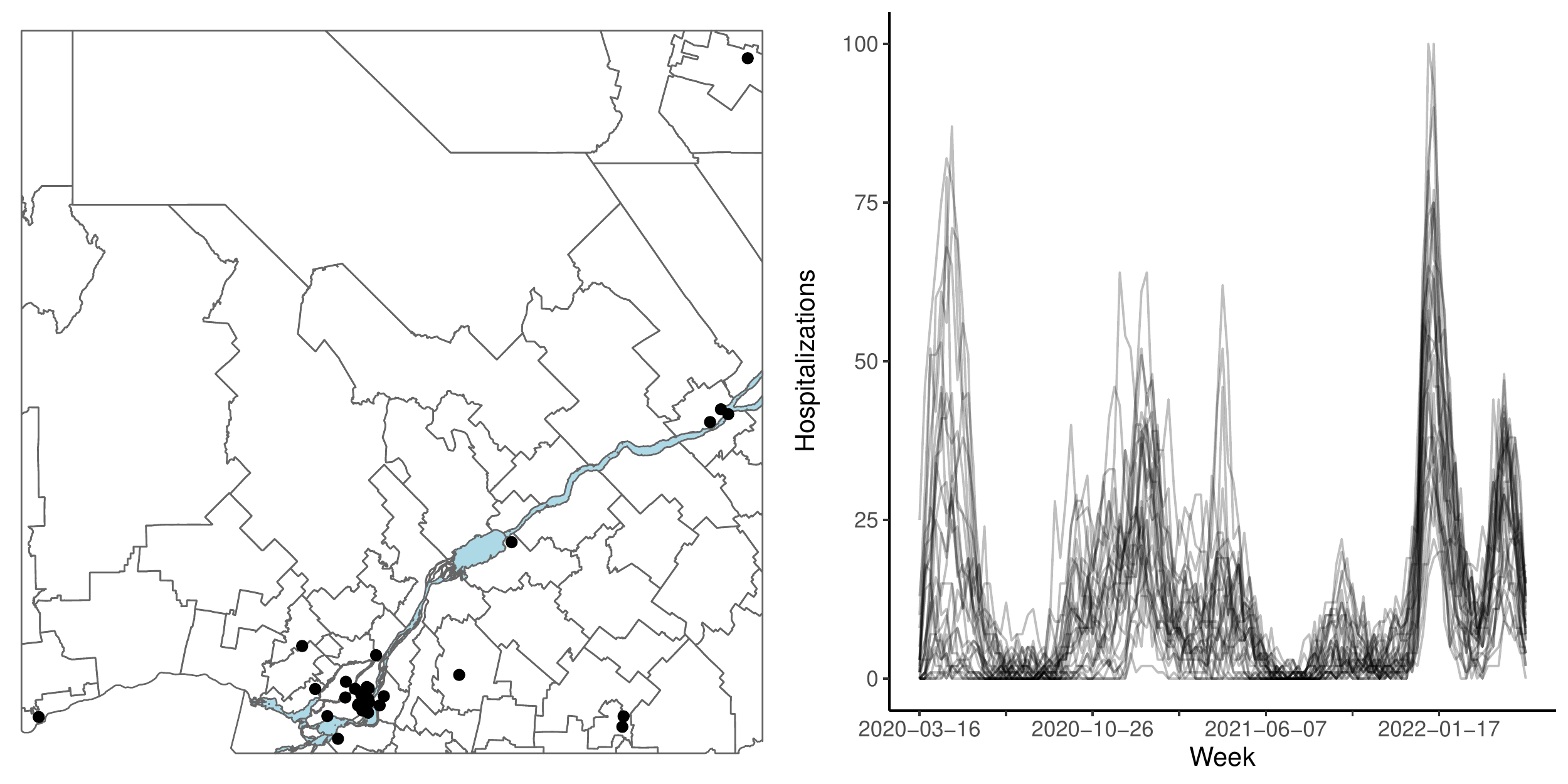}
	\caption{ The left graph shows a map of the part of Quebec where the 30 hospitals (solid circles) included in the study are located. Borders separate counties. Each line in the right graph gives the number of hospitalizations in one of the 30 hospitals included in the study. \label{fig:fig_intro}} 
\end{figure}

\begin{figure}[!t]
 	\centering
 	\includegraphics[width=.8\linewidth]{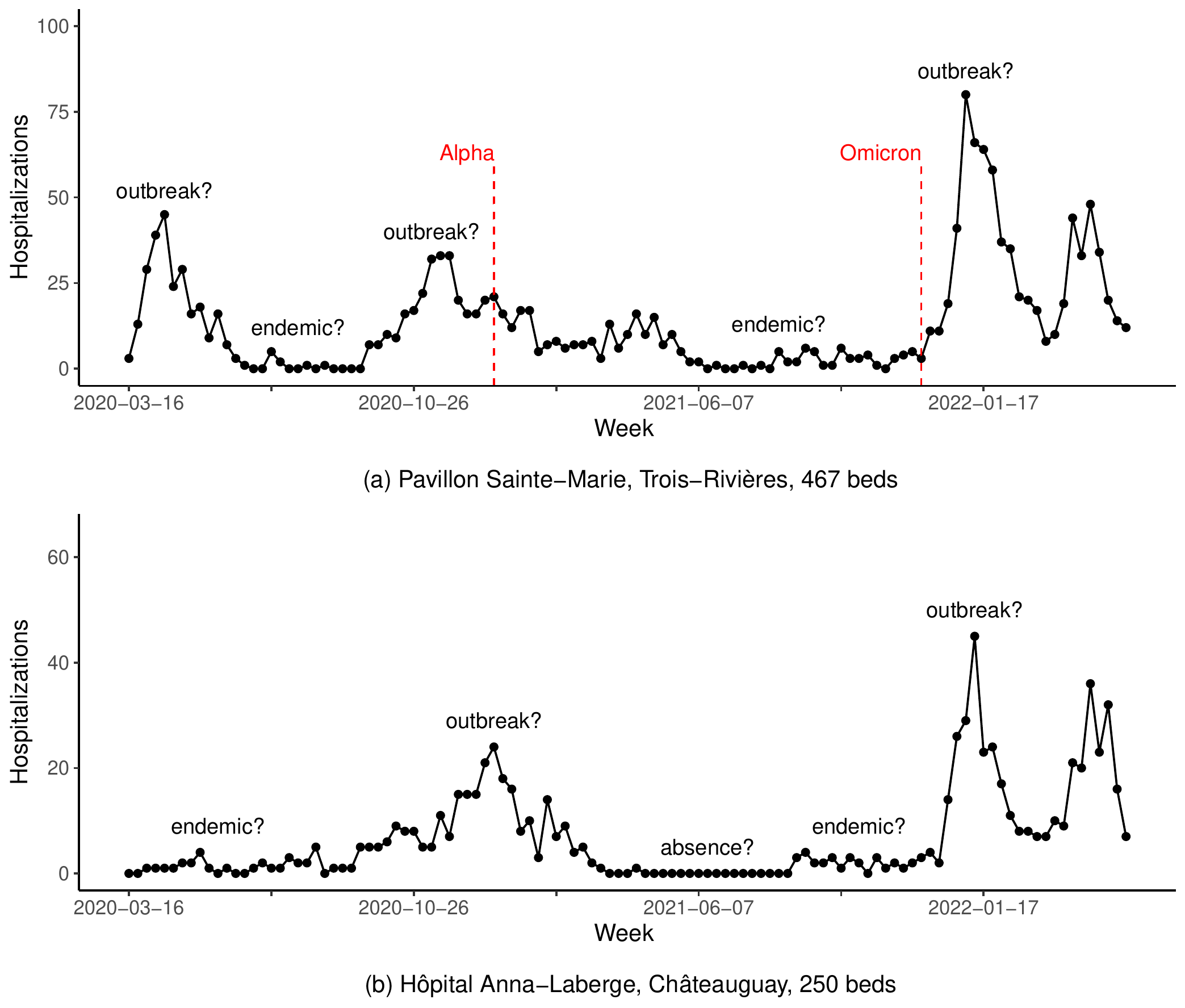}
	\caption{(a) An illustration of endemic and outbreak periods in the weekly COVID-19 hospitalizations for Pavillon Sainte-Marie in Trois--Rivières. (b) An illustration of absence, endemic and outbreak periods in the weekly COVID-19 hospitalizations for Hôpital Anna--Laberge in Châteauguay. The vertical {\color{black}red} lines are drawn at the introduction of the Alpha and Omicron variants for all of Quebec. The "?" reflects the fact that the periods are not fully observable in our framework. \label{fig:fig1}} 
\end{figure}

In this paper we analyze weekly COVID-19 hospital admissions in the 30 largest hospitals in Quebec, a map is given in Figure \ref{fig:fig_intro} along with the time series for each hospital. Figure \ref{fig:fig1}(a) shows the weekly COVID-19 hospitalizations for Pavillon Sainte-Marie in Trois--Rivières, an average-sized hospital among the 30. As illustrated in Figure \ref{fig:fig1}(a), and this can also be observed less clearly in Figure \ref{fig:fig_intro} across all hospitals, the hospital admissions can be characterized by a series of outbreak periods separated by quiescent endemic periods with low levels of hospitalizations. The outbreak periods represent COVID-19 outbreaks in the surrounding communities serviced by the hospital (catchment areas) as we do not include hospital-acquired infections. Clearly, most hospitalizations occur during the outbreak periods. However, this does not reflect the true burden of the outbreaks as the rapid rise in hospitalizations at the beginning of an outbreak can be difficult to adjust to, leading to shortages and a lack of beds \citep{shinglerOverwhelmedCOVIDQuebec2022a}. To gain a better understanding of how the outbreaks develop, a substantive aim of our analysis is to quantify how certain factors, such as mobility and the introduction of new variants (marked in Figure \ref{fig:fig1}(a), see Section \ref{section:spec} for more details), are associated with the emergence and persistence of the outbreaks. Additionally, we want to detect and forecast the outbreaks so that the hospitals can better plan the allocation of their medical resources.

{\color{black} Many methods have been proposed for modeling infectious disease counts in space and time, a subject that has gained increased attention in recent years partly due to the COVID-19 pandemic. Among the most popular are epidemiological compartmental models, such as susceptible-infectious-recovered (SIR) models \citep{keelingModelingInfectiousDiseases2007,bjornstad_epidemics_2023}. These approaches attempt to model individuals within the population moving through different disease compartments such as susceptible, infectious, recovered and quarantined \citep{crawfordImpactCloseInterpersonal2022}. Multivariate autoregressive count time series models, where the current expected counts are modeled as some function of past counts, are also popular \citep{ssentongo_pan-african_2021,bracher_endemic-epidemic_2022} and can be motivated from discrete-time SIR models \citep{bauerStratifiedSpaceTime2018,mizumoto_transmission_2020}. When fit with covariates, both of these approaches are often used to investigate associations between certain factors and some measure of disease transmission, e.g., the average number of new cases per each recent previous case \citep{bauerStratifiedSpaceTime2018,flaxman_estimating_2020,ssentongo_pan-african_2021}. However, they usually do not clearly distinguish between outbreak and non-outbreak periods. As mentioned above, we are interested in studying the transition between calm endemic periods and outbreak periods, and, therefore, we require a precise model definition of those transitions. Also, these methods are not well suited for outbreak detection and forecasting. While a compartmental model could forecast a likely increase in some epidemiological indicator, it cannot distinguish between an increase due to random endemic variation and an actual outbreak developing, which is an important problem in outbreak detection \citep{unkelStatisticalMethodsProspective2012}. Several other methods that are more focused on producing predictions, such as spline-based \citep{bauer_bayesian_2016} or machine learning \citep{rahimi_review_2021} methods, are also popular, however, they are often not appropriate for learning about the dynamics of the disease.}

{\color{black}In this paper} we take a Markov switching approach to modeling the occurrence of the outbreak periods. {\color{black}As we explain below, this approach allows us to study, detect and forecast transitions between endemic and outbreak periods.} Markov switching models assume a time series can be described by several submodels, usually called states or regimes, where switching between submodels is governed by a hidden first-order Markov chain \citep{hamiltonNewApproachEconomic1989}. Markov switching models that switch between endemic and outbreak states have a long history \citep{souza_forecasting_1982,amorosSpatiotemporalHierarchicalMarkov2020}. In this framework, the transitions between the endemic and outbreak states are usually modeled as a change in the parameters of an autoregressive process, i.e., a change in transmission intensity \citep{luProspectiveInfectiousDisease2010}. These transitions are assumed unobservable and must be inferred probabilistically from the time series, which is convenient as we usually do not observe exactly when the outbreaks start and end. The probabilities that govern the transitions between the endemic and outbreak states can depend on covariates \citep{dieboldRegimeSwitchingTimevarying1993}. This allows for investigating the association of various factors with the probabilities of outbreak emergence and persistence and can aid in forecasting the outbreaks \citep{nunesNowcastingInfluenzaEpidemics2013}. Additionally, Bayesian methods can be used to compute the posterior probability an outbreak is currently happening or will happen soon for the purpose of outbreak detection \citep{martinez-beneitoBayesianMarkovSwitching2008} and forecasting \citep{nunesNowcastingInfluenzaEpidemics2013}, respectively. Outbreak/endemic Markov switching models have become increasingly popular in epidemiology, especially for outbreak detection \citep{unkelStatisticalMethodsProspective2012}, with recent applications to influenza \citep{lytrasFluHMMSimpleFlexible2019}, cutaneous leishmaniasis \citep{rahmanianPredictingCutaneousLeishmaniasis2021} and salmonella \citep{zacherSupervisedLearningUsing2022}.

Despite their growing popularity, outbreak/endemic Markov switching models have not focused on the analysis of outbreaks in small areas, especially with many zeroes. Figure \ref{fig:fig1}(b) shows weekly COVID-19 hospitalizations for the Hôpital Anna--Laberge in Châteauguay, a relatively small hospital. As illustrated in Figure \ref{fig:fig1}(b), it may be more appropriate in smaller areas to describe switching between three periods: absence, endemic and outbreak. This has a strong epidemiological justification as it is well known that many infectious diseases frequently go extinct in small communities \citep{bartlettMeaslesPeriodicityCommunity1957,keelingModelingInfectiousDiseases2007}. Ignoring long periods of disease absence would end up likely assigning too many zeroes to the endemic state, biasing its mean towards zero. This is undesirable as it could lead to the outbreak state being too dominant at medium counts, leading to false alarms \citep{rathAutomatedDetectionInfluenza2003}. Another advantage of considering an absence state is that we can model the probabilities of disease extinction and reemergence, which are epidemiologically interesting \citep{douwes-schultzZerostateCoupledMarkov2022a}.

There has also not been much work done on spatio-temporal outbreak/endemic Markov switching models, where the focus is on the analysis of outbreaks in several different but connected areas across time. This is a focus of our analysis as many important decisions are made at the individual hospital level and clearly, there is some, though not perfect, synchronization of outbreaks between hospital catchment areas, compare Figures 
\ref{fig:fig1}(a) and \ref{fig:fig1}(b) for example. \cite{amorosSpatiotemporalHierarchicalMarkov2020} fit a spatio-temporal outbreak/endemic Markov switching model, but they only borrowed spatial strength in the observation component of the model. Individuals will mix between areas, causing outbreaks to spread geographically \citep{grenfellTravellingWavesSpatial2001a}. Therefore, an outbreak should be more likely to emerge or persist in an area if there are outbreaks likely occurring in neighboring areas. One way to achieve this is to use a coupled Markov switching model \citep{pohlePrimerCoupledStateswitching2021} where the states of neighbors are directly entered into the transition probabilities \citep{douwes-schultzZerostateCoupledMarkov2022a,touloupouScalableBayesianInference2020}. For outbreak detection and forecasting, considering evidence of outbreaks in neighboring areas could provide an early warning at the very beginning of an outbreak when there is still uncertain evidence within the area. \cite{heatonSpatiotemporalAbsorbingState2012} did let the probabilities of outbreak emergence depend on outbreaks in neighboring areas, however, they did not allow the transition probabilities to depend on covariates, and they used an absorbing state model, meaning it is difficult to apply their model to time series that contain multiple outbreaks. 

In our framework, we assume the disease switches between three states in each area: absence, endemic and outbreak. The epidemiological count in the absence state is assumed to always be 0, while it follows two autoregressive negative binomial processes in the outbreak and endemic states, distinguished by a lower level of transmission during the endemic period. Switching between the three states is governed by a first-order Markov chain where we focus on modeling the following four transition probabilities: absence to endemic (disease emergence), endemic to absence (disease extinction), endemic to outbreak (outbreak emergence) and outbreak to outbreak (outbreak persistence). Each transition probability can depend on covariates as well as a weighted sum of outbreak occurrence in neighboring areas. This allows us to investigate associations with important epidemiological transitions while also incorporating outbreak spread between areas. Nonhomogeneous Markov switching models can be sensitive to overfitting as the transition matrix may be non-persistent at some levels of the covariates, potentially leading to rapid switching between the states. Clearly, it is not realistic to rapidly switch between outbreak and endemic states, so we additionally introduce clone states \citep{kaufmannHiddenMarkovModels2018} into the model to enforce a minimum endemic and outbreak duration.

This paper is structured as follows. In Section \ref{section:model} we introduce our proposed model, a three-state coupled Markov switching model. In Section \ref{section:inferproc} we describe Bayesian inference using data augmentation and make use of the individual forward filtering backward sampling algorithm of \cite{touloupouScalableBayesianInference2020}.{ \color{black} In Section \ref{section:sim3} we evaluate outbreak detection and forecasting results from the model on simulated data where the exact start and end times of the outbreaks are known.} In Section \ref{section:motex} we apply the model to COVID-19 outbreaks across Quebec based on admissions in the 30 largest hospitals. We close with a general discussion in Section 
\ref{section:disc}.

\section{A Three-state Coupled Markov Switching Model} \label{section:model}

Let $y_{it}$ be an epidemiological count indicator, e.g., counts of hospitalizations, associated with area $i=1,\dots,N$ across $t=1,\dots,T$ time periods. Let $S_{it} \in \{1,2,3\}$ be an indicator for the epidemiological state of the disease, where $S_{it}=1$ if the disease is absent in area $i$ during time $t$, $S_{it}=2$ if the disease is in an endemic state and $S_{it}=3$ if the disease is in an outbreak state. 

We assume that the epidemiological count indicator is always 0 when the disease is absent and follows two distinct autoregressive negative binomial processes in the endemic and outbreak states, that is, 
\begin{align}
&y_{it} \mid S_{it},y_{i(t-1)}  \sim \begin{cases} 
        0, & \text{if $S_{it} =1$ (absence)}  \\[5pt]
     NB(\lambda_{it}^{EN},r^{EN}), &  \text{if $S_{it} =2$ (endemic)} \\[5pt]
     NB(\lambda_{it}^{OB},r^{OB}), &  \text{if $S_{it} =3$ (outbreak)},
   \end{cases} \label{eqn:y_spec} 
\end{align} where $\lambda_{it}^{EN}$ and $\lambda_{it}^{OB}$ are the means in the endemic and outbreak states respectively and $r^{EN}$ and $r^{OB}$ are the overdispersion parameters, so that, for example, $\text{Var}(y_{it}|S_{it}=3,y_{i(t-1)})=\lambda_{it}^{OB}(1+\lambda_{it}^{OB}/r^{OB})$. A zero count could be produced by all three states while a positive count can be produced by either the endemic or outbreak states, therefore, none of the states are observable and $S_{it}$ is a latent variable. Intuitively, we often do not know if an outbreak is occurring and there could be zeroes during an endemic period, or early outbreak period, due to a failure to detect the disease or a lack of severe cases for hospitalizations and deaths. {\color{black} As justification for the negative binomial distribution in (\ref{eqn:y_spec}), over a Poisson distribution, we found there was a high amount of overdispersion in the Quebec hospitalizations, see Table \ref{tab:Table 2}.}

As we are modeling an infectious disease we would expect the previous count $y_{i(t-1)}$ to affect the current expected count when the disease is present \citep{bauerStratifiedSpaceTime2018}. Therefore, we use log-linear autoregressive models \citep{liboschikTscountPackageAnalysis2017} for $\lambda_{it}^{EN}$ and $\lambda_{it}^{OB}$, 
\begin{align}
\begin{split}
\log(\lambda_{it}^{EN}) &= \beta_{0i}^{EN}+\bm{x}_{it}^T\bm{\beta}^{EN}+\rho^{EN}\log(y_{i(t-1)}+1) \\[5pt]
\log(\lambda_{it}^{OB}) &= \beta_{0i}^{OB}+\bm{x}_{it}^T\bm{\beta}^{OB}+\rho^{OB}\log(y_{i(t-1)}+1), \label{eqn:lambda}
\end{split}
\end{align} where $\beta_{0i}^{EN} \sim N\left(\beta_0^{EN},\left(\sigma^{EN}\right)^2\right)$ and $\beta_{0i}^{OB} \sim N\left(\beta_0^{OB},\left(\sigma^{OB}\right)^2\right)$ are random intercepts meant to account for between area differences and $\bm{x}_{it}$ is a vector of space-time covariates that may affect transmission of the disease within the endemic and outbreak periods. We allow the covariate effects, $\bm{\beta}^{EN}$ and $\bm{\beta}^{OB}$, to be different as outbreaks can lead to behavioral changes in hosts \citep{verelstBehaviouralChangeModels2016}. {\color{black} We also considered spatially correlated random effects \citep{amorosSpatiotemporalHierarchicalMarkov2020} for the area specific intercepts in (\ref{eqn:lambda}). However, we found for our motivating example that, possibly due to the amount of uncertainty about the underlying disease states, the spatial association of the intercepts could not be estimated precisely enough.}

To model the switching between absence, endemic and outbreak periods we assume that $S_{it}$ follows a three-state nonhomogeneous Markov chain within each area. In order to incorporate outbreak spread between areas we condition the transition matrix on $\bm{S}_{(-i)(t-1)}=(S_{1(t-1)},\dots,S_{(i-1)(t-1)},S_{(i+1)(t-1)},\dots,S_{N(t-1)})^T$, the vector of all state indicators excluding area $i$ at time $t-1$. We propose the following conditional transition matrix for the Markov chain, for $t=2,\dots,T$,
\begin{align}
\begin{split}
&\Gamma\left(S_{it}|\bm{S}_{(-i)(t-1)}\right) = \\
&\kbordermatrix{
\textbf{State}  & S_{it}=1 \, \textbf{(absence)} & &  S_{it}=2 \, \textbf{(endemic)} & &  S_{it}=3 \, \textbf{(outbreak)} \\[8pt]
  S_{i(t-1)}=1 \, \textbf{(absence)} & 1-p12_{it}& & p12_{it} & & 0 \\[5pt]
  S_{i(t-1)}=2 \, \textbf{(endemic)} & p21_{it} & & 1-p21_{it}-p23_{it} & & p23_{it} \\[5pt]
  S_{i(t-1)}=3 \, \textbf{(outbreak)} & 0 & & 1-p33_{it} & & p33_{it}
  }, \label{eqn:tm} 
\end{split}
\end{align} where $\Gamma\left(S_{it}|\bm{S}_{(-i)(t-1)}\right)_{lk}=P(S_{it}=k|S_{i(t-1)}=l,\bm{S}_{(-i)(t-1)})$ for $l,k=1,2,3$ and we have the following epidemiological interpretations of the transition probabilities, 
\begin{equation*}
  \begin{split}
    &p12_{it} = \text{ probability of disease emergence,}\\
    &p23_{it} = \text{ probability of outbreak emergence,}
  \end{split}
\quad
  \begin{split}
    &p21_{it} = \text{ probability of disease extinction,} \\ 
    &p33_{it} = \text{ probability of outbreak persistence.}
  \end{split}
\end{equation*} From (\ref{eqn:tm}), we assume it is not possible to move from absence to outbreak and vice versa in a single time step. For a time step of one week, like in our application, this is reasonable. However, for longer time steps this assumption should be examined.

We assume each transition probability in (\ref{eqn:tm}) can depend on a $p$-dimensional vector of space-time covariates $\bm{z}_{it}$ as well as a weighted sum of outbreak occurrence in neighboring areas during the previous time period, to model outbreak spread between areas. Starting with the probabilities of disease emergence, $p12_{it}$, and outbreak persistence, $p33_{it}$, we let
\begin{align}
\text{logit}(plk_{it})=\alpha_{lk,0}+\bm{z}_{it}^T\bm{\alpha}_{lk}+\alpha_{lk,p+1} \sum_{j \in NE(i)} \omega_{ji} I[S_{j(t-1)}=3], \label{eqn:logit}
\end{align} for $lk=12,33$, where $I[\bigcdot]$ is an indicator function, $NE(i)$ is the set of all neighboring areas of area $i$ and $\omega_{ji}$ is a fixed weight meant to reflect \emph{a priori} knowledge of the level of influence area $j$ has on area $i$. For example, one could use distance-based weights or weights based on trade between regions, see \cite{schrodleAssessingImpactMovement2012}. It is also possible to estimate connectivity in a coupled Markov switching model \citep{douwes-schultzZerostateCoupledMarkov2022a}. However, we do not consider this here due to the complexity and number of transitions in our model, the model used in \cite{douwes-schultzZerostateCoupledMarkov2022a} only had one general presence state and one absence state. For the probabilities of disease extinction, $p21_{it}$, and outbreak emergence, $p23_{it}$, a multinomial logistic regression is needed so that the second row of (\ref{eqn:tm}) sums to 1, that is,
\begin{align}
\log\left(\frac{plk_{it}}{1-p21_{it}-p23_{it}}\right) &=  \alpha_{lk,0}+\bm{z}_{it}^T\bm{\alpha}_{lk}+\alpha_{lk,p+1} \sum_{j \in NE(i)} \omega_{ji} I[S_{j(t-1)}=3], \label{eqn:multlogit}
\end{align} for $lk=21,23$. Therefore, $\bm{\alpha}_{21}$ and $\bm{\alpha}_{23}$ in (\ref{eqn:multlogit}) represent the effects of the covariates on the relative odds of transitioning to the absence and outbreak states respectively compared to remaining in the endemic state. 

Covariates will likely affect the transition probabilities in most cases. For example, dengue outbreaks are unlikely to occur during the winter when mosquito activity is low \citep{desclouxClimateBasedModelsUnderstanding2012b}, the introduction of new variants, such as Omicron, has likely played an important role in the development of new COVID-19 waves \citep{masloCharacteristicsOutcomesHospitalized2022} and disease extinction is less likely in smaller communities \citep{bartlettMeaslesPeriodicityCommunity1957}. {\color{black}Note it can often make sense to include the same covariate in both $\bm{x}_{it}$ and $\bm{z}_{it}$. For instance, as mentioned, population size should be related to disease extinction, however, it may also affect disease transmission during the endemic and outbreak periods as individuals are more likely to move through high-population areas \citep{grenfellTravellingWavesSpatial2001a}.} As for the inclusion of neighboring outbreak indicators in (\ref{eqn:logit})-(\ref{eqn:multlogit}), individuals will mix with those in other areas which can cause outbreaks to spread geographically, a phenomenon known as traveling waves \citep{grenfellTravellingWavesSpatial2001a}. Therefore, an outbreak in a neighboring area should affect the transition probabilities due to either direct spread or because it indicates spread from a common source, such as a large city. 

{\color{black}
\subsection{Identifiability constraints} \label{section:constraints}
Constraints are often placed on the parameters of a Bayesian mixture model to remove multimodality in the posterior distribution \citep{fruhwirth-schnatterFiniteMixtureMarkov2006}, which often occurs genuinely and due to the invariance of the likelihood function to permutations in the state labeling of the parameters \citep{jasraMarkovChainMonte2005}. The constraints are commonly chosen based on knowledge of the process being modeled \citep{martinez-beneitoBayesianMarkovSwitching2008,stonerAdvancedHiddenMarkov2020}. To help motivate constraints in our case, note that another way to express a log-linear model in (\ref{eqn:lambda}) is as, for example, $\lambda_{it}^{OB}=\exp(\beta_{0i}^{OB}+\bm{x}_{it}^T\bm{\beta}^{OB})\left(y_{i(t-1)}+1\right)^{ \rho^{OB}}$, which takes the form of a transmission rate multiplied by the previous counts. Disease transmission should always increase when moving from the endemic state to the outbreak state, especially if the model is to be used for issuing alarms for surveillance. Additionally, the autoregressive coefficients, $\rho^{EN}$ and $\rho^{OB}$, control the speed the disease moves through the local population \citep{minin_spatio-temporal_2019}, which should be higher during the outbreak periods. Therefore, we assume the following constraints,
\begin{align}
\begin{split}
&\beta_{0i}^{EN}+ \bm{x}_{it}^T\bm{\beta}^{EN}+.01 < \beta_{0i}^{OB} +\bm{x}_{it}^{T}\bm{\beta}^{OB}\text{ for $i=1,\dots,N$ and $t=2,\dots,T$, and,} \\[5pt]
&\rho^{EN} + .05 <  \rho^{OB}, \label{eqn:const}
\end{split}
\end{align} meaning we are assuming at least a one percent increase in transmission when moving from the endemic state to the outbreak state (in Section \ref{section:results} we show the results are not very sensitive to reasonably changing this one percent value). Initially, we had considered a simpler constraint on just the intercepts and the autoregressive coefficients, however, we found in a simulation study, in Section 2 of the supplementary materials (SM), that the simpler constraint does not ensure consistent convergence of our Markov chain Monte Carlo (MCMC) algorithm. Constraining the entirety of the transmission rates, as in (\ref{eqn:const}), greatly improved the convergence rate, did not introduce any significant bias into the inferential procedure and did not add a substantial amount of time to the model fitting.}

\subsection{Prior specification}

For the count part of the model, we specified wide independent normal and gamma priors for most lower-level elements of $\bm{\beta}$. We used $\text{Unif}(0,1)$ priors for $\rho^{EN}$ and $\rho^{OB}$ to meet the stability conditions of \cite{liboschikTscountPackageAnalysis2017} {\color{black}which ensures the counts are not expected to grow without bound.} For $r^{EN}$ and $r^{OB}$ we used $\text{Unif}(0,10)$ and $\text{Unif}(0,50)$ priors respectively. An upper limit of 10 was chosen for $r^{EN}$ as it is important for the endemic state distribution to have a long right tail to prevent frequent false alarms during outbreak detection \citep{rathAutomatedDetectionInfluenza2003}. To impose the constraints in (\ref{eqn:const}) we truncated the prior distribution of $\bm{\beta}$ \citep{kaufmannHiddenMarkovModels2018}. Some shrinkage to the null is generally recommended for logistic and multinomial logistic regression parameters to avoid separation issues and reduce bias away from 0 \citep{bullModifiedScoreFunction2002}. Therefore, following \cite{gelmanWeaklyInformativeDefault2008}, we used Cauchy priors with scale $2.5/2\cdot\text{sd}(z_{itq})$ for the effects of covariate $z_{itq}$ on the transition probabilities in (\ref{eqn:logit})-(\ref{eqn:multlogit}). For the effects of neighboring outbreaks on the transition probabilities, i.e., $\alpha_{lk,p+1}$ for $lk=12,21,23,33$ in (\ref{eqn:logit})-(\ref{eqn:multlogit}), we used more aggressive shrinkage and assigned $N(0,(.36/\text{max}\{\omega_{ji}\}_{\substack{j,i=1\\ j \neq i}}^{N})^2)$ priors. This states that with probability .95 \emph{a priori} we believe that an outbreak occurring in a single neighboring area should not more than double or less than halve the odds, or relative odds, of any epidemiological transition, e.g., should not more double the odds of an outbreak emerging relative to remaining in the endemic state \citep{wakefieldBayesianFrequentistRegression2013}. Our reasoning is that we do not want a single area to be given too much influence as we want the model to consider evidence of outbreaks across multiple neighboring areas. Additionally, while borrowing spatial strength can be important, we do not want it to overpower within-area information too strongly.

\subsection{Clone states} \label{section:clone}

As the transition probabilities in (\ref{eqn:tm}) depend both on covariates and latent neighboring states, the transition matrix may fluctuate between persistence and non-persistence. This could lead to time periods where there is rapid switching between the outbreak and endemic states, which is not realistic. \cite{heatonSpatiotemporalAbsorbingState2012} dealt with this by using an absorbing state model where the probability of outbreak persistence was fixed at 1, but that does not allow the analysis of multiple outbreaks or the study of outbreak persistence. Our solution is to introduce clone states into the model with determined transitions to enforce a minimum endemic and outbreak state duration, which is common in econometrics \citep{kaufmannHiddenMarkovModels2018}. We can introduce a new latent state indicator $S_{it}^{*} \in \{1,2,3,4,5,6,7\}$ such that $S_{it}=1$ if $S_{it}^*=1$, $S_{it}=2$ if $S_{it}^* \in \{2,3\}$ and $S_{it}=3$ if $S_{it}^* \in \{4,5,6,7\}$, and with the following conditional transition matrix, for $t=2,\dots,T$,
\begin{align}
\begin{split}
&\Gamma\left(S_{it}^*|\bm{S}_{(-i)(t-1)}\right) = \\
&\kbordermatrix{
\textbf{State} &S_{it}^*=1  & S_{it}^*=2 & S_{it}^*=3 &S_{it}^*=4&S_{it}^*=5&S_{it}^*=6&S_{it}^*=7   \\[8pt]
  S_{i(t-1)}^*=1, \, S_{i(t-1)}=1 \, \textbf{(absence)} & 1-p12_{it}&p12_{it} &0  &0 & 0 &0&0 \\[5pt]
  S_{i(t-1)}^*=2, \, S_{i(t-1)}=2 \, \textbf{(endemic)} &0& 0& 1 &0 & 0&0&0 \\[5pt]
   S_{i(t-1)}^*=3, \, S_{i(t-1)}=2 \, \textbf{(endemic)} & p21_{it} & 0 & 1-p21_{it}-p23_{it} & p23_{it} & 0 &0&0 \\[5pt]
   S_{i(t-1)}^*=4, \, S_{i(t-1)}=3 \, \textbf{(outbreak)} &0& 0& 0&0 & 1&0&0 \\[5pt]
   S_{i(t-1)}^*=5, \, S_{i(t-1)}=3 \, \textbf{(outbreak)} &0& 0& 0&0 & 0&1&0 \\[5pt]
   S_{i(t-1)}^*=6, \, S_{i(t-1)}=3 \, \textbf{(outbreak)} &0& 0& 0&0 & 0&0&1 \\[5pt]
   S_{i(t-1)}^*=7, \, S_{i(t-1)}=3 \, \textbf{(outbreak)} &0& 1-p33_{it}& 0&0 & 0&0&p33_{it}
  }, \label{eqn:tm_clone}
 \end{split}
\end{align} where $\Gamma\left(S_{it}^*|\bm{S}_{(-i)(t-1)}\right)_{lk}=P(S_{it}^*=k|S_{i(t-1)}^*=l,\bm{S}_{(-i)(t-1)})$ for $l,k=1,\dots,7$. The new transition matrix (\ref{eqn:tm_clone}) will prevent rapid switching between endemic and outbreak periods by imposing a minimum endemic duration of 2 weeks and a minimum outbreak duration of 4 weeks for our motivating example. A COVID-19 outbreak that lasts less than 4 weeks in Quebec is likely either a false alarm or too small to be concerning. {\color{black} Also, there should be at least some time between outbreaks for the susceptible population to replenish \citep{keelingModelingInfectiousDiseases2007}.} Clearly, the idea of (\ref{eqn:tm_clone}) could be used to impose any arbitrary minimum state durations, and so we assume (\ref{eqn:tm_clone}) throughout the rest of the paper w.l.o.g.. 

We will refer to the model defined by (\ref{eqn:y_spec})-(\ref{eqn:multlogit}), with (\ref{eqn:tm}) replaced by (\ref{eqn:tm_clone}), as the coupled Markov switching negative binomial model with 1 absence state, 2 endemic states and 4 outbreak states, i.e., the CMSNB(1,2,4) model. To finish model specification we also need to specify an initial state distribution for the Markov chain in each area, i.e., $p(S_{i1}^*)$ for $i=1,...,N$, which we assume does not depend on any unknown parameters. Note that moving from (\ref{eqn:tm}) to (\ref{eqn:tm_clone}) does not add any new parameters to the model, however, the restrictive transition matrix does slightly complicate the inferential procedure, which will now be discussed.

\section{Inferential Procedure} \label{section:inferproc}

Let $\bm{S}^*=(S_{11}^*,\dots,S_{1T}^*,\dots,S_{N1}^*,\dots,S_{NT}^*)^T$ be the vector of all state indicators, let $\bm{y}=(y_{11},\dots,y_{1T},\dots,y_{N1},\dots,y_{NT})^T$ be the vector of all counts, let $\bm{\beta}$ be the vector of all model parameters in the count part of the model, i.e., parameters in (\ref{eqn:y_spec})-(\ref{eqn:lambda}), let $\bm{\theta}$ be the vector of all model parameters in the Markov chain part of the model, i.e., parameters in (\ref{eqn:logit})-(\ref{eqn:multlogit}), and, finally, let $\bm{v}=(\bm{\beta},\bm{\theta})^T$ be the vector of all model parameters. Then the likelihood of $\bm{v}$ given $\bm{y}$ and $\bm{S}^*$ is given by, 
\begin{align}
\text{L}(\bm{y},\bm{S}^*|\bm{v}) = \prod_{i=1}^{N} \prod_{t=2}^{T} p(y_{it}|S_{it},y_{i(t-1)},\bm{\beta})\prod_{i=1}^{N} p(S_{i1}^*) \prod_{t=2}^{T} p(S_{it}^*|S_{i(t-1)}^*,\bm{S}_{(-i)(t-1)},\bm{\theta}). \label{eqn:like}
\end{align} Recall from the previous section that $\bm{S}^*$ is not observed. It is not possible to marginalize out $\bm{S}^*$ from (\ref{eqn:like}) as doing so would require matrix multiplication with a $7^N\times7^N$ matrix \citep{douwes-schultzZerostateCoupledMarkov2022a}. Additionally, we want to make inferences about $\bm{S}^{*}$ for the purpose of outbreak detection and forecasting as well as historical retrospection. Therefore, we estimate $\bm{S}^*$ along with $\bm{v}$ by sampling both from their joint posterior distribution which, from Bayes’ theorem, is proportional to,
\begin{align}
p(\bm{S}^*,\bm{v}|\bm{y}) \propto \text{L}(\bm{y},\bm{S}^*|\bm{v})p(\bm{v}), \label{eqn:post}
\end{align} where $p(\bm{v})$ is the prior distribution of $\bm{v}$.

As the joint posterior (\ref{eqn:post}) is not available in closed form, we resort to MCMC methods, in particular, we used a hybrid Gibbs sampling algorithm with some steps of the Metropolis–Hastings algorithm to sample from it. We sampled all elements of $\bm{v}$ without conjugate priors individually, using an adaptive random walk Metropolis step \citep{shabyExploringAdaptiveMetropolis2010}. It is easy to implement the sampling of each element of $\bm{S}^*$ one-at-a-time from $p(S_{it}^*|\bm{y},\bm{v},{\color{black}\{S_{jl}^*\}_{jl \neq it}})$ \citep{douwes-schultzZerostateCoupledMarkov2022a}. However, we found that one-at-a-time sampling mixed so slowly that it is not usable with our model. To see this, consider the following hypothetical state sequence for $S_{it}^*$, $456723$. Note it is not possible to sample any new single element of this sequence and in general we found that one-at-time sampling gets stuck in small regions of the parameter space. To avoid issues like this with one-at-a-time sampling when fitting Markov switching models, \cite{chibCalculatingPosteriorDistributions1996} proposed to sample all of $\bm{S}^*$ jointly from $p(\bm{S}^*|\bm{v},\bm{y})$. However, this is not possible with our model as it would involve matrix multiplication with $7^N\times7^N$ matrix \citep{douwes-schultzZerostateCoupledMarkov2022a}. As an alternative, we can block sample $\bm{S}^*$ with each block containing all the state indicators in a single location \citep{touloupouScalableBayesianInference2020}. Let $\bm{S}_i^*=(S_{i1}^*,\dots,S_{iT}^*)^T$ denote the vector of all state indicators in area $i$ and let $\bm{S}_{(-i)}^*$ be $\bm{S}^*$ with $\bm{S}_i^*$ removed. Then we can sample all of $\bm{S}_i^*$ jointly from its full conditional distribution, which is given by, 
\begin{align}
p(\bm{S}_i^*|\bm{v},\bm{S}_{(-i)}^*,\bm{y}) =p(S_{iT}^*|\bm{S}_{(-i)}^*,\bm{y},\bm{v}) \prod_{t=1}^{T-1}p(S_{it}^*|S_{i(t+1)}^*,\bm{S}_{(-i)(1:t+1)}^*,\bm{y}_{i(1:t)},\bm{v}),
\end{align} using an individual forward filter backward sampling (iFFBS) algorithm \citep{touloupouScalableBayesianInference2020}. More details are given in SM Section 1. It is also possible to block sample $\bm{S}^*$ in multi-location blocks \citep{douwes-schultzZerostateCoupledMarkov2022a} but we do not consider this here as it does not scale well with large transition matrices. 

Our hybrid Gibbs sampler was implemented using the R package Nimble \citep{valpineProgrammingModelsWriting2017}. Nimble comes with built-in Metropolis–Hastings and categorical (equivalent to one-at-a-time sampling) samplers. The iFFBS samplers were implemented using Nimble’s custom sampler feature. All Nimble R code, including for the custom iFFBS samplers, are provided on GitHub (\url{https://github.com/Dirk-Douwes-Schultz/CMSNB124_code}). Nimble was chosen as it is extremely fast (C++ compiled) and only requires the coding of new samplers. In the SM Section 2, we provide a simulation study, which shows that our proposed Gibbs sampler can recover the true parameters of a CMSNB(1,2,4) model that is specified like in our motivating example in Section 
\ref{section:motex}.

\subsection{Outbreak detection, forecasting and historical retrospection} \label{section:outdet}

Once a sample from the joint posterior (\ref{eqn:post}) has been obtained, the posterior probability that the disease is currently in epidemiological state $s$, for $s=1$ (absence), $s=2$ (endemic) and $s=3$ (outbreak), in area $i$ can be approximated as $P(S_{iT}=s|\bm{y}) \approx \frac{1}{Q-M} \sum_{m=M+1}^{Q}I[S_{iT}^{[m]}=s]$, where the superscript $[m]$ denotes a draw from the posterior distribution of the variable, $M$ is the size of the burn-in sample and $Q$ is the total MCMC sample size. The posterior probability $P(S_{iT}=3|\bm{y})$ represents the probability that an outbreak is currently happening in area $i$ given all observed data and can be used for the purpose of outbreak detection \citep{martinez-beneitoBayesianMarkovSwitching2008}. We recommend using $P(S_{iT}=3|\bm{y})$ along with other, external, pieces of information to help determine if an outbreak is likely occurring. While $P(S_{iT}=3|\bm{y})$ has no closed form, we can condition on the parameters and previous states to gain some intuition into how the model performs outbreak detection, 
\begin{align}
\begin{split}
p(S_{iT}|S_{i(T-1)}^*,\bm{S}_{(-i)(T-1)},\bm{v},\bm{y}) &\propto  \\ 
&p(y_{iT}|S_{iT},y_{i(T-1)},\bm{\beta})p(S_{iT}|S_{i(T-1)}^*,\bm{S}_{(-i)(T-1)},\bm{\theta}). \label{eqn:detectcond}
\end{split}
\end{align} From (\ref{eqn:detectcond}), the model generally weighs two factors for outbreak detection, the relative likelihood that the outbreak state generated the observed count and the probability of entering the outbreak state at the current time $P(S_{iT}=3|S_{i(T-1)}^*,\bm{S}_{(-i)(T-1)},\bm{\theta})$, which recall may depend on outbreaks in neighboring areas and covariates, see (\ref{eqn:logit})-(\ref{eqn:multlogit}) and (\ref{eqn:tm_clone}).

As for forecasting, we used a simulation procedure to draw realizations from the posterior predictive distributions \citep{fruhwirth-schnatterFiniteMixtureMarkov2006}. Algorithm 1 in the SM will obtain realizations from the posterior predictive distribution of the counts, $y^{[m]}_{i(T+k)} \sim p(y_{i(T+k)}|\bm{y})$, and the epidemiological state of the disease, $S^{[m]}_{i(T+k)} \sim p(S_{i(T+k)}|\bm{y})$, for $i = 1, \dots , N$, $k = 1, \dots , K$ and $m = M + 1, \dots , Q$. Then, the posterior probability that the disease will be in state $s$, $k$ time steps from now, in area $i$, can be approximated as $P(S_{i(T+k)}=s|\bm{y}) \approx \frac{1}{Q-M} \sum_{m=M+1}^{Q}I[S_{i(T+k)}^{[m]}=s]$. 

Finally, it is also important to examine the posterior probability that the disease was in state $s$ during past time periods $P(S_{it}=s|\bm{y}) \approx \frac{1}{Q-M} \sum_{m=M+1}^{Q}I[S_{it}^{[m]}=s]$ for $t=1,\dots,T-1$, $i=1,\dots,N$ and $s=1$ (absence), $s=2$ (endemic) and $s=3$ (outbreak) for several reasons. Firstly, it may be of historical interest to investigate the epidemiological history of the disease in various areas. Also, the estimation of $\bm{v}$ depends heavily on the estimates of the past states as indicated by the joint likelihood function (\ref{eqn:like}). If the classification of past epidemiological periods is not sensible, this would cast doubt on the estimates of $\bm{v}$ and may point to model misspecification.

{\color{black}
\subsection{Model comparison} \label{section:modelcompare}
To compare competing models, we use the widely applicable information criterion (WAIC) \citep{gelman_understanding_2014}. For state-space models, the WAIC is more accurate when the latent states are marginalized \citep{auger-methe_guide_2021}. For models without neighboring outbreak indicators in the transition probabilities, such as the Non-coupled Model in Section \ref{section:motex}, we can use the forward filter \citep{fruhwirth-schnatterFiniteMixtureMarkov2006} to calculate the marginal density $p(y_{it}|\bm{y}_{1:(t-1)},\bm{v})$, where $\bm{y}_{1:(t-1)}=(y_{11},\dots,y_{1(t-1)},\dots,y_{N1},\dots,y_{N(t-1)})^T$, and use it to calculate the WAIC. However, for models with neighboring outbreak indicators in the transition probabilities it is not computationally possible to completely marginalize $\bm{S}^{*}$ \citep{douwes-schultzZerostateCoupledMarkov2022a} and, therefore, we use the partially marginalized density $p(y_{it}|\bm{S}_{(-i)(1:t)}^*,\bm{y}_{1:(t-1)},\bm{v})$ instead. See SM Section 3.1 for more details. The model with the lowest WAIC is considered to have the best fit and as a rule of thumb, a difference of 5 or more in the WAIC is considered significant. A simulation study we conducted, in SM Section 3.2, suggests that when there is no significant difference in the WAIC to prefer the less complex model and when there is a significant difference the WAIC almost always chooses the correct model, between a spatial and non-spatial Markov switching model.

}

{
\section{\color{black}Simulation Study} \label{section:sim3}
 \color{black} In Sections \ref{section:retro} and \ref{section:real_time} below we attempt to evaluate the retrospective and real-time state estimates of a CMSNB(1,2,4) model (see Section \ref{section:outdet}) fitted to the Quebec hospitalizations. In this case, it is difficult to exactly quantify the accuracy of the state estimates as the true underlying states are not known, and we can only evaluate the estimates visually. One solution is to design a simulation study where the exact start and end times of the outbreaks are pre-determined, although this does have the disadvantage of using simulated and not real data \citep{buckeridge_outbreak_2007}. 
  
We designed a simulation study with 30 areas divided into 5 clusters of 6 areas each. Each area alternated between a 15-week endemic period and a 15-week outbreak period for a total of 4 of each period, 120 weeks total. For every area in a cluster, the start of each outbreak was randomized to occur within the first 4 weeks of the corresponding outbreak period. Therefore, we did not explicitly model outbreak spread within a cluster, we just assumed the outbreaks occurred around the same time. We also assumed a 40\% chance, taken roughly from our results in Section \ref{section:results}, that each endemic period contains a 7-week absence period inserted into the middle. The counts were simulated from a negative binomial distribution with overdispersion 10 and conditional means $e^{.1\text{beds}_i}(y_{i(t-1)}+1)^{.5}$ (endemic), $e^{.75+.05\text{beds}_i}(y_{i(t-1)}+1)^{.75}$ (outbreak) and $y_{it}=0$ (absence). The covariate $\text{beds}_i$ was taken from our real data example to produce some between-area heterogeneity in the transmission. These values are somewhat similar to those from Section \ref{section:results} and produced realistic looking simulations, SM Figure 4 shows some of the simulated time series. We only generated one simulation as each model must be fit 20 times to produce the needed state estimates, see below.

\begin{figure}[t]
 	\centering
 	\includegraphics[width=.7\linewidth]{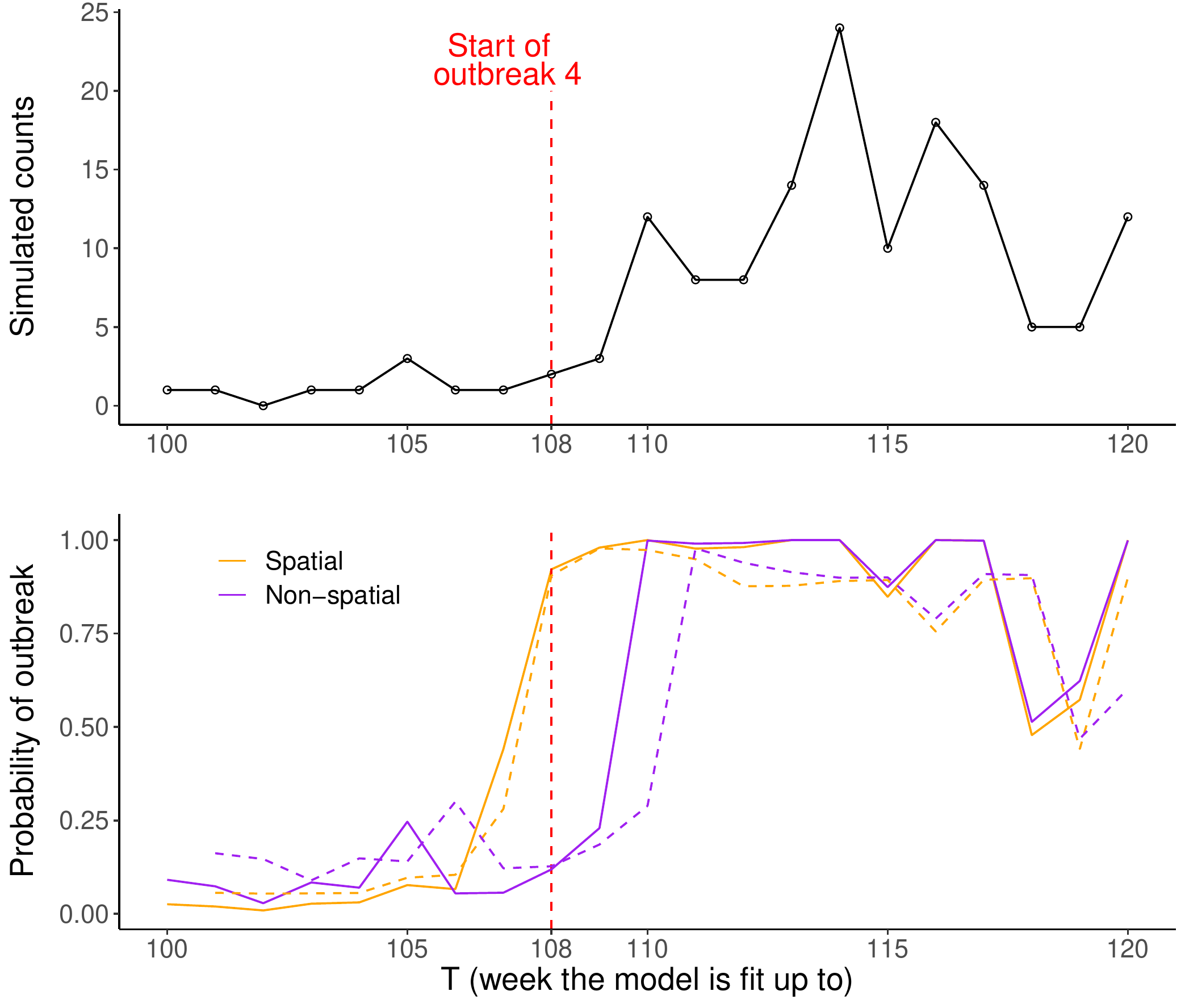}
	\caption{ \color{black} Top graph shows simulated counts from the last 20 weeks in one of the areas from the simulation study. Bottom graph solid lines show the real-time outbreak detection probabilities, $P(S_{iT}=3|\bm{y})$, versus $T$. Bottom graph dashed lines show the one-week ahead outbreak forecasts from the previous week, $P(S_{iT}=3|\bm{y}_{1:(T-1)})$, versus $T$. The {\color{black}red dashed vertical} line indicates the exact pre-determined start time of outbreak 4 in the area. The Spatial Model is in {\color{black}orange} and the Non-spatial Model is in {\color{black}purple}.} \label{fig:sim3_main}
\end{figure}

To fit to the simulated data, we considered a CMSNB(1,2,4) model with the count part correctly specified, including no random intercepts, and $\sum_{j \in NE(i)}I[S_{j(t-1)}=3]$ included only in the relative odds of outbreak emergence, where $NE(i)$ contained all areas in the same cluster as area $i$. We will call this the Spatial Model, and for comparison purposes, we also considered a homogeneous non-spatial model without $\sum_{j \in NE(i)}I[S_{j(t-1)}=3]$ in the transition probability. We focused on outbreak emergence in the simulation study, as it is the most important transition to capture in a surveillance setting. We evaluated three outbreak state estimates from the models: (1) retrospective probabilities, $P(S_{it}=3|\bm{y})$ from fitting the models to the full simulated data set (2) real-time detection probabilities, $P(S_{iT}=3|\bm{y})$ from fitting the models up to time $T$ for $T=100,\dots,120$ (20 separate fits) and (3) real-time one week ahead forecasts, $P(S_{iT}=3|\bm{y}_{1:(T-1)})$ from fitting the models up to time $T-1$ for $T=101,\dots,120$. We only evaluated the real-time state estimates on the last outbreak in each area to ensure stable model fitting. We evaluated the outbreak state estimates using the area under the ROC curve (AUC), sensitivity, specificity and timeliness \citep{buckeridge_outbreak_2007}, the last three calculated with a 50\% threshold. Timeliness for a single outbreak is simply the number of weeks into the outbreak, starting at one, when the outbreak state estimate first rises above 50\%, and the overall timeliness is the average timeliness across all outbreaks.

The results are summarized in Table 4 of the SM. The Spatial Model was able to recover the true underlying outbreak states well retrospectively, AUC of .995, and during outbreak detection, AUC of .983, and one week ahead forecasting, AUC of .965. The Spatial Model improved, over the Non-spatial Model, every criterion for each type of outbreak state estimate, especially for real-time outbreak detection and forecasting. For instance, the Spatial Model improved the sensitivity of the real-time outbreak detection probabilities by 6\% and reduced the timeliness of the one-week ahead outbreak forecasts by one week. Additionally, the Spatial Model showed a fair tradeoff between outbreak detection and forecasting, with the outbreak forecasts having two percent lower sensitivity but an improvement in timeliness of .43 weeks. In contrast, the forecasts from the Non-spatial Model reduced the sensitivity by 6.7\% and only improved the timeliness by .06 weeks. This suggests outbreak forecasting should only be attempted by nonhomogeneous models. These points are further illustrated in Figure~\ref{fig:sim3_main} which compares the real-time outbreak detection probabilities and the one-week ahead outbreak forecasts in one of the areas from the simulation study.
}

\section{Application to COVID-19 Outbreaks Across Quebec} \label{section:motex}

\subsection{Model specification and fitting} \label{section:spec}

We fitted a CMSNB(1,2,4) model to weekly COVID-19 hospitalizations across the 30 largest hospitals in Quebec ($i=1,\dots,N=30$), in terms of overall COVID-19 admissions, between  March 16th 2020 and May 9th 2022 ($t=1,\dots,T=113$). Hospitalizations typically lag infections by 1--2 weeks, \citep{wardUnderstandingEvolvingPandemic2021} and there are a few days of reporting delay. We did not include hospital-acquired infections so that the outbreak periods represent outbreaks in the catchment areas of the hospitals. A hospital's catchment area is typically difficult to exactly determine and can stretch disjointly across a broad geographical region \citep{gilmourIdentificationHospitalCatchment2010}. By examining the locations of a sample of patients from each hospital, we found that the catchment areas are mostly contained within, and sometimes adjacent to, the same county where the hospital is located. We assumed that the number of beds in a hospital is a reasonable proxy for the population size of the catchment area.

Increased mobility has been a major concern of the Quebec government during the epidemic due to a fear that it will lead to increased COVID-19 transmission, potentially overwhelming hospitals \citep{Wiltonclosures2020}. In general, epidemiologists have long theorized that mobility plays an important role in the development of recurring outbreaks \citep{soperInterpretationPeriodicityDisease1929}. While quantifying mobility is challenging, Google mobility metrics \citep{COVID19CommunityMobility} tend to give a reasonable approximation to more accurate, but only privately available, mobility measures based on close contact rates \citep{crawfordImpactCloseInterpersonal2022}. \cite{COVID19CommunityMobility} provides daily metrics that measure mobility as a percent change in the number of visits to certain venues, based on personal electronic devices, from days of the same weekday in January 2020. We used the retail+recreation mobility metric which measures changes in visits to places like restaurants, cafes, shopping centers, museums, libraries, and movie theaters as such venues have been heavily targeted by the Quebec government \citep{MeasuresAdoptedOrders2022}. In Quebec, Google mobility metrics are available at the county level, although the metrics are missing for around 10 percent of counties. Another major concern has been the introduction of new COVID-19 variants, especially the Alpha and Omicron variants \citep{Olivierlockdown2021,Stevensondshutschool2021}. New variants of COVID-19 can be more contagious and more resistant to existing vaccines compared to previous variants, although they can also be less deadly \citep{chenchulaCurrentEvidenceEfficacy2022}. 

Based on the above discussion, we considered ${\color{black}\bm{x}_{it}= \color{black} \bm{z}_{it}=(\text{beds}_{i},\text{mobility}_{\text{county}(i)(t-4)},} \\ \text{new\_variant}_t)^T$ as covariates possibly associated with {\color{black}transmission during the endemic and outbreak periods in Equation (\ref{eqn:lambda}) and} disease emergence, disease extinction, outbreak emergence and outbreak persistence in Equations (\ref{eqn:logit})-(\ref{eqn:multlogit}). Here, $\text{beds}_{i}$ is the number of beds in hospital $i$ and $\text{mobility}_{\text{county}(i)(t-4)}$ is the retail+recreation Google mobility metric for the county where hospital $i$ is located averaged across week $t-4$ (recall it is a daily metric). We lagged the mobility metrics by one month to account for delays between infection, hospitalization and reporting, as well as the time needed for the effects of a spike in mobility to trickle down into the general population. For counties with missing mobility metrics, we substituted the metrics from the nearest county with a similar urban makeup. The binary covariate $\text{new\_variant}_t$ was 1 if 3 months after the introduction of the Alpha variant in Quebec, introduced December 29, 2020, and the Omicron variant, introduced November 29, 2021, and 0 otherwise \citep{UpdatesCOVID19Variants2022}. The introduction of the variants are marked in Figure \ref{fig:fig1}(a) and we will in general refer to the outbreak that occurs in most hospitals around the introduction of Omicron as the Omicron outbreak. We included $\text{new\_variant}_t$ only in the relative odds of outbreak emergence {\color{black}and in the outbreak transmission rate} to capture the short-term effects of the introduction of a new variant on outbreak emergence {\color{black}and transmission}. We combined the Omicron and Alpha variants since there is likely not enough information in the data to estimate the effects of each separately. Additionally, having separate effects for the two variants can make it challenging to use the model in real-time, since we may not know if a future variant is more like Alpha or Omicron. Note, we did not have county-level information on the introduction of the new variants so $\text{new\_variant}_t$ is the same for all hospitals.

{\color{black}From a sample of patients from hospitals $i$ and $j$ grouped into $l=1,\dots,91$ neighborhoods, which divide Quebec, we calculated the spatial weight $\omega_{ji}$, in Equations (\ref{eqn:logit})-(\ref{eqn:multlogit}), as $$\omega_{ji}=\sum_{l=1}^{91}\sqrt{p_{jl}*p_{il}},$$ where, for example, $p_{jl}$ is the proportion of sampled patients from hospital $j$ who lived in neighborhood $l$, which is known as the Bhattacharyya coefficient \citep{biRoleBhattacharyyaDistance2019}.} The Bhattacharyya coefficient must be between 0 and 1 and is a way to measure the amount of overlap between two categorical samples. Therefore, hospitals with more overlap in their catchment areas were given a larger weight. We took the 5 nearest neighbors of hospital $i$, in terms of the largest weights, to form the neighborhood set $NE(i)$. {\color{black}Note $\omega_{ji}$ can be seen as a distance-based decay weight as $\omega_{ji}=\exp(-d_{ji}^{BD})$ where $d_{ji}^{BD}$ is the Bhattacharyya distance \citep{biRoleBhattacharyyaDistance2019}}. To finish model specification, we used a uniform initial state distribution in each area. 

We fit the CMSNB(1,2,4) model specified above to the Quebec hospitalization data using our proposed hybrid Gibbs sampler from Section \ref{section:inferproc}. We ran the Gibbs sampler for 200,000 iterations on three chains with an initial burn-in of 50,000 iterations. All sampling was started from random values in the parameter space to avoid convergence to local modes. Convergence was checked using the Gelman–Rubin statistic (all estimated parameters<1.05), the minimum effective sample size (>1000) and by visually examining the traceplots \citep{plummerCODAConvergenceDiagnosis2006}.

For comparison purposes we also fit a model without neighboring outbreak indicators in the transition probabilities, that is, with $\alpha_{lk,p+1}=0$ for $lk=12,21,23,33$ in (\ref{eqn:logit})-(\ref{eqn:multlogit}), which we will refer to as the Non-coupled Model. Additionally, we fit a two-state model without any absence or clone states, which we will refer to as the No Absence/Clone State Model. We chose these two models for comparison to examine the effects of our main contributions. We will sometimes refer to our CMSNB(1,2,4) model as the Full Coupled Model during comparison. {\color{black} Finally, we also compared to an endemic-epidemic (EE) model \citep{bracher_endemic-epidemic_2022}, which lies outside our Markov switching framework. The EE model is a state-of-the-art multivariate autoregressive count time series model which is popular for modeling spatio-temporal infectious disease counts \citep{bauerStratifiedSpaceTime2018,ssentongo_pan-african_2021} and is commonly used as a benchmark \citep{bauer_bayesian_2016,stojanovic_bayesian_2019}. The EE model is usually used for forecasting and for investigating associations between covariates and disease transmission \citep{ssentongo_pan-african_2021}. We compared to the latest iteration of the EE model \citep{bracher_endemic-epidemic_2022} which allows for multiple temporal lags, whose order we chose using the WAIC. More details, including the parameter estimates, are given in SM Section 4.

Table \ref{tab:WAICMM} shows the WAIC for the 4 considered models. The Full Coupled Model has the lowest WAIC which lends support to the inclusion of neighboring outbreak indicators in the transition probabilities and the addition of the absence/clone states. The Markov switching models all have a lower WAIC compared to the Endemic-epidemic model, however, we would argue the main advantage of the Markov switching models is that they offer more interesting and richer interpretations. For instance, the Endemic-epidemic model concludes that mobility had a strong positive association with overall disease transmission, see SM Table 3. From Section~\ref{section:results} below, the CMSNB(1,2,4) model broadly agrees with this conclusion but provides a deeper understanding of the effect; mobility mainly affects disease transmission during the outbreak and not endemic periods and mobility increases the risk of outbreaks emerging and persisting and reduces the chances of the disease going extinct in smaller areas. These richer covariate interpretations could help policymakers better identify useful interventions. Additionally, the Markov switching models can be utilized for retrospective and real-time state estimation/forecasting, as we illustrate in Sections \ref{section:retro} and \ref{section:real_time}.

\renewcommand{\arraystretch}{1.5}
\begin{table}[t]
\color{black}
\centering
\caption{\label{tab:WAICMM} Shows the WAIC of the 4 considered models from Section 5.1 fitted to the Quebec hospitalizations. The best fitting model, the one with the lowest WAIC, is bolded.}
\begin{tabular}{ll}
\hline
\textbf{Model}   & \textbf{WAIC} \\ \hline
\textbf{Full Coupled}     &      \textbf{17,516}         \\
Non-coupled      &      17,639         \\
No Absence/Clone &      17,545         \\
Endemic-epidemic &    17,845           \\ \hline
\end{tabular}
\end{table}

} 

\subsection{Results} \label{section:results}

\renewcommand{\arraystretch}{1.5}
\begin{table}[t]
\centering
\caption{\label{tab:Table 1} Posterior means and 95\% posterior credible intervals (in parentheses) for the estimated parameters from the Markov chain part of the fitted CMSNB(1,2,4) model. The intercept row shows the transition probabilities at an average level of mobility and beds, no new variant and assuming no neighboring outbreaks. Other rows show the (relative) odds ratios of the corresponding covariate. Units are given in parentheses after the covariates, for reference, .56 is the average weight $\omega_{ji}$ between two neighboring areas. {\color{black} The units for beds and mobility are equal to one standard deviation.} Odds ratios whose 95\% posterior credible intervals do not contain 0 are bolded.}

\begin{tabular}{lcccc} 
 \hline                                    & \multicolumn{4}{c}{\textbf{Probability or (relative) Odds Ratio}}            \\ \hline 
                             & \textbf{Disease} & \textbf{Disease} & \textbf{Outbreak} & \textbf{Outbreak} \\[-7pt] 
\textbf{Covariate}                             &  \textbf{Emergence} &  \textbf{Extinction} &  \textbf{Emergence} &  \textbf{Persistence} \\  \hline
Intercept  & {\color{black}.33}    & {\color{black}.03}   & {\color{black}.02}  & {\color{black}.88}  \\[-5pt]
           &  {\color{black}(.15, .61)} &  {\color{black}(.01, .06)} & {\color{black}(.01, .03)} & {\color{black}(.83, .93)} \\
$\text{beds}$ (100s)  & {\color{black}1.71}    & {\color{black}\textbf{.44}}   & {\color{black}1.15}  & {\color{black}1.13}  \\[-5pt]
  & {\color{black}(.82, 3.44)} & {\color{black}\textbf{(.21, .74)}} & {\color{black}(.86, 1.49)} & {\color{black}(.85, 1.48)} \\
$\text{mobility}$ ({\color{black} 20}\%)  & {\color{black}.77}    & {\color{black}\textbf{.53}}  & {\color{black}\textbf{1.73}} & {\color{black}\textbf{1.65}} \\[-5pt]
 & {\color{black}(.38, 1.36)} & {\color{black}\textbf{(.24, .98)}} & {\color{black}\textbf{(1.08, 2.69)}} & {\color{black}\textbf{(1.18, 2.26)}} \\
$\text{new variant}$ & --   & -- & {\color{black}\textbf{16.22}} & -- \\[-5pt] 
& & & {\color{black}\textbf{(6.91, 33.14)}}& \\
$\text{weighted neighborhood}$ & {\color{black}1.15} & {\color{black}1.01} & {\color{black}\textbf{1.98}} & {\color{black}\textbf{1.31}} \\[-5pt] 
$\text{outbreak sum (.56)}$& {\color{black}(.84, 1.53)}  & {\color{black}(.73, 1.35)} & {\color{black}\textbf{(1.58, 2.46)}}& {\color{black}\textbf{(1.14, 1.51)}} \\ \hline
\end{tabular}
\end{table}

\begin{figure}[t]
 	\centering
 	\includegraphics[width=.9\linewidth]{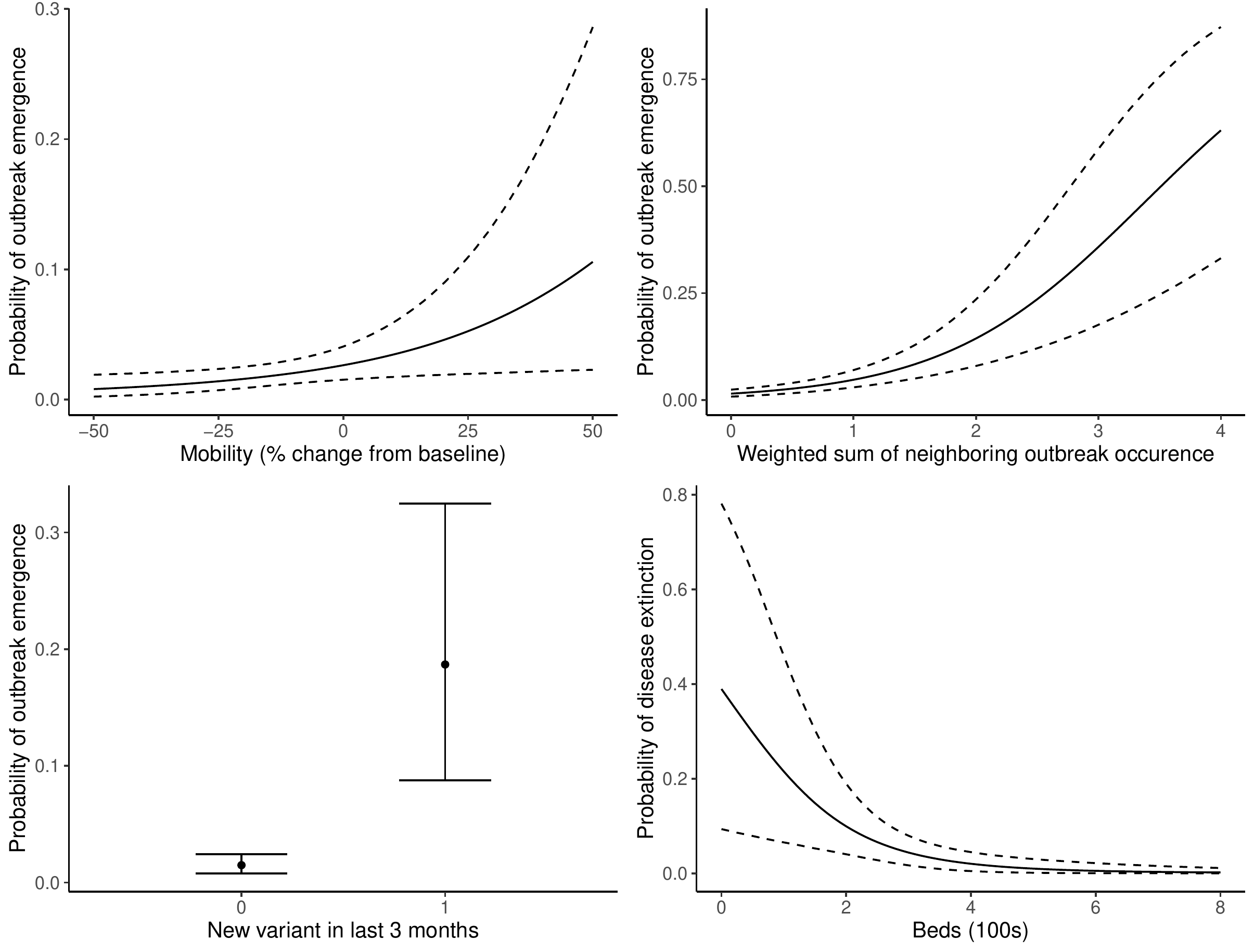}
	\caption{Posterior means (solid lines) and 95\% posterior credible intervals (dashed lines) of some of the estimated transition probabilities versus some of the covariates. Other covariates were fixed at either their average values, for beds and mobility, or at 0, for new variant and the weighted neighborhood outbreak sum.  \label{fig:fig_trans_probs}} 
\end{figure}

Table \ref{tab:Table 1} gives the estimated parameters from the Markov chain part of the fitted CMSNB(1,2,4) model. As relative odds ratios from multinomial logistic regression can be difficult to interpret, we also plot some of the transition probabilities versus some of the covariates in Figure \ref{fig:fig_trans_probs}. Smaller areas were associated with a higher probability of COVID-19 extinction, which follows well known theories from \cite{bartlettMeaslesPeriodicityCommunity1957} {\color{black} that state population size is inversely related to the rate of extinction of an infectious disease.} We did not find evidence that larger areas experience more frequent or longer outbreaks compared to smaller areas (although outbreaks in larger areas do tend to be more severe in terms of transmission, see Table \ref{tab:Table 2}). We found that mobility in retail and recreation venues had a positive association with both outbreak emergence and outbreak persistence {\color{black}and a negative association with disease extinction.} For example, we estimated that a {\color{black} 20} percent {\color{black}(one standard deviation)} increase in the number of visits to retail and recreation venues, from the baseline week, was associated with a {\color{black}73 (8, 169)} percent increase in the odds of an outbreak emerging relative to remaining in the endemic state. However, this result may not imply a causal relationship between mobility and outbreak emergence. Mobility could be high at the start of an outbreak because outbreaks typically begin after a long period of relative calm, so it is natural for people to start going out more to movie theaters, restaurants, etc. It is difficult to control for this in our framework, as we would have to account for the time since the last outbreak ended, making the model non-Markovian. {\color{black}Also, note the high amount of uncertainty around the odds ratios for mobility.}

The introduction of the Alpha and Omicron variants had a very strong association with the emergence of new COVID-19 outbreaks. We estimated that the probability of an outbreak emerging, at average levels of the other covariates and assuming no neighboring outbreaks, was {\color{black}.02 (.01, .03)} if no new variant had been introduced recently and {\color{black}.19 (.09, .32)} if a new variant had been introduced in the last 3 months. Emerging variants can trigger new COVID-19 waves due to them potentially being more contagious and more resistant to existing vaccines compared to previous variants \citep{masloCharacteristicsOutcomesHospitalized2022}. Finally, we found that the presence of outbreaks in neighboring areas was strongly associated with outbreak emergence and persistence. We estimated that an outbreak occurring in a single neighboring area of average connectivity was associated with a {\color{black}98 (58, 146)} percent increase in the odds of an outbreak emerging relative to remaining in the endemic state and a {\color{black}31 (14, 51)} percent increase in the odds of an outbreak persisting. This could be due to either direct disease spread from the neighboring area, the signaling of spread from a common source and, potentially, partly due to correlated missing covariates. Interestingly, we found no strong evidence of an association between neighboring outbreaks and disease emergence or disease extinction, despite there being strong epidemiological justification for a relationship due to outbreak spread \citep{grenfellTravellingWavesSpatial2001a}. Recall that the weighted sum of neighboring outbreak occurrence, $\sum_{j \in NE(i)} \omega_{ji} I[S_{j(t-1)}=3]$ in (\ref{eqn:logit})-(\ref{eqn:multlogit}), is a latent covariate and so there is likely not a lot of information about its effect on these rarer transitions (the disease only goes extinct in the smaller catchment areas).

\begin{table}[t]
\centering
\caption{\label{tab:Table 2} Posterior means and 95\% posterior credible intervals (in parentheses) from the count part of the fitted CMSNB(1,2,4) model. {\color{black}The intercepts
and covariate effects are exponentiated so that they represent rates and rate ratios. Rate
ratios whose 95\% posterior credible intervals do not contain 0 are bolded. The units for beds
and mobility are equal to one standard deviation.}}

\begin{tabular}{lccc} 
 \hline           &                         & \multicolumn{2}{c}{\color{black}\textbf{Rate Ratios}}            \\ \hline 
    \textbf{Covariate}          &  \textbf{Parameter}           & \textbf{Endemic} & \textbf{Outbreak}  \\  \hline
Intercept of random intercepts  & {\color{black}$e^{\beta_0}$}   & {\color{black}.99}   & {\color{black}2.18}  \\[-5pt]
           &  &  {\color{black}(.85, 1.13)} & {\color{black}(1.97, 2.41)}  \\
Std. dev of random intercepts & $\sigma$   & {\color{black}.31}   & {\color{black}.13} \\[-5pt]
           &  &  {\color{black}(.21, .44)} & {\color{black}(.1, .17)}  \\
beds (100s) & {\color{black}$e^{\beta_{\text{beds}}}$}   & {\color{black}\textbf{1.07}}  & {\color{black}\textbf{1.20}} \\[-5pt]
           &  &  {\color{black}\textbf{(1.02, 1.11)}} & {\color{black}\textbf{(1.09, 1.32)}} \\
{\color{black} mobility (20\%)} & {\color{black}$e^{\beta_{\text{mobility}}}$}   & {\color{black}\textbf{1.06}}  & {\color{black}\textbf{1.17}} \\[-5pt]
           &  &  {\color{black}\textbf{(1, 1.13)}} & {\color{black}\textbf{(1.14, 1.20)}} \\
{\color{black} new variant} & {\color{black}$e^{\beta_{\text{new\_variant}}}$}   & --  & {\color{black}1.01} \\[-5pt]
           &  &   & {\color{black}(.97, 1.06)} \\
autoregressive & $\rho$   & {\color{black}.69}  & {\color{black}.75} \\[-5pt]
           &  &  {\color{black}(.65, .72)} & {\color{black}(.72, .78)}  \\
overdispersion & $r$   & {\color{black}5.50}  & {\color{black}8.95} \\[-5pt]
           &  &  {\color{black}(4.13, 7.37)} & {\color{black}(7.89, 10.14)}  \\
           \hline
\end{tabular}
\end{table}

Table \ref{tab:Table 2} gives the estimated parameters from the count part of the model. Transmission during the outbreak periods was on average {\color{black}121 (91, 157)} percent higher than during the endemic periods. {\color{black} Mobility in retail and recreation venues and population size were both positively associated with transmission during the outbreak and endemic periods, and outbreaks amplified the effects of population size and mobility on transmission. Note, as we are modeling hospitalizations, an effect in terms of transmission in the hospitalizations could reflect an effect on transmission in the actual cases and/or an effect on the severity of the disease. Mobility and beds should not affect disease severity; however, new variants often do \citep{chenchulaCurrentEvidenceEfficacy2022}. This could partly explain why we found the introduction of a new variant likely did not have a large effect on transmission in the hospitalizations, Omicron was much more transmissible person to person but less severe \citep{UpdatesCOVID19Variants2022}. Also, we combined the Alpha and Omicron variants, and, as can be seen in Figure \ref{fig:fig1}, Alpha does not appear to have had a large impact on transmission in the hospitalizations. Finally, for sensitivity analysis, we also fit the CMSNB(1,2,4) model using .05 and .1 in place of .01 in the constraint (\ref{eqn:const}). The only posterior that changed noticeably was for the effect of mobility on transmission in the endemic period ($e^{\beta_{\text{mobility}}^{EN}}$) which increased from 1.06 (1, 1.13) to 1.08 (1.02, 1.15) going from .01 to .1 in the constraint.}

\subsection{Retrospective evaluation and comparison} \label{section:retro}

\begin{figure}[!t]
 	\centering
 	\includegraphics[width=\linewidth]{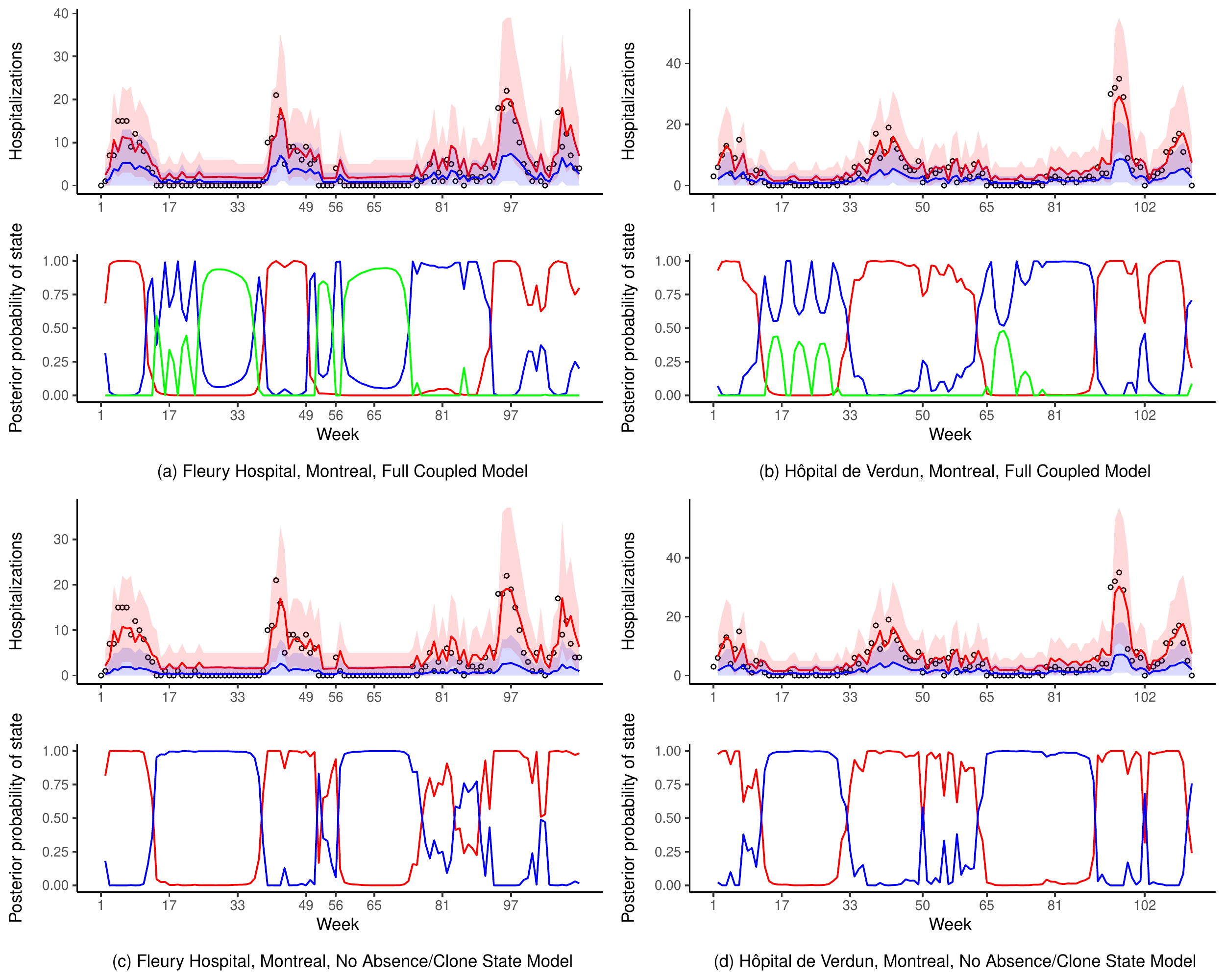}
	\caption{Top graphs show the posterior means (solid lines) and 95\% posterior credible intervals (shaded areas) of the endemic state distribution (blue) and the outbreak state distribution (red) versus the observed hospitalizations (points). The bottom graphs show the posterior probability that the disease was in the absence state (green line), the endemic state (blue line) and the outbreak state (red line) during each week of the study period, that is, $P(S_{it}=s|\bm{y})$ for $t=1,\dots,T$ and $s=1$ (absence) in green, $s=2$ (endemic) in blue and $s=3$ (outbreak) in red, see Section \ref{section:outdet}. \label{fig:retro_compare}}
\end{figure}

As mentioned in Section \ref{section:outdet} it is important to examine the retrospective state estimates of an epidemiological Markov switching model, that is, the posterior probability that the disease was in each state during each week of the study period, to ensure the models' estimates of the epidemiological history of the disease are sensible. Starting with the bottom graphs of Figures \ref{fig:retro_compare} (a) and (c), we compare the retrospective state estimates for Fleury Hospital in Montreal, one of the smaller hospitals, between models with, (a), and without, (c), an absence state, as it is a good example of how accounting for long periods of disease absence in smaller areas can be important. The top graphs show the posteriors of the endemic and outbreak state distributions, which can be helpful for better understanding the state estimates. To draw from the posterior of the state distributions we drew from $p(y_{it}|S_{it}=s,y_{i(t-1)},\bm{\beta}^{[m]})$, see (\ref{eqn:y_spec})-(\ref{eqn:lambda}), for $m=M+1,\dots,Q$, $t=1,\dots,T$, $s=2$ (endemic) and $s=3$ (outbreak). The Full Coupled Model identifies 3, unlikely 4, outbreak periods for Fleury Hospital which are separated by a few endemic and absence periods. Long sequences of zeroes in Fleury Hospital are generally assigned to absence periods by the Full Coupled Model. In contrast, the No Absence/Clone State Model assigns the long strings of zeroes to endemic periods, bringing the endemic state distribution much closer to 0 compared to the Full Coupled Model. This causes some issues in the state estimation for the No Absence/Clone State Model, as the outbreak state becomes too dominant. Firstly, the No Absence/Clone State Model classifies a very small increase in hospitalizations, {\color{black} around week 56}, as likely a short outbreak period, which is not realistic. Secondly, the No Absence/Clone State Model {\color{black} identifies a smaller outbreak, around week 81, before the final Omicron outbreak}. {\color{black} In contrast, the Full Coupled Model only identifies the Omicron outbreak here, which appears more realistic} as the number of hospitalizations in the few weeks before Omicron never becomes high enough to be very concerning {\color{black}and, by checking the other plots, there is no strong evidence of outbreaks in any neighboring areas}.

In Figures \ref{fig:retro_compare} (b) and (d) we compare the retrospective states estimates for {\color{black}Hôpital de Verdun} between models with, (b), and without, (d), clone states. {\color{black}The No Absence/Clone State Model shows rapid switching between the outbreak and endemic states around weeks 50 and 102.} {\color{black} During week 50, it is more realistic that the one-week dip in the hospitalizations is due to random variation, which is what the Full Coupled Model identifies, and that there are not multiple outbreaks occurring. The Full Coupled Model does place some weight on an endemic period between weeks 102 and 103. A two-week dip into the endemic period seems more realistic than a one-week dip, however, we could prevent two-week dips by increasing the number of endemic clone states.} For the Full Coupled Model, we looked through the plots in Figure \ref{fig:retro_compare} for each hospital and found the retrospective state estimates to be sensible.

\subsection{Real-time evaluation and comparison} \label{section:real_time}

\begin{figure}[!t]
 	\centering
 	\includegraphics[width=\linewidth]{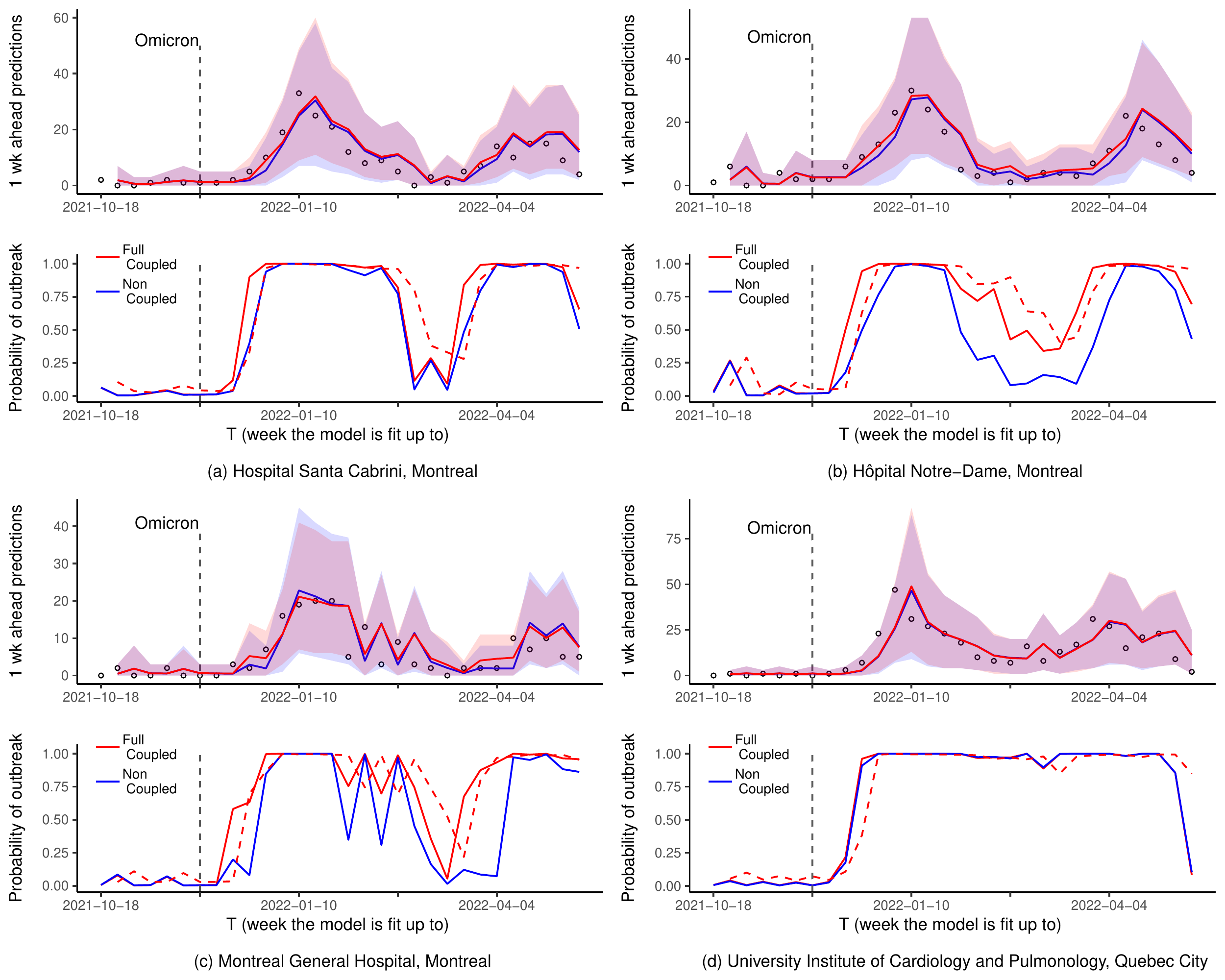}
	\caption{{\color{black}Top graphs show the posterior means (solid lines) and 95\% posterior credible
intervals (shaded areas) of the one-week ahead posterior predictive distributions, that is, $p(y_{iT}|\bm{y}_{1:(T-1)})$, versus $T$.} Bottom graphs solid lines show the posterior probabilities that an outbreak is currently happening, that is, $P(S_{iT}=3|\bm{y})${\color{black}, versus $T$}. Bottom graphs dashed line shows the one-week ahead outbreak forecasts from the previous week, that is, $P(S_{iT}=3|\bm{y}_{1:(T-1)})${\color{black}, versus $T$}. The Full Coupled Model is in {\color{black}red} and the Non-coupled Model is in {\color{black}blue}. {\color{black}The {\color{black}grey dashed} lines are drawn at the introduction
of the Omicron variant for all of Quebec.}\label{fig:prosp_compare}}
\end{figure}

To evaluate the real-time performance of the models we fit the Full Coupled, Non-coupled {\color{black}and EE} models up to 6 weeks before the introduction of Omicron and then up to every week after, that is, we fit the models up to time $T$ for $T=84=\text{2021-10-18},...,113=\text{2022-05-09}$, producing 30 sets of posterior samples for each model. We did not include $\text{new\_variant}_t$ as a covariate since many of the fitted models are fit before, or just after, Omicron and so there is likely not enough information to estimate the effect for many of the fits. {\color{black} Additionally, we found after discarding 27\% of the data for the real-time evaluation that MCMC chains would sometimes get stuck in local modes in regions of the parameter space that only used the absence state and one of the count states. This is not an uncommon problem in complex Bayesian mixture modeling and, following the discussion and recommendations in Section 4.2.3 of \cite{fruhwirth-schnatterFiniteMixtureMarkov2006}, we applied mild shrinkage to the transition probabilities by assigning $N(0,2.5^2)$ priors to the intercepts in Equations (\ref{eqn:logit})-(\ref{eqn:multlogit}), which stabilized model fitting.} Figure \ref{fig:prosp_compare} compares the posterior probabilities that an outbreak is currently happening (bottom graphs solid lines) between the Full Coupled Model, in {\color{black}red}, and the Non-coupled Model, in {\color{black}blue}{\color{black}, for four hospitals}. Recall from Section \ref{section:outdet} that the posterior probability that an outbreak is currently happening is given by $P(S_{iT}=3|\bm{y})$ and would be used for the purpose of outbreak detection in a real-time scenario. The Full Coupled Model gives an earlier warning of the Omicron outbreak in the first three hospitals, (a)-(c), but not in the final hospital, (d). In (d) the hospitalizations rise very rapidly and so there is a lot of within area information about the Omicron outbreak, making the coupling less useful for providing an early warning. {\color{black}SM Figure 5} (a) shows that, when averaged across all hospitals, early on during the Omicron wave the Full Coupled Model gives a {\color{black}.15-.2} higher posterior probability that an outbreak is currently happening, an earlier warning compared to the Non-coupled Model, while it does not show an increased risk of an outbreak prior to Omicron. The Omicron wave was highly synchronized between the hospitals and, therefore, as can be seen from {\color{black}SM Figure 5} (b), the probability of outbreak emergence would have been higher for the Full Coupled Model during the start of the wave as evidence of outbreaks accumulated in neighboring hospitals. In general, {\color{black}SM Figure 5} shows that the Full Coupled Model tends to enforce the status quo, making outbreak occurrence more (less) likely if there are (are not) outbreaks occurring in neighboring areas.

The dashed line in the bottom graphs of Figure \ref{fig:prosp_compare} show the one-week ahead outbreak forecasts $P(S_{iT}=3|\bm{y}_{1:(T-1)})$, from the previous week, for the Full Coupled Model. Note that, we typically observe $P(S_{iT}=3|\bm{y})>P(S_{iT}=3|\bm{y}_{1:(T-1)})>P(S_{i(T-1)}=3|\bm{y}_{1:(T-1)})$ around the start of the Omicron wave. We looked through the plots in Figure \ref{fig:prosp_compare} for each hospital and found the real-time state estimates/forecasts to be sensible in most hospitals. We did find evidence of false alarms being thrown in 2/30 hospitals during the real-time evaluation, shown in the SM Section 7.3. The false alarms appear to have been mostly caused by hospitalizations spiking during periods of high outbreak risk, meaning the model has no information that an outbreak starting should be unlikely to counter the false alarm. Still, the model corrected itself quickly, by the next week in most cases. 

{\color{black} The top graphs of Figure \ref{fig:prosp_compare} summarize the one-week ahead predictions, from the previous week, of the hospitalizations from the Full Coupled and Non-coupled models. SM Figure 9 summarizes the one-week ahead predictions from the EE model for the same hospitals. In panels (a)-(c) of Figure \ref{fig:prosp_compare}, there is a much larger discrepancy between the state estimates of the two models than between the predictions of the counts. This indicates that differences in predictive performance may be a poor proxy for differences in the accuracy or, especially, timeliness of the state estimates. Outbreaks often emerge at low counts where there is a lot of overlap between the state distributions, see Figure \ref{fig:retro_compare}. Therefore, a large difference in the state estimates during the start of an outbreak may not translate to a large difference in predictions. We used the multivariate log score \citep{bracher_endemic-epidemic_2022}, see SM Section 7.4, to compare the one-week ahead forecasts from the three models. The Full Coupled Model obtained the lowest mean score, indicating the best performance on average across the entire Omicron outbreak, at 3.20. This was followed by the Non-coupled, 3.25, and EE, 3.27, models. Permutation tests \citep{paul_predictive_2011} showed moderate evidence of a true difference in the overall mean scores, with two-sided p-values of .033, Full Coupled versus Non-coupled, and .052, Full Coupled versus EE. 
To investigate local differences in forecasting performance, we plot the multivariate log scores across time in SM Figure 10. The Full Coupled Model obtained a much lower score, indicating superior predictive performance, around the emergence of the Omicron wave, which is the most important period to capture for disease surveillance \citep{buckeridge_outbreak_2007}. There are some important limitations to using the multivariate log scores to compare the models. We can only evaluate the models over the Omicron wave, where model fitting is stable, which includes just 29 weeks and is missing some important features of the full data set such as long strings of zeroes. In contrast, the WAIC evaluates the models over the entire data set. Finally, an important advantage of the Markov switching models for forecasting, over the EE model, is that they can determine whether a forecasted increase in hospitalizations will likely be due to an outbreak emerging, as opposed to random endemic variation. See the bottom graphs in Figure \ref{fig:prosp_compare}. If an increase in the counts is due to endemic variation it will be much less concerning to health authorities \citep{unkelStatisticalMethodsProspective2012}.}

Finally, the map in {\color{black}SM Figure 6} shows the likely state in the catchment area of each hospital during the last week of the study period, March 5th, 2022, from the Full Coupled Model with new variant as a covariate. {\color{black} According to the map, most catchment areas were in the outbreak state at this time, especially in/around Montreal.}

\section{Discussion} \label{section:disc}

We have proposed a three-state coupled nonhomogeneous Markov switching model for the general analysis of spatio-temporal outbreak occurrence. The model can be used for investigating associations between various factors, including geographical outbreak spread, and outbreak emergence, outbreak persistence, disease extinction and disease emergence, as well as for detecting and forecasting the outbreaks. We made three main contributions to the existing \citep{amorossalvadorBayesianTemporalSpatiotemporal2017} endemic/outbreak Markov switching literature. Firstly, to account for long periods of disease absence in smaller areas, \citep{bartlettMeaslesPeriodicityCommunity1957} we added an absence state to our model in addition to the more traditional endemic and outbreak states. Secondly, to incorporate geographical outbreak spread \citep{grenfellTravellingWavesSpatial2001a}, we allowed the transition probabilities to depend on whether outbreaks were occurring in neighboring areas. Previous two-state approaches \citep{heatonSpatiotemporalAbsorbingState2012} have allowed the probabilities of outbreak emergence to depend on neighboring outbreaks, but they did not allow outbreak emergence to depend on covariates, and they fixed the probability of outbreak persistence at one to prevent rapid switching between endemic and outbreak periods, meaning their method could not be applied to multiple outbreaks. Finally, to allow for the analysis of multiple outbreaks, we instead introduced clone states \citep{kaufmannHiddenMarkovModels2018} into the model, which prevents rapid switching between endemic and outbreak periods by enforcing a minimum endemic and outbreak duration.

We applied our model to the analysis of COVID-19 outbreaks across Quebec based on admissions in the 30 largest hospitals. We found that mobility in retail and recreation venues, the introduction of new variants and the occurrence of outbreaks in neighboring areas all had positive associations with the emergence or persistence of new COVID-19 outbreaks in Quebec. {\color{black} Additionally, mobility and population size had positive associations with transmission during the outbreak and endemic periods, and the effects were amplified during the outbreak periods.} As for disease extinction and emergence, we only found evidence that the disease is more likely to go extinct in smaller areas {\color{black}and at lower levels of mobility.} Disease extinction and emergence are rarer transitions, so more data needs to be collected to study them with high precision. 

Regarding model performance, we found our model gave realistic estimates of past epidemiological states {\color{black} for the Quebec hospitalizations} and {\color{black} appeared to} perform outbreak detection/forecasting well in a real-time evaluation of the Omicron wave, absent a small number of false alarms during the real-time evaluation. {\color{black} However, on the Quebec hospitalizations it was difficult to exactly quantify the accuracy of the state estimates, e.g., calculate AUC, sensitivity, etc., as the true underlying states were not known.} {{\color{black} Therefore, we also conducted a simulation study with known outbreaks and found our model achieved good sensitivity, specificity and timeliness both retrospectively and during real-time outbreak detection and forecasting.}

Regarding model comparison {\color{black}among different Markov switching models.} {\color{black}In our simulation study, where we assumed outbreaks in neighboring areas occurred around the same time, including neighboring outbreak indicators in the transition probabilities greatly improved sensitivity, specificity and timeliness during real-time outbreak detection and forecasting. As for our analysis of the Quebec hospitalizations, model comparison using WAIC supported the inclusion of absence/clone states and the addition of neighboring outbreak indicators in the transition probabilities.} We also found for the Quebec analysis that the addition of an absence state can be important in smaller areas for reducing bias in the endemic state distribution towards 0, a bias that we observed often leads to unrealistic state estimates. {\color{black} We further observed that clone states are important for preventing rapid switching between endemic and outbreak periods, making the state estimates more realistic.}  Finally, in a real-time evaluation we found that the incorporation of outbreak spread led to an earlier warning of the Omicron wave in Quebec as the wave was highly synchronized across the hospitals. Ultimately, our contributions show a lot of promise for improving state estimation both retrospectively and in real-time, especially when there are small areas and highly spatially synchronized outbreaks. Although we applied our model to spatio-temporal counts of hospitalizations, it could also be applied to spatio-temporal counts of disease cases, which are popular to model for outbreak occurrence analysis \citep{knorr-heldHierarchicalModelSpace2003b,watkinsDiseaseSurveillanceUsing2009}, or deaths.

{\color{black}Now we will address the question of how our approach fits into the broader popular literature on spatio-temporal infectious disease modeling, including compartmental models \citep{bauerStratifiedSpaceTime2018} and multivariate time series models \citep{ssentongo_pan-african_2021}. One aspect of our approach that we believe is unique is the ability to quantify associations, through odds ratios, between covariates and certain epidemiological transitions, such as outbreak emergence or disease extinction. Also, our model can provide the probability that a spike in an epidemiological indicator is due to an actual outbreak developing as opposed to just random endemic variation, which is valuable for outbreak detection \citep{unkelStatisticalMethodsProspective2012}. On the Quebec hospitalizations, we found our model fit better and had better interpretations compared to the Endemic-epidemic model \citep{bracher_endemic-epidemic_2022}, a popular alternative.}

There are also some limitations with our approach. {\color{black} Conditional on the states, we only consider first-order autoregression, see the discussion at the end of SM Section 4. Given weekly data and a disease with a short serial interval, such as COVID-19, this is likely not a major limitation \citep{bracher_endemic-epidemic_2022}. However, in some cases, especially with daily data, it can be important to account for higher-order temporal dependencies. As another limitation,} we assume that the probability of an outbreak emerging does not depend on the amount of time since the last outbreak has ended. After an outbreak ends, the susceptible population should increase over time due to demographic changes, waning immunity, and other factors, leading to an increased outbreak risk \citep{keelingModelingInfectiousDiseases2007}. In a hidden semi-Markov model \citep{langrockHiddenMarkovModels2011}, the transition probabilities can depend on the amount of time that has been spent in the state, which can be pursued in further work. {\color{black}Additionally}, given the large number of transitions in our model, the number of covariate effects grows quickly with the number of covariates (4 times). Shrinkage could be applied to the covariate effects in the transition probabilities to eliminate many unimportant effects \citep{wangBayesianNonHomogeneousHidden2022}. {\color{black} Finally, we did not include vaccination data in the model. While data on the percentage of individuals with at least two doses is available in Quebec \citep{inspqCOVID19DataQuebec2022}, there is very little variation in the covariate. Vaccination coverage increased rapidly and was only greater than zero during the final Omicron outbreak. Also, there are likely interactions between vaccination and new variants \citep{chenchulaCurrentEvidenceEfficacy2022}. Therefore, we decided there was not enough information in the available data to capture the complicated effects of vaccination with the model. Not including vaccination information could have consequences on the estimated effects of mobility and new variants. For instance, there were fewer mobility restrictions on vaccinated individuals \citep{Cabrerapassport2021} who were also less likely to contract, especially severe, cases of the disease \citep{johnson_covid-19_2022}.}

\section*{Acknowledgements}

This research was part of a larger project of INESSS (Institut national d'excellence en santé et en services sociaux), whose objective was to produce bed occupancy projections for COVID-19 patients. The access to data was made possible through a tripartite agreement between the MSSS, the RAMQ and INESSS. This work is part of the PhD thesis of D. Douwes-Schultz under the supervision of A. M. Schmidt in the Graduate Program of Biostatistics at McGill University, Canada. Douwes-Schultz is grateful for financial support from IVADO and the Canada First Research Excellence Fund/Apogée (PhD Excellence Scholarship 2021-9070375349). Schmidt is grateful for financial support from the Natural Sciences and Engineering Research Council (NSERC) of Canada (Discovery Grant RGPIN-2017-04999).  Shen is grateful for financial support from the Fonds de recherché du Québec–Santé (FRQS) (Doctoral training award 313602). Buckeridge is supported by a Canada Research Chair in Health Informatics and Data Science (950-232679. This research was enabled in part by support provided by Calcul Québec (www.calculquebec.ca) and Compute Canada (www.computecanada.ca).

\bibliography{ms}

\end{document}


\maketitle

\tableofcontents

\section{The Individual Forward Filtering Backward Sampling (iFFBS) Algorithm}

In this section, we describe how $\bm{S}^*$ is sampled in our hybrid Gibbs sampling algorithm. We will borrow all notation from the main text. First, to obtain valid initial values for the hidden Markov chain we sample $\bm{S}_i^{*[1]}$ from a 6 state Markov chain with transition Matrix given by Equation (7) in the main text without the absence state. The transition probabilities are fixed at .8 for remaining in a state and .2 for transitioning out of a state since we expect persistence. We do not include the absence state in the initial values since it might cause the initial joint likelihood function, Equation (8) in the main text, to be evaluated at 0 as the absence state cannot produce a positive count. After initialization, the following steps are repeated for $m=2,\dots,Q$, where $Q$ is the total number of iterations for the Gibbs sampler,
\begin{enumerate}
    \item Sample $\bm{v}^{[m]}$ from $p(\bm{v}|\bm{S}^{*[m-1]},\bm{y})$
    \item Sample $\bm{S}_i^{*[m]}$ from $p(\bm{S}_i^{*}|\bm{S}_1^{*[m]},\dots,\bm{S}_{i-1}^{*[m]},\bm{S}_{i+1}^{*[m-1]},...,\bm{S}_{N}^{*[m-1]},\bm{v}^{[m]},\bm{y})$ for $i=1,...,N$.
\end{enumerate} As mentioned in the main text, in the first step elements of $\bm{v}$ without conjugate priors are sampled individually using an adaptive random walk Metropolis step \citep{shabyExploringAdaptiveMetropolis2010}. Here we will provide the individual forward filtering backward sampling (iFFBS) algorithm for sampling from $p(\bm{S}_i^*|\bm{S}_{(-i)}^*,\bm{v},\bm{y})$ needed for step 2. The algorithm was originally proposed by \cite{touloupouScalableBayesianInference2020}. In this Section we will sometimes use the subscript $t_1{:}t_2$ to denote a temporally indexed vector subsetted to the interval $t_1$ to $t_2$, e.g., $\bm{y}_{i(1:t)}=(y_{i1},\dots,y_{it})^T$.

First note that,
\begin{align}
p(\bm{S}_i^*|\bm{S}_{(-i)}^*,\bm{v},\bm{y}) =p(S_{iT}^*|\bm{S}_{(-i)}^*,\bm{y},\bm{v}) \prod_{t=1}^{T-1}p(S_{it}^*|S_{i(t+1)}^*,\bm{S}_{(-i)(1:t+1)}^*,\bm{y}_{i(1:t)},\bm{v}), \label{eqn:full_dens}
\end{align} and that, from Bayes' Theorem,
\begin{align}
p(S_{it}^*|S_{i(t+1)}^*,\bm{S}_{(-i)(1:t+1)}^*,\bm{y}_{i(1:t)},\bm{v}) \propto p(S_{i(t+1)}^*|S_{it}^{*},\bm{S}_{(-i)t},\bm{\theta})p(S_{it}^*|\bm{S}_{(-i)(1:t+1)}^*,\bm{y}_{i(1:t)},\bm{v}). \label{eqn:full_dens2}
\end{align} The density $p(S_{i(t+1)}^{*}|S_{it}^{*},\bm{S}_{(-i)t},\bm{\theta})$ in (\ref{eqn:full_dens2}) is simply the appropriate transition probability of the Markov chain in area $i$, which can be obtained from Equation (7) of the main text. Therefore, if we can calculate $P(S_{it}^*=s^{*}|\bm{S}_{(-i)(1:t+1)}^*,\bm{y}_{i(1:t)},\bm{v})$ for $s^{*}=1,\dots,7$ and $t=1,\dots,T$, called the filtered probabilities, then $\bm{S}_i^{*}$ can be sampled backward using Equations (\ref{eqn:full_dens}) and (\ref{eqn:full_dens2}).

Starting with $t=1$ we have that, 
\begin{align}
\begin{split}
p(S_{i1}^*|\bm{S}_{(-i)(1:2)}^*,y_{i1},\bm{v}) &\propto p(S_{i1}^*|y_{i1}) p(\bm{S}_{(-i)2}^*|S_{i1}^*,\bm{S}_{(-i)1}^*,\bm{\theta})\\
& \propto p(S_{i1}^*|y_{i1}) \prod_{j \, : \, i \,\in \, NE(j)}p(S_{j2}^*|S_{j1}^*,\bm{S}_{(-j)1},\bm{\theta}). \label{eqn:filter1}
\end{split}
\end{align} Here $p(S_{i1}^*|y_{i1})$ is the initial state distribution, which is fixed by the modeler, and \\ $p(S_{j2}^*|S_{j1}^*,\bm{S}_{(-j)1},\bm{\theta})$ is a transition probability of the Markov chain in area $j$. Note that \\ $ \prod_{j \, : \, i \,\in \, NE(j)}p(S_{j2}^*|S_{j1}^*,\bm{S}_{(-j)1},\bm{\theta})$ only depends on whether area $i$ is in the outbreak state or not so only 2 values need to be calculated. Also note that since $S_{i1}^*$ can only take seven values it is straightforward to derive the filtered probabilities using Equation (\ref{eqn:filter1}),
\begin{align*}
P(S_{i1}^*=s^{*}|\bm{S}_{(-i)(1:2)}^*,y_{i1},\bm{v}) = \frac{P(S_{i1}^*=s^*|y_{i1}) \prod\limits_{\substack{j \, : \, i \,\in \, NE(j) \\ S_{i1}^*=s^*}}p(S_{j2}^*|S_{j1}^*,\bm{S}_{(-j)1},\bm{\theta})}{\sum_{k=1}^{7}P(S_{i1}^*=k|y_{i1}) \prod\limits_{\substack{j \, : \, i \,\in \, NE(j) \\ S_{i1}^*=k}}p(S_{j2}^*|S_{j1}^*,\bm{S}_{(-j)1},\bm{\theta})}, 
\end{align*} for $s^{*}=1,\dots,7$. 

For $t=2,\dots,T-1$ we have that,
\begin{align*}
p(S_{it}^*|\bm{S}_{(-i)(1:t+1)}^*,\bm{y}_{i(1:t)},\bm{v}) &\propto p(y_{it}|S_{it},y_{i(t-1)},\bm{\beta})p(S_{it}^*|\bm{S}_{(-i)(1:t)}^*,\bm{y}_{i(1:t-1)},\bm{v}) \\  & \,\,\,\, \times \prod_{j \, : \, i \,\in \, NE(j)}p(S_{j(t+1)}^*|S_{jt}^*,\bm{S}_{(-j)t},\bm{\theta}).
\end{align*} Here $p(y_{it}|S_{it},y_{i(t-1)},\bm{\beta})$ is given by Equation (1) of the main text. Note that,
\begin{align*}
P(S_{it}^*&=s^*|\bm{S}_{(-i)(1:t)}^*,\bm{y}_{i(1:t-1)},\bm{v})= \\
&\sum_{k=1}^{7} P(S_{it}^*=s^*|S_{i(t-1)}^*=k,\bm{S}_{(-i)(t-1)},\bm{\theta})P(S_{i(t-1)}^*=k|\bm{S}_{(-i)(1:t)}^*,\bm{y}_{i(1:t-1)},\bm{v}),
\end{align*} where $P(S_{i(t-1)}^*=k|\bm{S}_{(-i)(1:t)}^*,\bm{y}_{i(1:t-1)},\bm{v})$ is the previous filtered probability. It then follows that, 
\begin{align}\label{eqn:iFF}
\begin{split}
&P(S_{it}^*=s^{*}|\bm{S}_{(-i)(1:t+1)}^*,\bm{y}_{i(1:t)},\bm{v}) =  \\[5pt] 
&\frac{p(y_{it}|S_{it}^*=s^*,y_{i(t-1)},\bm{\beta})P(S_{it}^*=s^*|\bm{S}_{(-i)(1:t)}^*,\bm{y}_{i(1:t-1)},\bm{v}) \prod\limits_{\substack{j \, : \, i \,\in \, NE(j) \\ S_{it}^*=s^*}}p(S_{j(t+1)}^*|S_{jt}^*,\bm{S}_{(-j)t},\bm{\theta})}{\sum_{k=1}^{7} p(y_{it}|S_{it}^*=k,y_{i(t-1)},\bm{\beta}) P(S_{it}^*=k|\bm{S}_{(-i)(1:t)}^*,\bm{y}_{i(1:t-1)},\bm{v}) \prod\limits_{\substack{j \, : \, i \,\in \, NE(j) \\ S_{it}^*=k}}p(S_{j(t+1)}^*|S_{jt}^*,\bm{S}_{(-j)t},\bm{\theta})}, 
\end{split}
\end{align} for $s^{*}=1,\dots,7$.

The logic for $t=T$ is similar but there is no forward product term, 
\begin{align*}
P(S_{iT}^*=s^{*}|\bm{S}_{(-i)}^*,\bm{y},\bm{v}) =  
\frac{p(y_{iT}|S_{iT}^*=s^*,y_{i(T-1)},\bm{\beta})P(S_{iT}^*=s^*|\bm{S}_{(-i)}^*,\bm{y}_{i(1:T-1)},\bm{v})}{\sum_{k=1}^{7} p(y_{iT}|S_{iT}^*=k,y_{i(T-1)},\bm{\beta})P(S_{iT}^*=k|\bm{S}_{(-i)}^*,\bm{y}_{i(1:T-1)},\bm{v})}, 
\end{align*} for $s^{*}=1,\dots,7$.

Once the filtered probabilities have been calculated $\bm{S}_i^{*}$ can be sampled backward using Equations (\ref{eqn:full_dens}) and (\ref{eqn:full_dens2}). Firstly, $S_{iT}^{*[m]}$ is drawn from $p(S_{iT}^*|\bm{S}_{(-i)}^*,\bm{y},\bm{v})$. Then, for $t=T-1,\dots,1$, $S_{it}^{*[m]}$ is drawn from the density defined by,
\begin{align*}
&P(S_{it}^*=s^*|S_{i(t+1)}^*=S_{i(t+1)}^{*[m]},\bm{S}_{(-i)(1:t+1)},\bm{y}_{i(1:t)},\bm{v}) = \\[5pt]
&\frac{P(S_{i(t+1)}^*=S_{i(t+1)}^{*[m]}|S_{it}^{*}=s^*,\bm{S}_{(-i)t},\bm{\theta})P(S_{it}^*=s^*|\bm{S}_{(-i)(1:t+1)}^*,\bm{y}_{i(1:t)},\bm{v})}{\sum_{k=1}^{7}P(S_{i(t+1)}^*=S_{i(t+1)}^{*[m]}|S_{it}^{*}=k,\bm{S}_{(-i)t},\bm{\theta})P(S_{it}^*=k|\bm{S}_{(-i)(1:t+1)}^*,\bm{y}_{i(1:t)},\bm{v})},
\end{align*} for $s^{*}=1,\dots,7$.

As mentioned in the main text, all Nimble code for the iFFBS samplers is provided on GitHub (\url{https://github.com/Dirk-Douwes-Schultz/CMSNB124_code}). Note that the only calculations that separate the iFFBS sampler from a traditional FFBS sampler for Markov switching models \citep{chibCalculatingPosteriorDistributions1996,fruhwirth-schnatterFiniteMixtureMarkov2006} are the forward product terms  $\prod_{j \, : \, i \,\in \, NE(j)}p(S_{j(t+1)}^*|S_{jt}^*,\bm{S}_{(-j)t},\bm{\theta})$ which are needed to account for between chain dependencies. Therefore, if an area is not a neighbor of any other area the iFFBS sampler reduces to the FFBS sampler which, being computationally simpler, should be used instead. As such, in our code, we assign all areas that are not neighbors of any other areas FFBS samplers, which we also custom code and provide on GitHub.

\subsection{Validating the algorithm}

\begin{figure}[t]
 	\centering
 	\includegraphics[width=.5\textwidth]{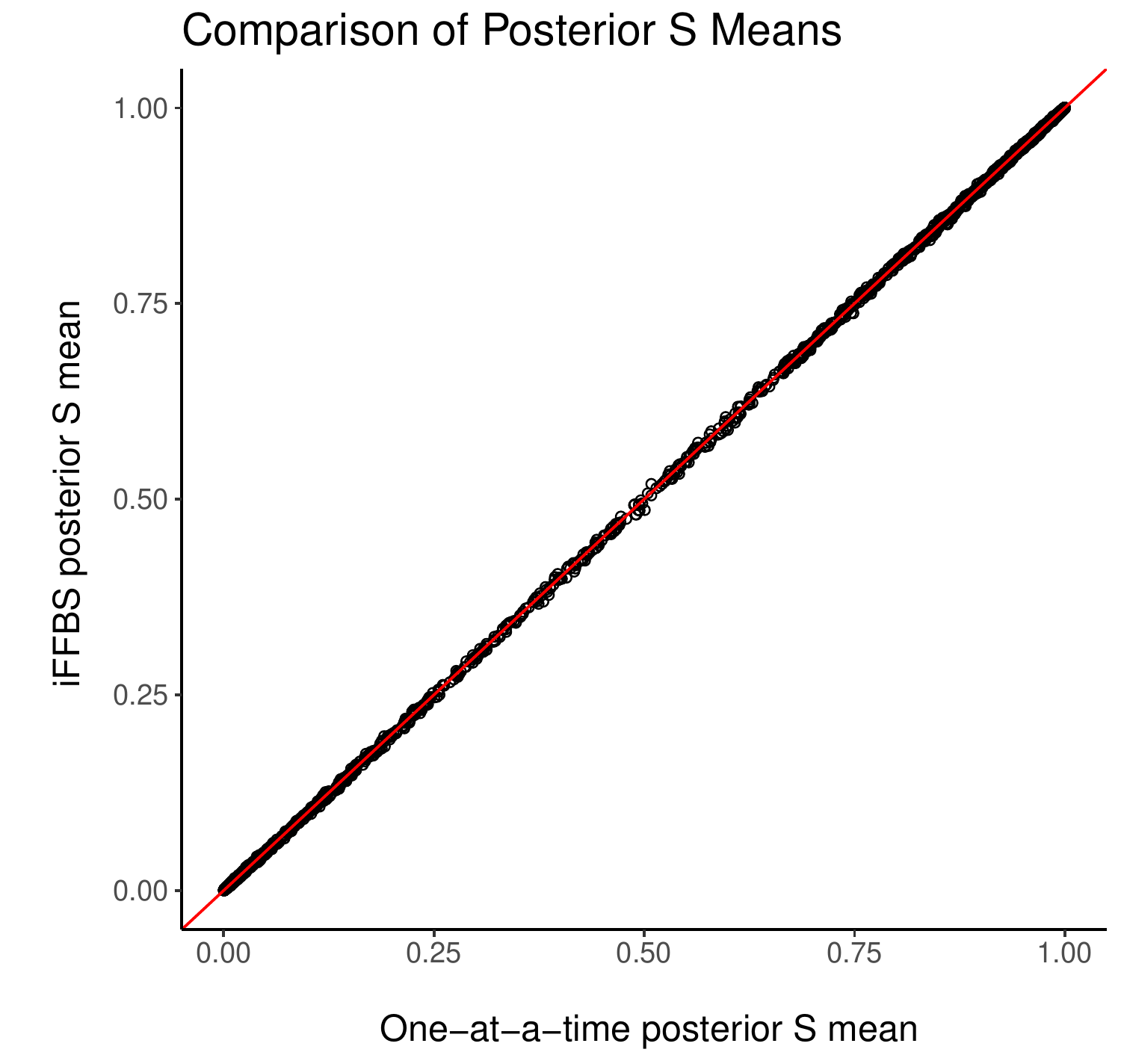}
	\caption{Comparison of the posterior means of $\bm{S}$ produced by the one-at-a-time and iFFBS
samplers when applied to {\color{black} the No Absence/Clone State Model} from Section 5 of the main text. {\color{black}$S_{it}=0$ indicates the endemic state and $S_{it}=1$ indicates the outbreak state.}} \label{fig:compare_samps}
\end{figure}

One way to validate a Markov chain Monte Carlo (MCMC) sampler is to compare the posterior distributions produced by the sampler to those produced by a more established or simpler sampler. Different MCMC sampling algorithms should return the same posterior distributions. Figure \ref{fig:compare_samps} compares the posteriors means of $\bm{S}$ produced by the one-at-a-time, described in Section 3 of the main text, and iFFBS samplers. These were compared on the COVID-19 hospitalization data from the main text with the {\color{black} No Absence/Clone State Model} specified in Section 5.1 of the main text.  We {\color{black} compared the samplers using the No Absence/Clone State Model} as the one-at-a-time samplers do not converge when clone states are present. As can be seen from the figure, both samplers produce the same posterior means for $\bm{S}$ within reasonable Monte Carlo error. Additionally, we compared the posterior means and 95\% posterior credible intervals for all elements of $\bm{v}$ (not shown) and there were no meaningful differences between the two samplers.

\section{Simulation Study {\color{black}to Assess Parameter Recovery}}

\begin{figure}[t]
 	\centering
 	\includegraphics[width=.9\textwidth]{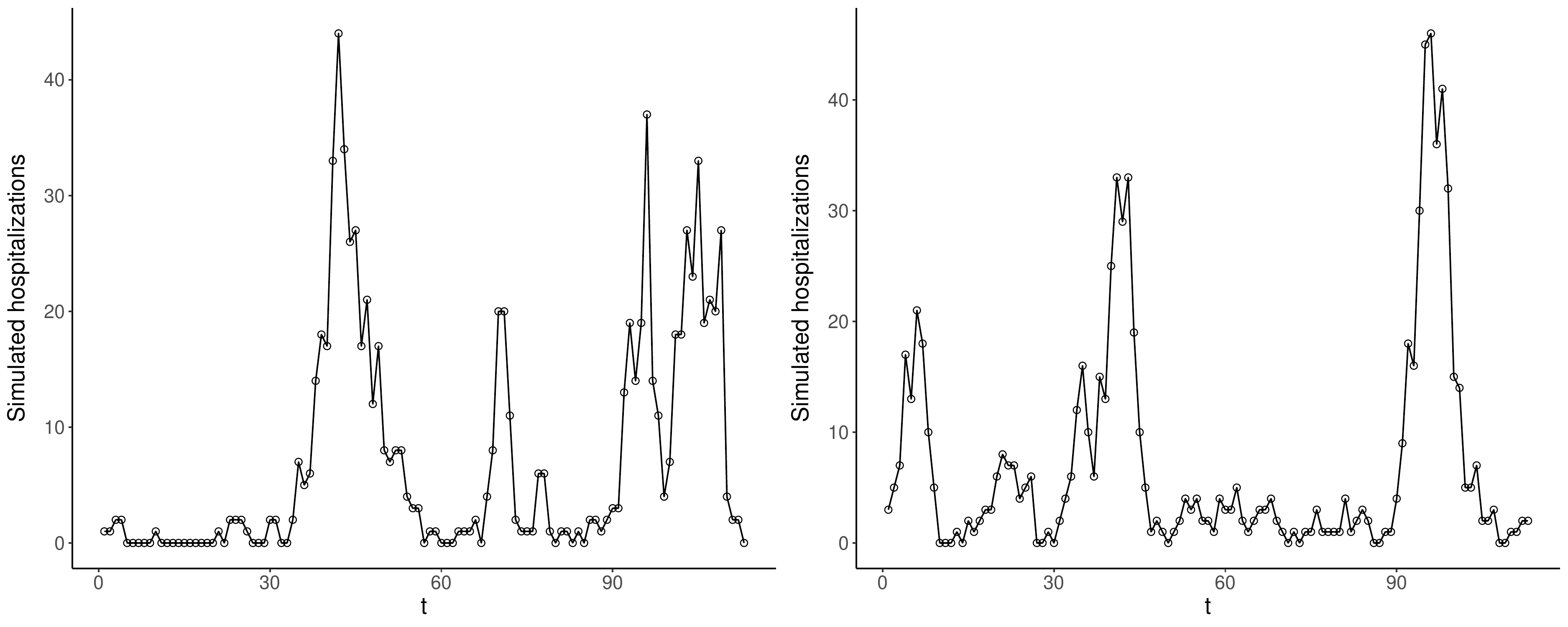}
	\caption{Shows simulated hospitalizations in 2 areas from a single replication of (5)\label{fig:sim_hosp}.} 
\end{figure}

We designed a simulation study to ensure our hybrid Gibbs sampling algorithm could recover the true parameters of the CMSNB(1,2,4) model. We simulated data from a slightly simplified version of the CMSNB(1,2,4) model specified in Section 5.1 of the main text. More specifically, we removed the random intercepts, we removed some insignificant effects from the Markov chain {\color{black}and the outbreak and endemic transmission rates}, and we assumed that the overdispersion in the hospitalizations was the same during the endemic and outbreak periods. This was done to reduce the number of parameters in the model as we need to run many simulations each of which is computationally costly. We generated data from the following CMSNB(1,2,4) model,
\begin{align}
\begin{split}
\log(\lambda_{it}^{EN}) &= \beta_{0}^{EN}+\beta_{\text{beds}}^{EN}\text{beds}_i + {\color{black} \beta_{\text{mob}}^{EN}\text{mobility}_{\text{county}(i)(t-4)}}+\rho^{EN}\log(y_{i(t-1)}+1) \\
\log(\lambda_{it}^{OB}) &= \beta_{0}^{OB}+\beta_{\text{beds}}^{OB}\text{beds}_i+{\color{black} \beta_{\text{mob}}^{OB}\text{mobility}_{\text{county}(i)(t-4)}}+\rho^{OB}\log(y_{i(t-1)}+1) \\
r^{EN}&=r^{OB}=r \\
\text{logit}(p12_{it}) &= \alpha_{12,0}+\alpha_{12,\text{beds}}\text{beds}_i \\
\log\left(\frac{p21_{it}}{1-p21_{it}-p23_{it}}\right) &=  \alpha_{21,0}+\alpha_{21,\text{beds}}\text{beds}_i+{\color{black}\alpha_{21,\text{mob}} \text{mobility}_{\text{county}(i)(t-4)}} \\
\log\left(\frac{p23_{it}}{1-p21_{it}-p23_{it}}\right) &=  \alpha_{23,0}+\alpha_{23,\text{mobi}}\text{mobility}_{\text{county}(i)(t-4)}+\alpha_{23,\text{newv}}\text{new\_variant}_t \\ & \,\,\,\, +\alpha_{23,\text{spat}} \sum_{j \in NE(i)} \omega_{ji} I[S_{j(t-1)}=3] \\
\text{logit}(p33_{it}) &= \alpha_{33,0}+\alpha_{33,\text{mobi}}\text{mobility}_{\text{county}(i)(t-4)} \\ & \,\,\,\, +\alpha_{33,\text{spat}} \sum_{j \in NE(i)} \omega_{ji} I[S_{j(t-1)}=3], \label{eqn:sim_model}
\end{split}
\end{align} for $i=1,\dots,30$ and $t=2,\dots,113$, and with the following true parameter values $\bm{v}=(\beta_{0}^{EN},\beta_{\text{beds}}^{EN},{\color{black}\beta_{\text{mob}}^{EN}},\rho^{EN},\beta_{0}^{OB},\beta_{\text{beds}}^{OB},{\color{black}\beta_{\text{mob}}^{OB}},\rho^{OB},r, \alpha_{12,0},\alpha_{12,\text{beds}},\alpha_{21,0},\alpha_{21,\text{beds}},{\color{black}\alpha_{21,\text{mob}}},\alpha_{23,0}, \alpha_{23,\text{mobi}}, \\ \alpha_{23,\text{newv}}, \alpha_{23,\text{spat}},\alpha_{33,0},\alpha_{33,\text{mobi}},\alpha_{33,\text{spat}})^T={\color{black}(0,.17,.003,.65,.78,,06,.007,.75,10,-.76,.45,-3.6,} \\ {\color{black} -.9,-.035,-4.15,.025,2.5,1.15,2,.025,.45)^T}$. The true parameter values were chosen to be similar to those estimated in our motivating example. In (\ref{eqn:sim_model}) $\text{beds}_i$, $\text{mobility}_{\text{county}(i)(t-4)}$, $\text{new\_variant}_t$, $NE(i)$ and $\omega_{ji}$ are all the same as in our motivating example. Finally, we assumed a uniform initial state distribution for the Markov chain in each area. Figure \ref{fig:sim_hosp} shows the simulated hospitalizations in 2 areas from a single replication of (\ref{eqn:sim_model}) and they appear somewhat realistic.

{\color{black}We considered two sets of constraints for fitting the CMSNB(1,2,4) model to simulations of (\ref{eqn:sim_model}). Firstly, we considered a constraint on just the intercepts and the autoregressive coefficients, $\beta_0^{EN}+.1<\beta_0^{OB}$ and $\rho^{EN}+.05<\rho^{OB}$, which we will refer to as the weak constraints. We also considered constraining the entirety of the transmission rates, $$\beta_{0}^{EN}+\beta_{\text{beds}}^{EN}\text{beds}_i + \beta_{\text{mob}}^{EN}\text{mobility}_{\text{county}(i)(t-4)}+.01<\beta_{0}^{OB}+\beta_{\text{beds}}^{OB}\text{beds}_i + \beta_{\text{mob}}^{OB}\text{mobility}_{\text{county}(i)(t-4)}$$ for $i=1,\dots,30$ and $t=2,\dots,113$, and $\rho^{EN}+.05<\rho^{OB}$, which we will refer to as the strong constraints. We chose minimum differences of .1 for the weak constraints on the intercepts and .01 for the strong constraints on the entirety of the transmission rates as the strong constraints constrain the minimum difference in transmission while the weak constraints constrain the average difference in transmission (we center all covariates) which should be a larger difference.} {\color{black} With our hybrid Gibbs sampling algorithm we fit the CMSNB(1,2,4) model (correctly specified) to 250 replications of (\ref{eqn:sim_model}) using the strong constraints and to another 700 replications of (\ref{eqn:sim_model})  (due to the low convergence rate, see below) using the weak constraints.} We mostly assumed the same prior distribution for $\bm{v}$ as specified in Section 2.2 of the main text. The only exception is that we used wider priors for $\alpha_{23,\text{spat}}$ and $\alpha_{33,\text{spat}}$ as our goal was not to shrink these effects but to recover the true parameter values. We ran our Gibbs sampler for 200,000 iterations on 3 chains, started from random values in the parameter space, with an initial burn-in of 50,000 iterations. For each replication convergence of the Gibbs sampler was checked using the minimum effective sample size ($>$1000) and the maximum Gelman-Rubin statistic ($<$1.05)\citep{plummerCODAConvergenceDiagnosis2006}.

\begin{figure}[t]
 	\centering
 	\includegraphics[width=\textwidth]{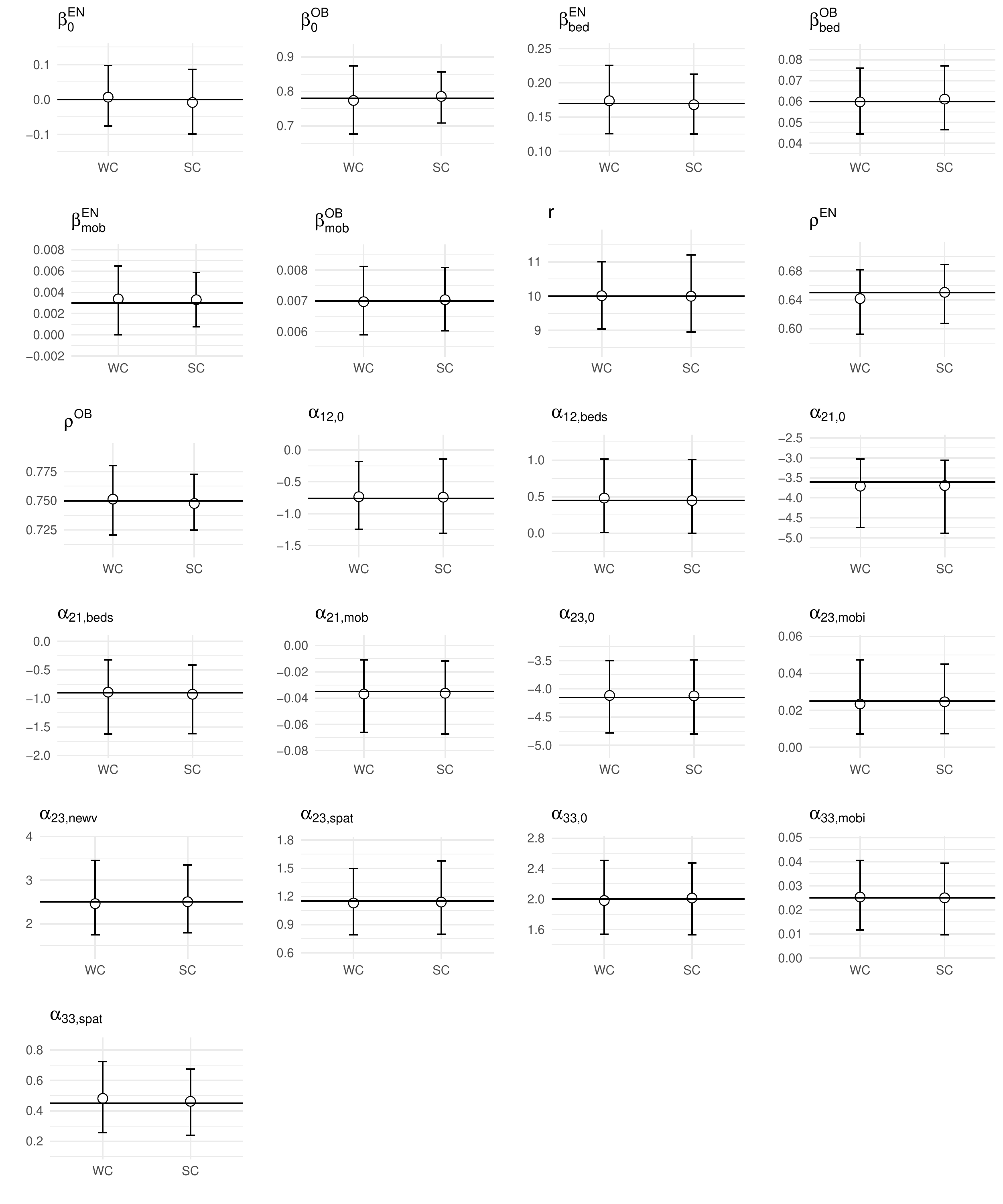} 
	\caption{{\color{black}Shows the sample mean (circles) and 95\% quantile (caps) of the posterior medians from fitting 237 replications of (5) using the strong constraints (SC) (see text around figure) and an additional 284 replications of (5) using the weak constraints (WC), with our hybrid Gibbs sampling algorithm.} The horizontal lines are drawn at the true parameter values.  \label{fig:sim_both}} 
\end{figure}

{\color{black}When using the weak constraints, our Gibbs sampler passed the convergence checks for only $284/700 = 40.6\%$ of the replications. When using the strong constraints, the convergence rate was much higher with our Gibbs sampler converging for $237/250=94.8\%$ of the replications.} {\color{black}Figure \ref{fig:sim_both} gives the sample mean and 95\% quantile of the posterior medians for the 237/250 replications that converged using the strong constraints and the 284/700 replications that converged using the weak constraints.} {\color{black} Figure \ref{fig:sim_both} shows that when our Gibbs sampler converges, using either the weak or the strong constraints, the posterior distributions are centered close to the true parameter values on average.} {\color{black}Additionally, the average coverage of the 95\% credible intervals was .95 with a minimum coverage of .91 and a maximum coverage of .98, when using either constraint, showing good coverage of the true parameter values.}  

{\color{black}In conclusion, regardless of whether the strong or weak constraints were used our hybrid Gibbs sampling algorithm was able to recover the true parameter values of the CMSNB(1,2,4) model well when it converged. However, the convergence rate was low when using the weak constraints. The strong constraints greatly improved the convergence rate and, from Figure \ref{fig:sim_both}, did not introduce any significant bias into the inferential procedure. Additionally, checking the strong constraints did not add a substantial amount of time to the model fitting (less than 1 hour). Therefore, we would recommend constraining the entirety of the transmission rates in practice and we do so throughout the main manuscript. We believe these constraints are reasonable in most applications since transmission should always increase when moving from the endemic state to the outbreak state.}

{\color{black} We found in our simulation study that even when the strong constraints are used our Gibbs sampler can still run into rare convergence issues, around 5\% of the time.} By examining a few of the non-converged MCMC samples{\color{black}, when the strong constraint was used,} we found the convergence issues were caused by genuine multimodality (the presence of multiple non-symmetric modes) in the posterior distribution, a common occurrence in Bayesian mixture modeling \citep{stephensDealingLabelSwitching2000,jasraMarkovChainMonte2005}. Multimodality in the posterior can be an important issue in Bayesian inference as most standard MCMC sampling algorithms, such as Metropolis-Hastings, do not mix well between the modes and, therefore, may not explore all important regions of the parameter space \citep{yaoStackingNonmixingBayesiana}. We cannot guarantee the absence of genuine multimodality in the {\color{black}237/250 or 284/700} replications where our Gibbs sampler passed convergence checks, as all chains could have been stuck in the same mode (although we do start the chains from random values in the parameter space making this less likely). However, when the Gibbs sampler passed convergence checks in our simulation study, we were able to recover the true parameters well. {\color{black}This implies that if there were extra modes in those posteriors they were too minor to affect inference significantly.} To double-check for genuine multimodality in our motivating example in Section 5 of the main text, we ran an additional {\color{black} 9} chains for the Full Coupled Model, started from random values in the parameter space, and they all converged to the same mode.

{\color{black} In all non-converged MCMC samples that we checked, when using the strong constraints, there was a mode centered close to the true parameter values and an extra mode centered on a submodel of the CMSNB(1,2,4) model that only used the absence state and the endemic state (based on the locations of the different MCMC chains). That is, the posterior probability of outbreak emergence and persistence, at average covariate values, was essentially 0 for the extra mode. It is not uncommon for the likelihood of a complex mixture model to contain multiple local modes in regions of the parameter space where some of the states have close to 0 probability of being visited, see the discussion in Section 4.2.3 of \cite{fruhwirth-schnatterFiniteMixtureMarkov2006}. (Note our strong constraints would not prevent this kind of multimodality as if the outbreak state is never visited then the transmission rate in the outbreak state does not contribute to the likelihood.) If such multimodality is encountered, it is recommended by \cite{fruhwirth-schnatterFiniteMixtureMarkov2006} to bound the posterior away from these regions. For instance, in our model, one could assign priors to the intercepts $\alpha_{lk,0}$ for $lk=12,21,23,33$, in Equations (4)-(5) of the main text, that shrink them towards 0 so that all states are likely to have a non-negligible chance of being visited. Another possible solution is to treat it as a model comparison problem, fitting the reduced model, that only uses an absence and one count state, and discarding the chains from the full model associated with the reduced model if it has a higher widely applicable information criterion (WAIC) (the WAIC is discussed in Section 3 below). Other possible solutions to consider if multimodality is encountered include stacking \citep{yaoStackingNonmixingBayesiana} (weighting the chains associated with different modes based on some comparison criteria) and tempered MCMC \citep{jasraMarkovChainMonte2005} (a type of MCMC algorithm that is often successful in sampling multimodal posteriors). It is difficult to test any of these possible solutions formally as, when using the strong constraints, we only ran into convergence issues 5\% of the time, and so many simulations would have to be run to see if they could recover the true parameters despite the non-convergence of our Gibbs sampler.}

\section{\color{black}Widely Applicable Information Criterion (WAIC)}

\subsection{\color{black}Formulation}

{\color{black}
As mentioned in the main text, the WAIC for a state-space model is more accurate when the latent states are marginalized \citep{auger-methe_guide_2021}. Starting with models that do not contain neighboring outbreak indicators in
the transition probabilities, such as the Non-coupled Model from Section 5 of the main text, we can use the forward filtering part of the FFBS algorithm \citep{fruhwirth-schnatterFiniteMixtureMarkov2006} to  calculate the marginalized density $p(y_{it}|\bm{y}_{1:(t-1)},\bm{v})$, where $\bm{y}_{1:(t-1)}$
is the vector of all counts in all areas through $t-1$. Then, following \cite{gelman_understanding_2014}, the WAIC can be calculated as,
\begin{align}\label{eqn: WAIC}
    \begin{split}
   &\text{lpdd} = \sum_{i=1}^{N}\sum_{t=2}^T \log\left(\frac{1}{Q-M}\sum_{m=M+1}^Q p(y_{it}|\bm{y}_{1:(t-1)},{\bm{v}}^{[m]})\right), \\
   &\text{pwaic} =\sum_{i=1}^{N}\sum_{t=2}^T Var_{m=M+1}^Q \log\left(p(y_{it}|\bm{y}_{1:(t-1)},{\bm{v}}^{[m]})\right),\\
   &\text{WAIC} =-2(\text{lpdd}-\text{pwaic}),
   \end{split}
\end{align} where the superscript $[m]$ denotes a draw from the posterior of the variable and $Var$ denotes the sample variance.

For models with neighboring outbreak indicators in the transition probabilities, such as the Full Coupled Model from Section 5 of the main text, we cannot completely marginalize $\bm{S}^{*}$ and calculate $p(y_{it}|\bm{y}_{1:(t-1)},\bm{v})$ \citep{douwes-schultzZerostateCoupledMarkov2022a}. As an alternative we could condition on $\bm{S}^{*}$ and use $p(y_{it}|\bm{S}^{*[m]},\bm{y}_{1:(t-1)},\bm{v}^{[m]})=p(y_{it}|S_{it}^{[m]},y_{i(t-1)},\bm{\beta}^{[m]})$, which is given by Equation (1) in the main text, in place of $p(y_{it}|\bm{y}_{1:(t-1)},{\bm{v}}^{[m]})$ in Equation (\ref{eqn: WAIC}). For state-space models this is sometimes called the conditional WAIC, however, it has been shown to be inaccurate \citep{auger-methe_guide_2021}. As a compromise we marginalize as much of $\bm{S}^{*}$ as is computationally possible from $p(y_{it}|\bm{S}^{*},\bm{y}_{1:(t-1)},\bm{v})$. Note, from Equation (\ref{eqn:iFF}) in Section 1 above, it is possible to use the forward filtering part of the iFFBS algorithm to calculate the partially marginalized density,
\begin{align*}
p(y_{it}|\bm{S}_{(-i)(1:t)}^*,\bm{y}_{1:(t-1)},\bm{v}) &= p(y_{it}|\bm{S}_{(-i)(1:t)}^*,\bm{y}_{i(1:t-1)},\bm{v}) \\
& =\sum_{k=1}^{7} p(y_{it}|S_{it}^*=k,y_{i(t-1)},\bm{v})p(S_{it}^*=k|\bm{S}_{(-i)(1:t)}^*,\bm{y}_{i(1:t-1)},\bm{v}),
\end{align*} meaning we can marginalize all of $\bm{S}^{*}$ that is from the area and all future states in neighboring areas. Therefore, we use  $p(y_{it}|\bm{S}_{(-i)(1:t)}^{*[m]},\bm{y}_{1:(t-1)},\bm{v}^{[m]})$ in place of $p(y_{it}|\bm{y}_{1:(t-1)},{\bm{v}}^{[m]})$ in Equation (\ref{eqn: WAIC}). In the next subsection, we show that the WAIC calculated in this manner can distinguish the correct model.

\subsection{\color{black}Simulation study}

We designed a simulation study to ensure the WAIC, as formulated in the previous subsection, can choose the true data generating model. We focused on a comparison between models with/without neighboring outbreak indicators in the transition probabilities as this is one of the most important comparisons in the main text. 

Firstly, we considered a CMSNB(1,2,4) model without neighboring outbreak indicators in the Markov chain,
\begin{align} 
\begin{split} \label{eqn:nonspat}
\log(\lambda_{it}^{EN}) &= \beta_{0}^{EN}+\beta_{\text{beds}}^{EN}\text{beds}_i + \beta_{\text{mob}}^{EN}\text{mobility}_{\text{county}(i)(t-4)}+\rho^{EN}\log(y_{i(t-1)}+1) \\
\log(\lambda_{it}^{OB}) &= \beta_{0}^{OB}+\beta_{\text{beds}}^{OB}\text{beds}_i+ \beta_{\text{mob}}^{OB}\text{mobility}_{\text{county}(i)(t-4)}+\rho^{OB}\log(y_{i(t-1)}+1) \\
r^{EN}&=r^{OB}=r \\
\text{logit}(p12_{it}) &= \alpha_{12,0}+\alpha_{12,\text{beds}}\text{beds}_i \\
\log\left(\frac{p21_{it}}{1-p21_{it}-p23_{it}}\right) &=  \alpha_{21,0}+\alpha_{21,\text{beds}}\text{beds}_i \\
\log\left(\frac{p23_{it}}{1-p21_{it}-p23_{it}}\right) &=  \alpha_{23,0}+\alpha_{23,\text{mobi}}\text{mobility}_{\text{county}(i)(t-4)}+\alpha_{23,\text{newv}}\text{new\_variant}_t \\
\text{logit}(p33_{it}) &= \alpha_{33,0}+\alpha_{33,\text{mobi}}\text{mobility}_{\text{county}(i)(t-4)}, 
\end{split}
\end{align} for $i=1,\dots,30$ and $t=2,\dots,113$, and with the following true parameter values $\bm{v}=(\beta_{0}^{EN},\beta_{\text{beds}}^{EN},\beta_{\text{mob}}^{EN},\rho^{EN},\beta_{0}^{OB},\beta_{\text{beds}}^{OB},\beta_{\text{mob}}^{OB},\rho^{OB},r, \alpha_{12,0},\alpha_{12,\text{beds}},\alpha_{21,0},\alpha_{21,\text{beds}},\alpha_{23,0}, \alpha_{23,\text{mobi}}, \\ \alpha_{23,\text{newv}},\alpha_{33,0},\alpha_{33,\text{mobi}})^T=(0,.1,0,.5,.75,.05,.007,.75,10,-1,.5,-3,-1,-3.5,.04,1, \\ 2.5,.02)^T$. We will refer to the model defined by Equation (\ref{eqn:nonspat}) as the Non-spatial Model. 

We also considered a spatial extension of the Non-spatial Model with neighboring outbreak indicators in each transition probability, 
\begin{align} 
\begin{split} \label{eqn:spat}
\log(\lambda_{it}^{EN}) &= \beta_{0}^{EN}+\beta_{\text{beds}}^{EN}\text{beds}_i + \beta_{\text{mob}}^{EN}\text{mobility}_{\text{county}(i)(t-4)}+\rho^{EN}\log(y_{i(t-1)}+1) \\
\log(\lambda_{it}^{OB}) &= \beta_{0}^{OB}+\beta_{\text{beds}}^{OB}\text{beds}_i+ \beta_{\text{mob}}^{OB}\text{mobility}_{\text{county}(i)(t-4)}+\rho^{OB}\log(y_{i(t-1)}+1) \\
r^{EN}&=r^{OB}=r \\
\text{logit}(p12_{it}) &= \alpha_{12,0}+\alpha_{12,\text{beds}}\text{beds}_i +\alpha_{12,\text{spat}} \sum_{j \in NE(i)} \omega_{ji} I[S_{j(t-1)}=3] \\
\log\left(\frac{p21_{it}}{1-p21_{it}-p23_{it}}\right) &=  \alpha_{21,0}+\alpha_{21,\text{beds}}\text{beds}_i +\alpha_{21,\text{spat}} \sum_{j \in NE(i)} \omega_{ji} I[S_{j(t-1)}=3] \\
\log\left(\frac{p23_{it}}{1-p21_{it}-p23_{it}}\right) &=  \alpha_{23,0}+\alpha_{23,\text{mobi}}\text{mobility}_{\text{county}(i)(t-4)}+\alpha_{23,\text{newv}}\text{new\_variant}_t \\
& \,\,\,\, +\alpha_{23,\text{spat}} \sum_{j \in NE(i)} \omega_{ji} I[S_{j(t-1)}=3] \\
\text{logit}(p33_{it}) &= \alpha_{33,0}+\alpha_{33,\text{mobi}}\text{mobility}_{\text{county}(i)(t-4)} +\alpha_{33,\text{spat}} \sum_{j \in NE(i)} \omega_{ji} I[S_{j(t-1)}=3], 
\end{split}
\end{align} for $i=1,\dots,30$ and $t=2,\dots,113$, and with the following true parameter values $\bm{v}=(\beta_{0}^{EN},\beta_{\text{beds}}^{EN},\beta_{\text{mob}}^{EN},\rho^{EN},\beta_{0}^{OB},\beta_{\text{beds}}^{OB},\beta_{\text{mob}}^{OB},\rho^{OB},r, \alpha_{12,0},\alpha_{12,\text{beds}},\alpha_{12,\text{spat}},\alpha_{21,0},\alpha_{21,\text{beds}},\alpha_{21,\text{spat}} \\ ,\alpha_{23,0}, \alpha_{23,\text{mobi}},\alpha_{23,\text{newv}},\alpha_{23,\text{spat}},\alpha_{33,0},\alpha_{33,\text{mobi}},\alpha_{33,\text{spat}})^T=(0,.1,0,.5,.75,.05,.007,.75,10,\\-1,.5,.25,-3,.-1,-.25,-4,.04, 1,1.2,2,.02,.5)^T$. We will refer to the model defined by Equation (\ref{eqn:spat}) as the Spatial Model. 

Like in Section 2 above we simplified the models slightly compared to the main manuscript, such as by removing the random intercepts, to reduce the computational burden of running many simulations. The true parameter values were chosen to be like those estimated in our motivating example. In (\ref{eqn:nonspat}) and (\ref{eqn:spat}) $\text{beds}_i$, $\text{mobility}_{\text{county}(i)(t-4)}$, $\text{new\_variant}_t$, $NE(i)$ and $\omega_{ji}$ are all the same as in our motivating example. Finally, we assumed a uniform initial state distribution for the Markov chain in each area.

We simulated 25 replications of the Non-spatial Model and fit the Spatial and Non-spatial Model to each replication. Then we simulated 25 replications of the Spatial Model and again fit the Spatial and Non-spatial Model to each replication. This resulted in 100 models being fit in total. Table \ref{tab:SMTable 1} gives the results of the model comparison using WAIC across all replications. There are 24 replications in each row as 2 of the models did not converge and were removed from the table, this is in line with the convergence rate reported in Section 2 above. From Table \ref{tab:SMTable 1}, when the Non-spatial Model generated the data the WAIC was split between choosing the correct Non-spatial Model and showing no significant difference between the models, and the WAIC only once chose the incorrect Spatial Model. For the replication where the WAIC chose the wrong model one of the spatial effects was highly significant, which can occur through random chance even when there is no real effect. When the Spatial Model generated the data the WAIC always chose the correct Spatial Model. These results suggest that when the WAIC shows no significant difference to prefer the less complex model and that the WAIC rarely selects the incorrect model when there is a significant difference.

\renewcommand{\arraystretch}{1.5}
\begin{table}[t]
\color{black}
\centering
\caption{\label{tab:SMTable 1} {\color{black}Shows the results of model comparison using the WAIC for 24 replications of the Non-spatial Model and 24 replications of the Spatial Model. Each row represents the true model from which the 24 replications were produced.}}
\begin{tabular}{@{}llll@{}}
\toprule
\textbf{True Model} & \textbf{\begin{tabular}[c]{@{}l@{}}Non-spatial \\[-5pt] Model Chosen\end{tabular}} & \textbf{\begin{tabular}[c]{@{}l@{}}Spatial\\[-5pt] Model Chosen\end{tabular}} & \textbf{\begin{tabular}[c]{@{}l@{}}No Significant Difference \\[-5pt] $(|\bm{\Delta}\textbf{WAIC}|\bm{<5})$\end{tabular}} \\ \midrule
Non-spatial         &       13                                                                       &   1                                                                      &   10                                                                                                              \\
Spatial             &      0                                                                        &    24                                                                     &     0                                                                                                            \\ \bottomrule
\end{tabular}
\end{table}

A limitation is that we only compared a spatial and non-spatial model. The simulations are very computationally costly as one needs to run 2 models for every replication and there are multiple replications per model. Comparison between spatial and non-spatial models is arguably the most important comparison problem in the main text so we focused on it here.

\section{\color{black}Endemic-epidemic Model}

Following \cite{bracher_endemic-epidemic_2022}, the Endemic-epidemic (EE) framework assumes the counts follow a multivariate autoregressive negative binomial model,
\begin{align*}
y_{it}|\bm{y}_{1:(t-1)} \sim NB(\lambda_{it},\psi),
\end{align*} where $\bm{y}_{1:(t-1)}$ is the vector of all counts in all areas through $t-1$. The conditional mean $\lambda_{it}$ is decomposed as,
\begin{align} \label{eqn:EE_lambda}
\lambda_{it} = \lambda_{it}^{EP}\sum_{d=1}^{p}\sum_{j \in NE(i)}\lfloor u_d\rfloor \lfloor \omega_{ji} \rfloor y_{j,t-d} + \lambda_{it}^{BL}.
\end{align} Here we will assume $NE(i)$ is the same as in Section 5.1 of the main text but also includes area $i$. The first term on the right-hand side of (\ref{eqn:EE_lambda}) accounts for expected new incidence due to local and neighboring disease transmission. The effects of covariates on transmission can be captured by the epidemic component $\lambda_{it}^{EP}$ which is modeled in a log-linear fashion, in our case we assume,
\begin{align*}
\log(\lambda_{it}^{EP}) =\gamma_{0i}^{EP} + \gamma_{1}^{EP}\text{beds}_i+ \gamma_{2}^{EP}\text{mobility}_{\text{county}(i)(t-4)}+\gamma_3^{EP}\text{newv}_t,
\end{align*} where $\gamma_{0i}^{EP} \sim N(\gamma_0^{EP},\left(\sigma^{EP}\right)^2)$ is a normal random intercept. In (\ref{eqn:EE_lambda}) the spatial relationship between areas is described by the weights $\omega_{ji}>0$ which are normalized $\lfloor \omega_{ji} \rfloor=\omega_{ji}/\sum_{h: j \in NE(h)} \omega_{jh}$. The spatial weights are often assumed to follow a function, with unknown parameters, that decays with some measure of distance between areas \citep{bracher_endemic-epidemic_2022}. Following the main text Section 5.1, we assume $\omega_{ji}=\exp(-\phi d_{ji}^{BD})$, where $\phi>0$ is decay parameter to be estimated and given a $U(0,1000)$ prior (note that $d_{ii}^{BD}=0$). 

In (\ref{eqn:EE_lambda}) the relationship in the counts across time is described by the weights $u_d>0$ which are normalized $\lfloor u_d \rfloor=u_d/\sum_{g=1}^{p}u_g$ for identifiability \citep{bracher_endemic-epidemic_2022}. We use a geometric specification \citep{bracher_endemic-epidemic_2022} for the weights, $u_d=(1-\kappa)^{d-1}\kappa$, where $0<\kappa<1$ is to be estimated and given a $U(0,1)$ prior. Geometric weights decay over time. This is a reasonable assumption in our application as the serial interval, i.e., the time between the appearance of symptoms in successive generations, of COVID-19 is likely less than a week \citep{nishiura_serial_2020} implying most of the weight should be placed on the initial lag \citep{bracher_endemic-epidemic_2022}. 

In (\ref{eqn:EE_lambda}) $\lambda_{it}^{BL}$ is the baseline component, sometimes known as the endemic component \citep{bracher_endemic-epidemic_2022} (we use baseline to avoid confusion with $\lambda_{it}^{EN}$ in the main text), which captures contributions to incidence not directly related to local and neighboring disease transmission. For example, the baseline component might help capture imported cases from outside the study area. Like the epidemic component, the baseline component follows a log-linear model, in our case we assume,
\begin{align*}
\log(\lambda_{it}^{BL}) =\gamma_{0i}^{BL} + \gamma_{1}^{BL}\text{beds}_i+ \gamma_{2}^{BL}\text{mobility}_{\text{county}(i)(t-4)}+\gamma_3^{BL}\text{newv}_t,
\end{align*} where $\gamma_{0i}^{BL} \sim N(\gamma_0^{BL},\left(\sigma^{BL}\right)^2)$ is a normal random intercept. We found models with covariates in the baseline component fit better, according to the WAIC, compared to models without covariates.

To fit the EE models, we used standard MCMC methods in Nimble with wide priors, the code is available on GitHub \url{https://github.com/Dirk-Douwes-Schultz/CMSNB124_code}. We used the WAIC to choose the maximum temporal lag $p$ and to compare the EE model to the Markov switching models from the main text (see Section 5.1 in the main text). To compare models using the WAIC they must be fit to the same data. The Markov switching models conditioned on the first observation and were fit to $y_{it}$ for $i=1,\dots,30$ and $t=2,\dots,113$. Therefore, to remain consistent we introduced times $t=0$ and $t=-1$ and fit the EE models across $t=2,\dots,113$ conditioning on $t=1,0,-1$ (depending on $p$). Then to calculate the WAIC for the EE models we used Equation \ref{eqn: WAIC} in Section 3.1 above with $p(y_{it}|\bm{y}_{1:(t-1)},\bm{v}^{[m]})$ substituted with $p(y_{it}|\bm{y}_{(-1):(t-1)},\bm{v}^{[m]})=NB(y_{it}|\lambda_{it}^{[m]},\psi^{[m]})$, where $NB(y|\lambda,\psi)$ denotes a negative binomial density evaluated at $y$ with mean $\lambda$ and overdispersion $\psi$. Before time $t=-1$ there were no COVID-19 hospitalizations, so we only considered EE models with $p$ up to 3. This should not be a major limitation as, from Table \ref{tab:WAICEE}, the WAIC increased going from $p=2$ to $p=3$. From Table \ref{tab:WAICEE}, we decided on an EE model with $p=2$ and we will only discuss the results from that model in this section. 

\renewcommand{\arraystretch}{1.5}
\begin{table}[t]
\color{black}
\centering
\caption{\color{black} \label{tab:WAICEE} Shows the WAIC of Endemic-epidemic models with different maximum lags $p$. The best fitting model, the one with the lowest WAIC, is bolded.}
\begin{tabular}{ll}
\hline
\textbf{Maximum Lag ($p$)}   & \textbf{WAIC} \\ \hline
1  &      17,884         \\
$\bm{2}$     &      $\bm{17,845}$       \\
3 &    17,848          \\ \hline
\end{tabular}
\end{table}

We estimated the decay parameter for the spatial weights with distance, $\phi$, as 79.47 (67.13, 93.81) (posterior mean and 95\% posterior credible interval), which is difficult to interpret given the normalization of the weights. From examining the posteriors of the normalized weights $\lfloor \omega_{ji} \rfloor$, for most areas 85-90 percent of the weight was placed on the home area ($\lfloor \omega_{ii} \rfloor$) and the remaining 10-15 percent was largely distributed over the closest 1-2 neighbors. This implies most transmission occurred locally and from close neighbors. We estimated  $\kappa$ as .85 (.80, .90) and 87 (83, 91) percent of the temporal weight was placed on the first lag ($\lfloor u_{1} \rfloor$) and 13 (9, 17) percent was placed on the second lag ($\lfloor u_{2} \rfloor$). A large portion of the temporal weight was placed on the previous week which is sensible considering the short serial interval of COVID-19.

Table \ref{tab:EEest} shows the estimates from the baseline and epidemic components. The ``Epidemic'' column of the table shows the effect of each covariate on local and neighboring disease transmission. Mobility had a strong positive association with local and neighboring transmission while new variant and beds mainly affected the disease process through the baseline component. Note, as we are modeling hospitalizations, an effect in terms of transmission in the hospitalizations could reflect an effect on transmission in the actual cases and/or an effect on the severity of the disease. Mobility and beds should not affect disease severity; however, new variants often do \citep{chenchulaCurrentEvidenceEfficacy2022}. Therefore, the effects of new variant need to be interpreted with caution. It is interesting to compare the estimates in Table \ref{tab:EEest} to those in Tables 2 and 3 in the main text. Both models largely agree that all three covariates are important for the disease process. However, we would argue the CMNSB(1,2,4) model offers a deeper understanding of the covariate effects as it breaks down the effect of each covariate into its effect on transmission during the endemic and outbreak periods and on the epidemiological transitions. By breaking down the effects it could make it easier for policy makers to implement suitable interventions.

\begin{table}[t]
\color{black}
\centering
\caption{\color{black}\label{tab:EEest} Posterior means and 95\% posterior credible intervals (in parenthesis) from the baseline and epidemic components of the fitted Endemic-epidemic model. The intercepts and covariate effects are exponentiated so that they represent rates and rate ratios. Rate ratios whose 95\% posterior
credible intervals do not contain 0 are bolded. The units for beds and mobility are equal to one standard deviation.}

\begin{tabular}{lccc} 
 \hline           &                         & \multicolumn{2}{c}{\textbf{Rate Ratios}}            \\ \hline 
    \textbf{Covariate}          &  \textbf{Parameter}           & \textbf{Baseline} & \textbf{Epidemic}  \\  \hline
Intercept of random intercepts  & $e^{\gamma_0}$   & .84  & .93 \\[-5pt]
           &  &  (.71, .99) & (.88, .99)  \\
Std. dev of random intercepts & $\sigma$   & .66   & .13 \\[-5pt]
           &  &  (.42, .99) & (.09, .17)  \\
beds (100s) & $e^{\gamma_{1}}$   & \textbf{1.18}  & 1.02 \\[-5pt]
           &  &  \text{(1.01, 1.36)} & (.98, 1.06)  \\
mobility (20\%) & $e^{\gamma_2}$   & \textbf{1.25}  & \textbf{1.14} \\[-5pt]
           &  &  \textbf{(1.05, 1.49)} & \textbf{(1.11, 1.18)}  \\
new variant & $e^{\gamma_3}$   & \textbf{1.46}  & \textbf{1.06} \\[-5pt]
           &  &  \textbf{(1.20, 1.75)} & \textbf{(1.01, 1.11)}  \\
           \hline
\end{tabular}
\end{table}

There are also some advantages of the EE models. The EE model only took one hour to fit while the CMSNB(1,2,4) model took eight hours. Also, the EE model can be fit using standard MCMC methods while the CMSNB(1,2,4) model requires the coding of custom iFFBS samplers. The CMSNB(1,2,4) model additionally requires implementing constraints to ensure consistent convergence. Finally, the EE models allow one to better understand the temporal and spatial structure of the disease counts by estimating the weights $u_d$ and $\omega_{ji}$. It is difficult to implement spatial and temporal weighting of the disease counts in the CMSNB(1,2,4) model due to the complexity of those models. Appropriate constraints would have to be considered as well as how the weight structure might switch between the states.
}

\section{Temporal Predictions}

In this section, we are interested in the posterior predictive distribution of the counts $p(y_{i(T +k)}|\bm{y})$ and the epidemiological state of the disease $p(S_{i(T +k)}^*|\bm{y})$  for $i = 1, \dots, N$ and $k = 1, \dots , K$. The posterior predictive distribution of the counts is given by,
\begin{align}
\begin{split}
p(y_{i(T+k)}|\bm{y}) &=\int p(y_{i(T+k)}|S_{i(T+k)},y_{i(T+k-1)},\bm{\beta})p(S_{i(T+k)}^*|S_{i(T+k-1)}^*,\bm{S}_{(-i)(T+k-1)},\bm{\theta}) \\
& \qquad \times \prod_{j=1}^{N} p(y_{j(T+k-1)}|S_{j(T+k-1)},y_{j(T+k-2)},\bm{\beta})p(S_{j(T+k-1)}^*|S_{j(T+k-2)}^*,\bm{S}_{(-j)(T+k-2)},\bm{\theta}) \\ 
& \qquad  \dotsc \times \prod_{j=1}^{N} p(y_{j(T+1)}|S_{j(T+1)},y_{jT},\bm{\beta})p(S_{j(T+1)}^*|S_{jT}^*,\bm{S}_{(-j)T},\bm{\theta}) \\
& \qquad  \times p(\bm{S}_{(1:N)T}^*,\bm{v}|\bm{y}) \, d S_{i(T+k)}^* d\bm{y}_{(1:N)(T+1:T+k-1)} d\bm{S}_{(1:N)(T:T+k-1)}^* d\bm{\beta}d\bm{\theta}, \label{eqn:predy_int}
\end{split}
\end{align}
and the posterior predictive distribution for the state of the disease in area $i$ is given by,
\begin{align}
\begin{split}
p(S_{i(T+k)}^*|\bm{y}) &= \int p(S_{i(T+k)}^*|S_{i(T+k-1)}^*,\bm{S}_{(-i)(T+k-1)},\bm{\theta}) \\
& \qquad \times \prod_{j=1}^{N} p(S_{j(T+k-1)}^*|S_{j(T+k-2)}^*,\bm{S}_{(-j)(T+k-2)},\bm{\theta}) \\ 
& \qquad  \dotsc \times \prod_{j=1}^{N} p(S_{j(T+1)}^*|S_{jT}^*,\bm{S}_{(-j)T},\bm{\theta}) \\
& \qquad  \times p(\bm{S}_{(1:N)T}^*,\bm{v}|\bm{y}) \, d\bm{S}_{(1:N)(T:T+k-1)}^* d\bm{\beta}d\bm{\theta}. \label{eqn:preds_int}
\end{split}
\end{align}

The integrals (\ref{eqn:predy_int}) and (\ref{eqn:preds_int}) are intractable, however, they can be approximated by Monte Carlo integration,
\begin{align}
p(y_{i(T+k)}|\bm{y}) & \approx \frac{1}{Q-M} \sum_{m=M+1}^{Q} p(y_{i(T+k)}|S_{i(T+k)}^{*[m]},y_{i(T+k-1)}^{[m]},\bm{\beta}^{[m]}), \label{eqn:montey}
\end{align} and 
\begin{align}
p(S_{i(T+k)}^*|\bm{y}) & \approx \frac{1}{Q-M} \sum_{m=M+1}^{Q} p(S_{i(T+k)}^*|S_{i(T+k-1)}^{*[m]},\bm{S}_{(-i)(T+k-1)}^{[m]},\bm{\theta}^{[m]}),   \label{eqn:montes}
\end{align} where, like in the main text, the superscript $[m]$ denotes a draw from the posterior of the variable.

We can use a simulation procedure to draw realizations from the posterior predictive distributions \citep{fruhwirth-schnatterFiniteMixtureMarkov2006}. Algorithm 1  will obtain realizations from the posterior predictive distribution of the counts, $y^{[m]}_{i(T+k)} \sim p(y_{i(T+k)}|\bm{y})$, and the epidemiological state of the disease, $S^{*[m]}_{i(T+k)} \sim p(S_{i(T+k)}^*|\bm{y})$, for $i = 1, \dots , N$, $k = 1, \dots , K$ and $m = M + 1, \dots , Q$. Then the realizations from the posterior predictive distributions can be substituted into Equations (\ref{eqn:montey}) and (\ref{eqn:montes}) if the posterior predictive distributions themselves need to be calculated. Although, as $S_{i(T+k)}^*$ can only take 7 values it is easier to approximate $p(S_{i(T+k)}^*|\bm{y})$ with the frequency distribution of $S^{*[m]}_{i(T+k)}$. Note that in Algorithm 1, the first step draws a new state for the disease from the Markov chain in each area and then step 2 draws new counts conditional on the new states.

\begin{algorithm}[t]
\caption{Posterior Predictive Simulation}
\label{alg:alg1}
\SetAlgoLined
\For{$m$ in  $M+1:Q$}{
\For{$k$ in  $1:K$}{
\For{$i$ in $1:N$}{
1. Draw $S_{i(T+k)}^{*[m]}$ from $p(S_{i(T+k)}^*|S_{i(T+k-1)}^{*[m]},\bm{S}_{(-i)(T+k-1)}^{[m]},\bm{\theta}^{[m]})$. \\[10pt]
2. Draw $y_{i(T+k)}^{[m]}$ from $ p(y_{i(T+k)}|S_{i(T+k)}^{*[m]},y_{i(T+k-1)}^{[m]},\bm{\beta}^{[m]})$, where $y_{iT}^{[m]}=y_{iT}$.
}}}
\end{algorithm}

{\color{black}
\section{\color{black} Further Results from the Simulation Study to Quantify and Compare State Estimation (Section 4 of the Main Text)}

\subsection{\color{black}Simulated counts from 2 areas (SM Figure 4)}

Figure~\ref{fig:fig_sim3_ex} shows simulated counts in 2 areas from the simulation study described in Section 4 of the main text. 

\begin{figure}[t]
 	\centering
 	\includegraphics[width=.8\linewidth]{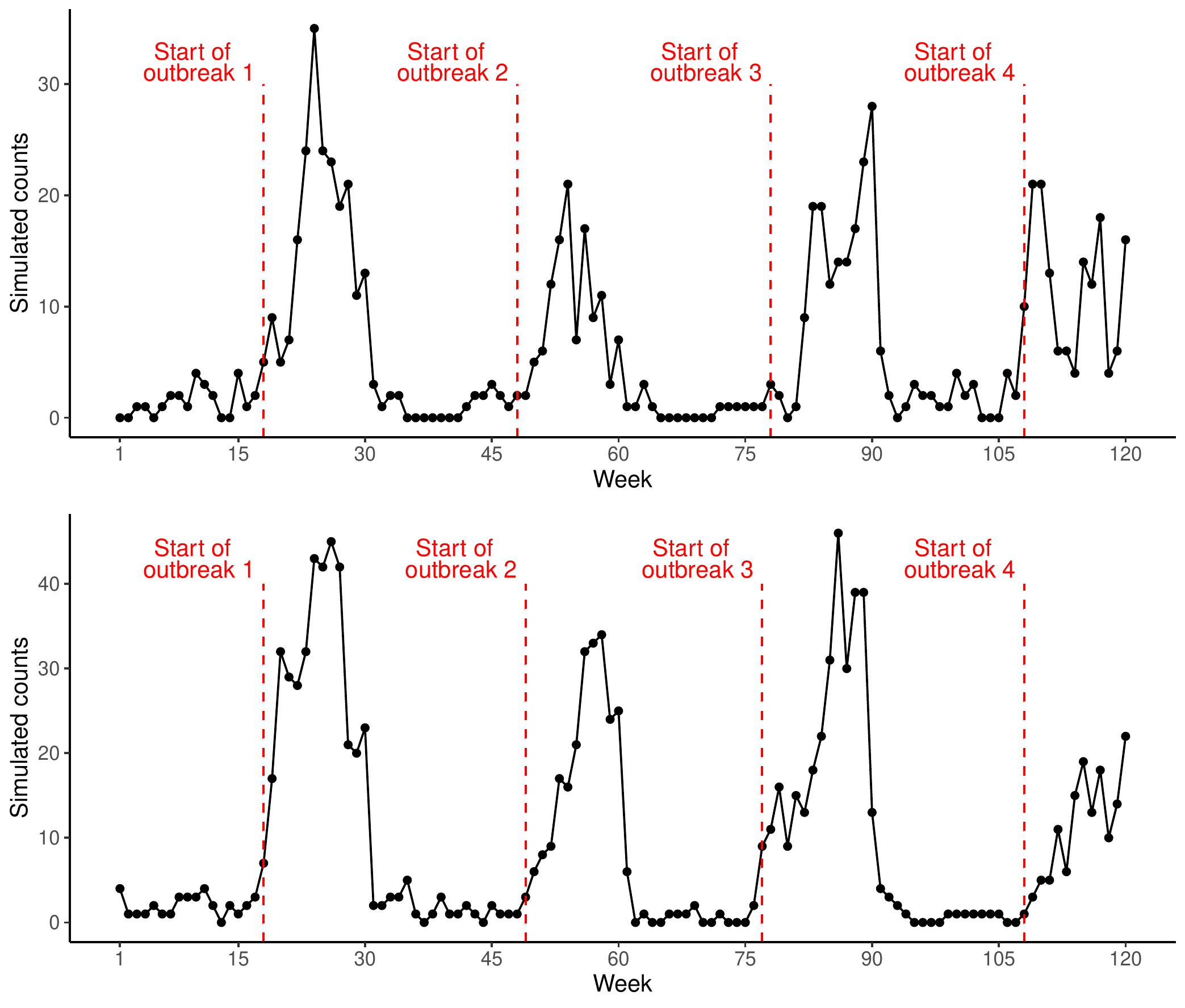}
	\caption{ \color{black}Shows simulated counts in 2 areas from the simulation study described in Section 4 of the main text. The red dotted lines indicate the exact pre-determined start time of each outbreak in the area. \label{fig:fig_sim3_ex}} 
\end{figure}

\subsection{\color{black}Primary results from the simulation study (SM Table 4)}

Table~\ref{tab:sim3SM} summarizes the primary results of the simulation study described in Section 4 of the main text.

\renewcommand{\arraystretch}{1.5}
\begin{table}[t]
\color{black}
\centering
\caption{\label{tab:sim3SM} \color{black} Shows the area under the ROC curve (AUC), sensitivity, specificity and timeliness (last 3 calculated with a 50 percent threshold) of three outbreak state estimates: (a) retrospective probabilities, $P(S_{it}=3|\bm{y})$ from fitting the models to the full simulated data set (b) real-time detection probabilities, $P(S_{iT}=3|\bm{y})$ from fitting the models up to time $T$ for $T=100,\dots,120$ (20 separate fits) and (c) real-time one week ahead forecasts, $P(S_{iT}=3|\bm{y}_{1:(T-1)})$ from fitting the models up to time $T-1$ for $T=101,\dots,120$. All outbreak state estimates were evaluated on the simulated data set described in Section 4 of the main text. The best criteria for each outbreak state estimate are bolded.}
\begin{tabular}{lcccccc}
\hline
\multicolumn{1}{c}{\textbf{}} & \multicolumn{2}{c}{\textbf{(a) Retrospective}} & \multicolumn{2}{c}{\textbf{\begin{tabular}[c]{@{}c@{}}(b) Real-time\\[-5pt] Detection\end{tabular}}} & \multicolumn{2}{c}{\textbf{\begin{tabular}[c]{@{}c@{}}(c) Real-time\\[-5pt] Forecast\end{tabular}}} \\ \hline
\textbf{Criterion}             & \textbf{Spatial}    & \textbf{Non-spatial}    & \textbf{Spatial}                            & \textbf{Non-spatial}                            & \textbf{Spatial}                            & \textbf{Non-spatial}                           \\ \hline
AUC                           & \textbf{.995}       & .992                    & \textbf{.983}                               & .960                                            & \textbf{.965}                               & .919                                           \\
Sensitivity                   & \textbf{.944}       & .935                    & \textbf{.894}                               & .835                                            & \textbf{.874}                               & .768                                           \\
Specificity                   & \textbf{.981}       & .967                    & \textbf{.982}                               & .964                                            & \textbf{.985}                               & .964                                           \\
Timeliness                    & \textbf{1.48}       & 1.62                    & \textbf{1.90}                               & 2.53                                            & \textbf{1.47}                               & 2.47                                           \\ \hline
\end{tabular}
\end{table}

}

{\color{black}

\section{\color{black} Further Results for the Application to COVID-19 Outbreaks Across Quebec}

\subsection{\color{black} Difference in the posterior probability that an outbreak is currently happening between the Full Coupled and Non-coupled models averaged across all hospitals (SM Figure 5)}

Figure \ref{fig:spread_compare} shows the difference in the posterior probability that an outbreak is currently happening between the Full Coupled and Non-coupled models averaged across all hospitals and was discussed in Section 5.4 of the main text.

\begin{figure}[!t]
 	\centering
 	\includegraphics[width=\linewidth]{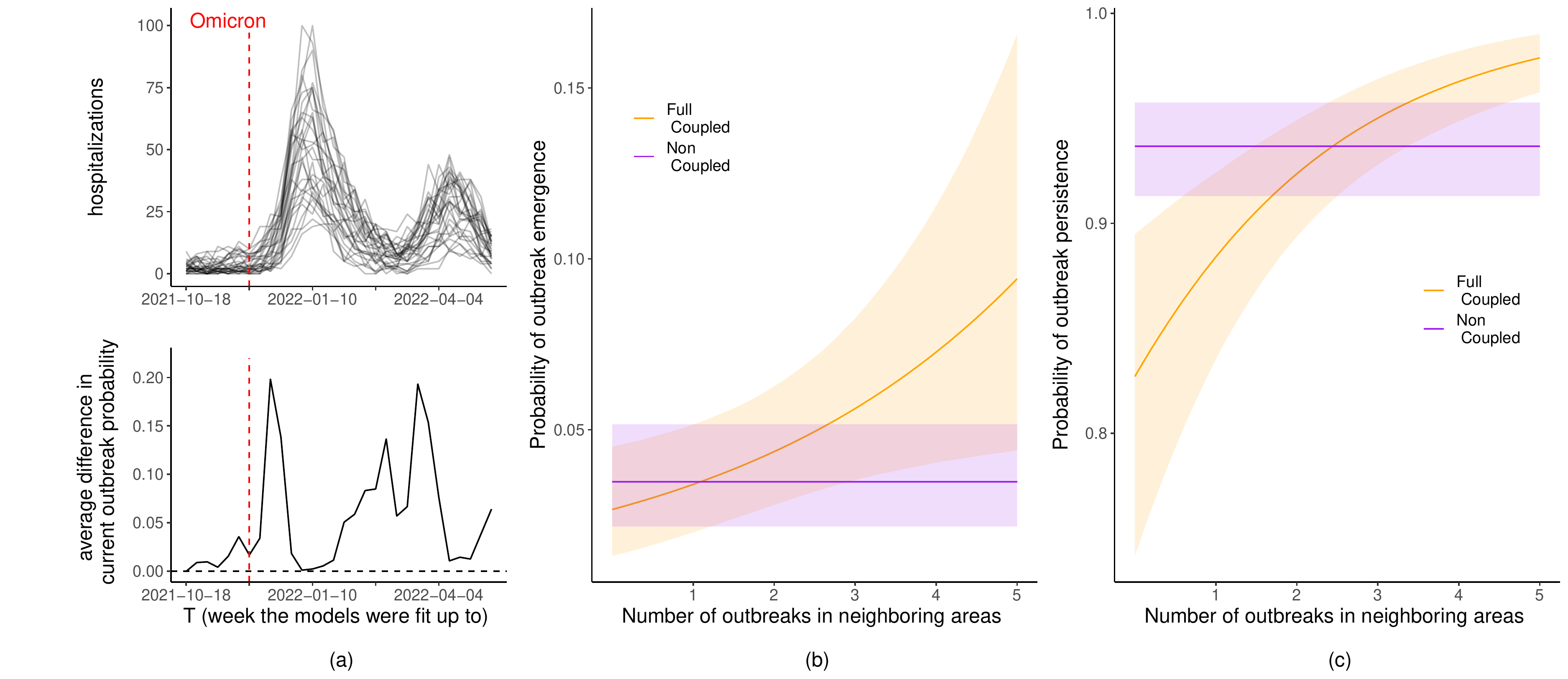}
	\caption{(a)(top) Each line gives the number of hospitalizations in one of the 30 hospitals included in the study. (a)(bottom) shows the difference in the posterior probability that an outbreak is currently happening between the Full Coupled and Non-coupled models averaged across all hospitals, that is,  $1/30\sum_{i=1}^{30}P(S_{iT}=3|\bm{y},\text{Full Coupled Model})-P(S_{iT}=3|\bm{y},\text{Non-coupled Model})${\color{black}, versus $T$}. (b) and (c) show posterior means (solid lines) and 95\% posterior credible intervals (shaded areas) of the probabilities of outbreak emergence, (b), and outbreak persistence, (c), versus the number of outbreaks in neighboring areas assuming average connectivity and other covariates fixed at their average values. For (b) and (c), the Full Coupled Model is in orange, the Non-coupled Model is in purple, and the models were fit up to $T=84$. \label{fig:spread_compare}}
\end{figure}
\subsection{\color{black}Map of the likely state in each catchment area during the last week of the study period (SM Figure 6)}

Figure \ref{fig:map1} shows a map of the likely state in each catchment area during the last week of the study period and was discussed in Section 5.4 of the main text.

\begin{figure}[!t]
 	\centering
 	\includegraphics[width=.75\linewidth]{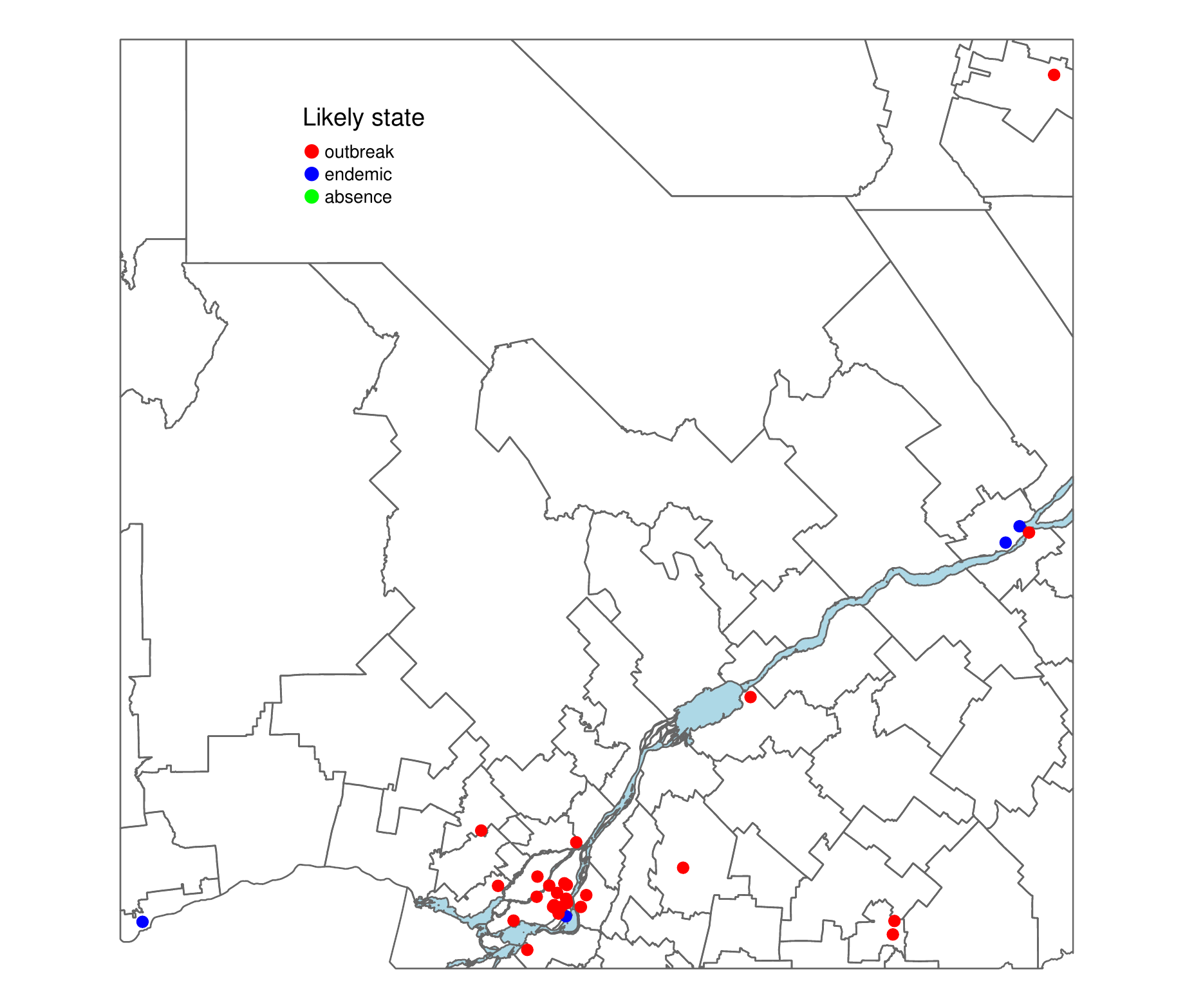}
	\caption{A map of the part of Quebec where the 30 hospitals (points) included in the study are located. Borders separate counties. The color of the points represents the likely state in the catchment area of the hospital during the last week of the study period, that is, red if $P(S_{iT}=3|\bm{y})>{\color{black}.5}$, blue if $P(S_{iT}=2 |\bm{y})>{\color{black}.5}$ and green if $P(S_{iT}=1|\bm{y})>{\color{black}.5}$, where $T=113=\text{2022-05-09}$. From the Full Coupled Model with new variant as a covariate. \label{fig:map1}}
\end{figure}

}
\subsection{\color{black}Analysis of real-time false alarms}

\begin{figure}[!t]
 	\centering
 	\includegraphics[width=\textwidth]{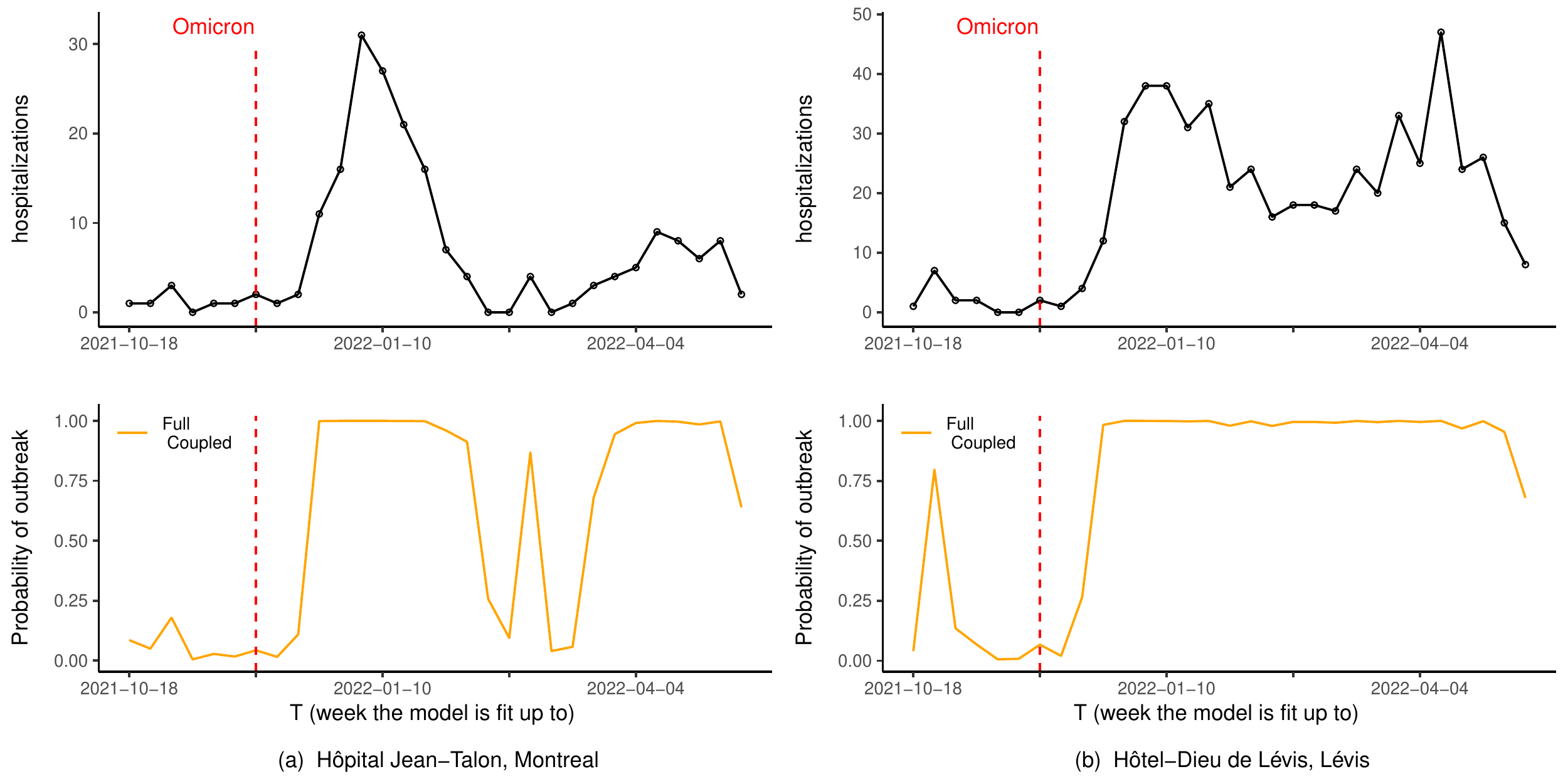} \vspace{-.5cm}
	\caption{{\color{black}Top graphs show the hospitalizations for the last 30 weeks of the study period where we conducted the real-time evaluation.} Bottom graphs: solid lines show the posterior probabilities that an outbreak is currently happening, that is, $P(S_{iT}=3|\bm{y})${\color{black}, versus $T$}.  {\color{black}The dotted red lines are drawn at the introduction
of the Omicron variant for all of Quebec.} Shows results from the Full Coupled Model for the two hospitals where we found evidence of false alarms during the real-time evaluation. \label{fig:SM_prosp_compare}}
\end{figure}

\begin{figure}[!t]
 	\centering
 	\includegraphics[width=\textwidth]{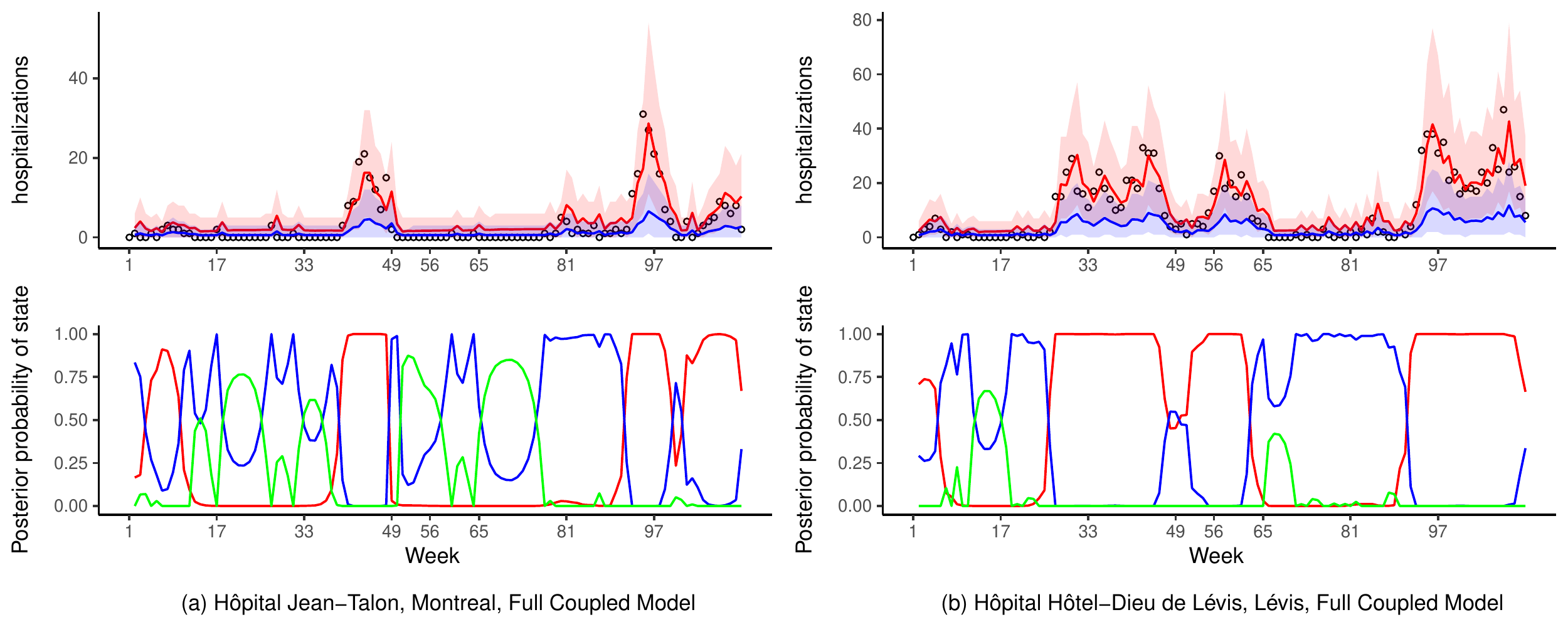} 
	\caption{Shows the retrospective state estimates from the Full Coupled Model, without new variant as a covariate, for the two hospitals where we found evidence of false alarms during the real-time evaluation. Follows the same graphical structure {\color{black}as Figure 5 from the main text.} \label{fig:SM_prosp_compare2}}
\end{figure}

Figure \ref{fig:SM_prosp_compare}, similar in structure to Figure 6 of the main text, shows the posterior probabilities that an outbreak is currently happening (bottom graphs) for the Full Coupled Model in the two hospitals where we found evidence of false alarms during our real-time evaluation. Recall in the real-time evaluation the model was fit up to time $T$ for $T=84=\text{2021-10-18},...,113=\text{2022-05-09}$ and then the summaries in Figure \ref{fig:SM_prosp_compare} were calculated for each $T$. We decided there was evidence of a false alarm being triggered at the time $T$ if $P(S_{iT}=3|\bm{y})>{\color{black}.5}$ and it did not appear that an outbreak had occurred at the time $T$ in hindsight. As can be seen in Figure \ref{fig:SM_prosp_compare}, in both hospitals there is a sharp increase in $P(S_{iT}=3|\bm{y})$ that appears to not be associated with any outbreaks. The model corrects itself quickly, however, with $P(S_{iT}=3|\bm{y})$ typically declining sharply 1-2 weeks after the false alarm. Figure \ref{fig:SM_prosp_compare2}, following the same graphical structure as {\color{black}Figure 5 in the main text}, shows the retrospective state estimates in the same two hospitals as Figure \ref{fig:SM_prosp_compare}. Note, we did not include new variant as a covariate in the model used to produce Figure \ref{fig:SM_prosp_compare2} so that the model would be the same as the one used in the real-time evaluation.  The false alarms do not show up in the retrospective state estimates, again showing that the model corrects itself after gathering further data. We found no evidence in any areas of false alarms in the retrospective state estimates.

As to the cause of the false alarms in Figure \ref{fig:SM_prosp_compare}, the risk of an outbreak was high in both hospitals when the false alarms were triggered. In Hôpital Jean-Talon mobility was high and in Hôpital Hôtel-Dieu de Lévis there was strong evidence of outbreaks in several neighboring areas. Compared to the Non-coupled Model (not shown), the same false alarm was triggered in Hôpital Hôtel-Dieu de Lévis but not in Hôpital Jean-Talon. This is likely because outbreak risk was elevated in Hôpital Jean-Talon, when the false alarm was triggered, due to evidence of outbreaks in neighboring areas. In this case, a single false alarm seems worth the tradeoff for the earlier warning of the Omicron outbreak across the hospitals.

\subsection{{\color{black}Real-time predictions of the hospitalizations}}

{\color{black}The real-time predictions in Figure 6 of the main text were obtained by fitting the Non-coupled and Full Coupled models up to time $T-1$, for $T=85,\dots,113$, and then running Algorithm \ref{alg:alg1} to draw from the posterior predictive distribution $p(y_{iT}|\bm{y}_{1:(T-1)})$. The EE model was also fit up to time $T-1$, for $T=85,\dots,113$, and a similar simulation method was used to produce the one-week ahead predictions in Figure \ref{fig:EEvsFC} below.

\begin{figure}[!t]
 	\centering
 	\includegraphics[width=\textwidth]{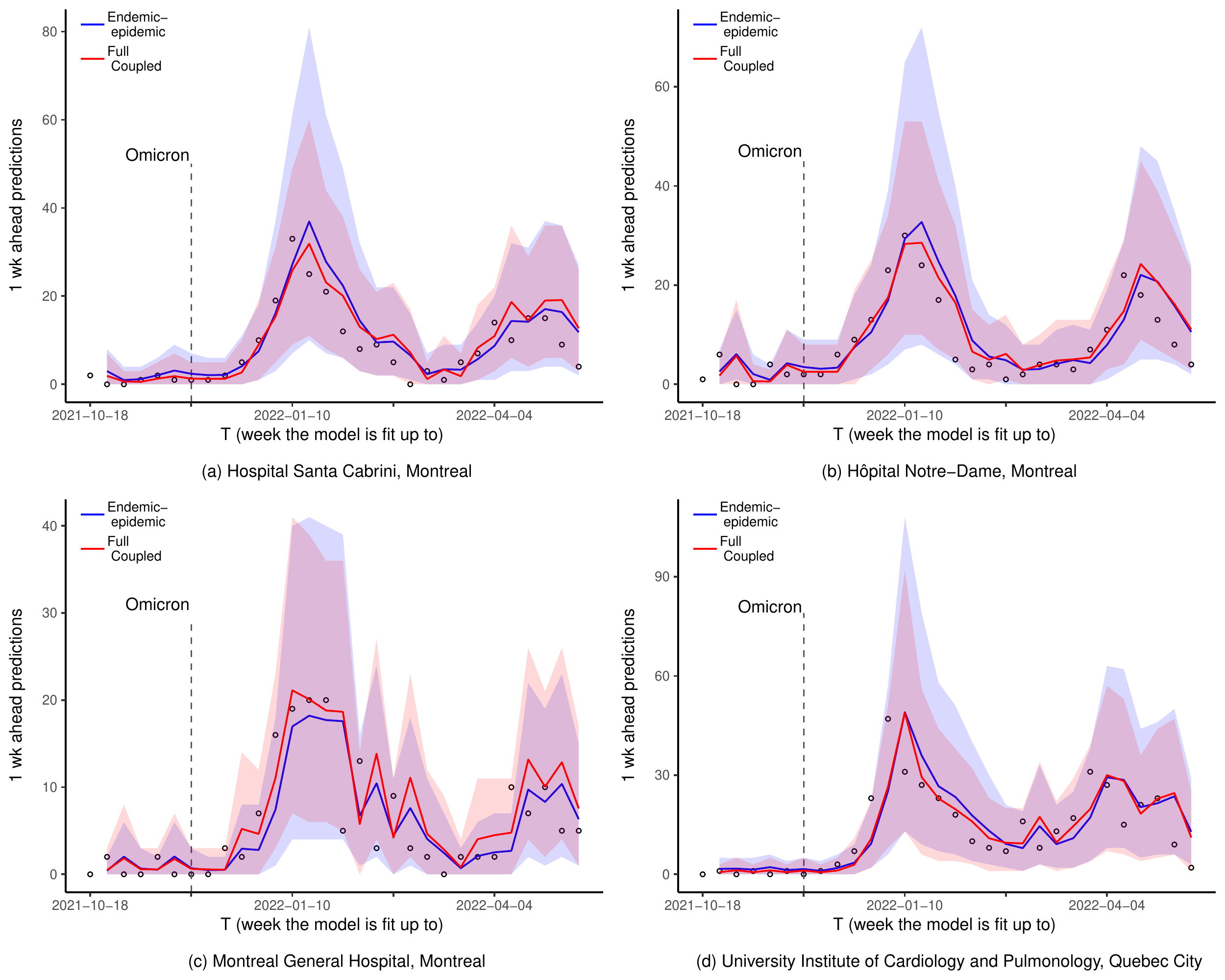} 
	\caption{{\color{black}Shows the posterior means (solid lines) and 95\% posterior credible
intervals (shaded areas) of the one-week ahead posterior predictive distributions, that is, $p(y_{iT}|\bm{y}_{1:(T-1)})$, versus $T$. The Full Coupled Model is in {\color{black} red} and the EE Model is in {\color{black}blue}. The {\color{black}dashed grey} lines are drawn at the introduction
of the Omicron variant for all of Quebec.\label{fig:EEvsFC}}}
\end{figure}

Suppose one of the models is fit up to time $T-1$ and we wish to evaluate the one-week ahead predictions for time $T$. The multivariate log score \citep{bracher_endemic-epidemic_2022} is given by,
\begin{align} \label{eqn:multiLog}
\text{MultiLogS}_{T}&=\frac{-\log\left(p(\bm{y}_{T}^{\text{obs}}|\bm{y}_{1:(T-1)})\right)}{N},
\end{align} where $\bm{y}_{T}^{\text{obs}}=(y_{1T}^{\text{obs}},\dots,y_{NT}^{\text{obs}})^T$ is the vector of all observed hospitalizations at time $T$ and $N=30$ in our example. A lower score indicates superior predictive performance. The multivariate log score is a strictly proper scoring rule \citep{czado_predictive_2009}, meaning it is expected to be minimized if and only if the predictive distribution is equal to the true data-generating distribution. For the Full Coupled and Non-coupled models, we can approximate $p(\bm{y}_{T}^{\text{obs}}|\bm{y}_{1:(T-1)})$ as,
\begin{align*}
p(\bm{y}_{T}^{\text{obs}}|\bm{y}_{1:(T-1)})& \approx \frac{1}{Q-M}\sum_{m=M+1}^{Q} p(\bm{y}_{T}^{\text{obs}}|\bm{S}_T^{[m]},\bm{\beta}^{[m]},\bm{y}_{T-1}) \\ 
&= \frac{1}{Q-M}\sum_{m=M+1}^{Q} \prod_{i=1}^{N}p(y_{iT}^{\text{obs}}|S_{iT}^{[m]},\bm{\beta}^{[m]},\bm{y}_{T-1}) \\
&= \frac{1}{Q-M}\sum_{m=M+1}^{Q} \exp\left(\sum_{i=1}^{N}\log\left(p(y_{iT}^{\text{obs}}|S_{iT}^{[m]},\bm{\beta}^{[m]},\bm{y}_{T-1})\right)\right),
\end{align*} and then substitute this approximation into Equation (\ref{eqn:multiLog}) to approximate $\text{MultiLogS}_{T}$. The multivariate log score for the EE model can be approximated similarly. Note we can use Algorithm \ref{alg:alg1} to obtain the draws $S_{iT}^{[m]}$.

Figure \ref{fig:multiLog} below shows the multivariate log scores, $\text{MultiLogS}_{T}$, across $T=85,\dots,113$ for the three models. Recall, that a lower score indicates superior predictive performance. From the figure, the Full Coupled Model better captured the emergence of the initial Omicron wave, while the EE model better captured the trough between the two waves and the decline of the second wave. Finally, we can calculate the mean scores, which are also strictly proper \citep{bracher_endemic-epidemic_2022}, by averaging $\text{MultiLogS}_{T}$ across $T=85,\dots,113$ for each model, see Section 5.4 of the main text. It is important to note that the mean scores only measure the average forecasting performance across the entire Omicron wave. Some models may better capture specific aspects of the wave such as the emergence or decline, as discussed above. Therefore, the mean scores may miss important local differences in forecasting performance and it is also important to examine the scores across time as in Figure \ref{fig:multiLog}.

\begin{figure}[!t]
 	\centering
 	\includegraphics[width=\textwidth]{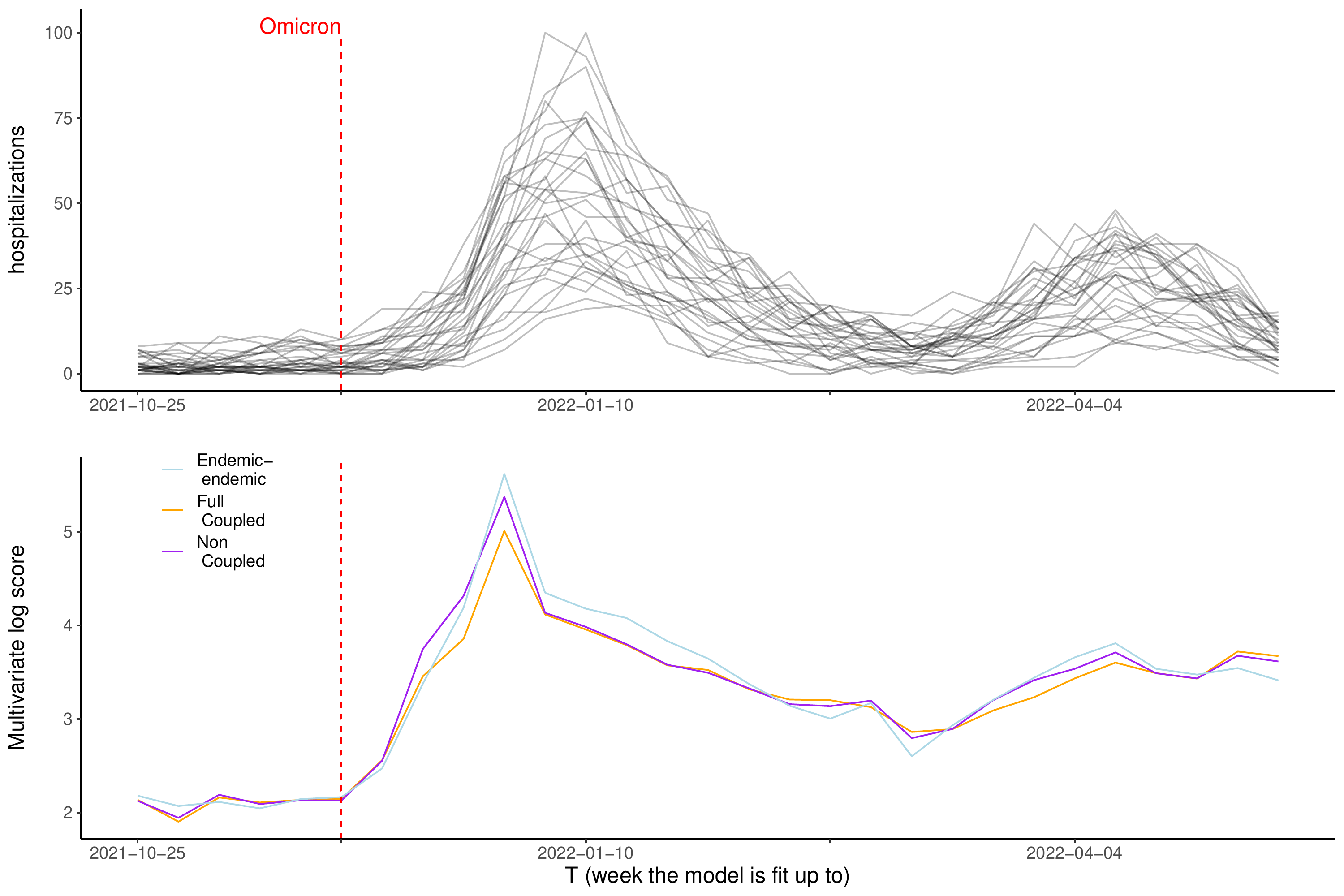} 
	\caption{{\color{black} Each line in the top graph gives the number of hospitalizations in one of the 30 hospitals
included in the study. The bottom graph shows the multivariate log scores, that is, $\text{MultiLogS}_{T}$, versus $T$. The Full Coupled Model is in orange, the Non-coupled Model is in purple and the EE model is in light blue. The dotted red lines are drawn at the introduction
of the Omicron variant for all of Quebec.\label{fig:multiLog}}}
\end{figure}

}

\clearpage

\section*{Acknowledgements}

This research was part of a larger project of INESSS (Institut national d'excellence en santé et en services sociaux), whose objective was to produce bed occupancy projections for COVID-19 patients. The access to data was made possible through a tripartite agreement between the MSSS, the RAMQ and INESSS. This work is part of the PhD thesis of D. Douwes-Schultz under the supervision of A. M. Schmidt in the Graduate Program of Biostatistics at McGill University, Canada. Douwes-Schultz is grateful for financial support from IVADO and the Canada First Research Excellence Fund/Apogée (PhD Excellence Scholarship 2021-9070375349). Schmidt is grateful for financial support from the Natural Sciences and Engineering Research Council (NSERC) of Canada (Discovery Grant RGPIN-2017-04999) and Institut de valorisation des données (IVADO) (PRF-2019-6839748021). Shen is grateful for financial support from the Fonds de recherché du Québec–Santé (FRQS) (Doctoral training award 313602). Buckeridge is supported by a Canada Research Chair in Health Informatics and Data Science (950-232679). This research was enabled in part by support provided by Calcul Québec (www.calculquebec.ca) and Compute Canada (www.computecanada.ca).

 \clearpage

\bibliography{supplement}